\documentclass[twoside,12pt]{article}
\usepackage{epsfig}
\usepackage{epsf}
\usepackage{graphics}

\newcommand{\beq}{\begin{equation}}
\newcommand{\eeq}{\end{equation}}
\newcommand{\bea}{\begin{eqnarray}}
\newcommand{\eea}{\end{eqnarray}}
\newcommand{\non}{\nonumber}
\renewcommand{\d}{\delta}
\renewcommand{\l}{\lambda}
\renewcommand{\L}{\Lambda}
\renewcommand{\b}{\beta}
\renewcommand{\a}{\alpha}

\renewcommand{\ni}{\noindent}
\newcommand{\xp}{\vec{x}_{\perp}}
\newcommand{\n}{\nu}
\newcommand{\m}{\mu}
\newcommand{\D}{\Delta}
\newcommand{\p}{\phi}
\renewcommand{\r}{\rho}
\newcommand{\s}{\sigma}
\renewcommand{\S}{\Sigma}
\renewcommand{\th}{\theta}
\newcommand{\tth}{\widetilde{\theta}}
\newcommand{\E}{{\cal E}}
\newcommand{\A}{{\cal A}}
\newcommand{\tA}{\widetilde{A}}
\newcommand{\e}{\epsilon}

\newcommand{\oh}{\frac{1}{2}}
\newcommand{\dg}{\dagger}
\renewcommand{\t}{\tau}
\newcommand{\rf}[1]{(\ref{#1})}
\newcommand{\ra}{\rightarrow}
\newcommand{\pa}{\partial}
\newcommand{\rra}{\right\rangle}
\newcommand{\lla}{\left\langle}
\newcommand{\tr}{{\rm Tr}\,}
\newcommand{\U}{{\cal U}}

\newcommand{\tU}{\widetilde{U}}
\newcommand{\V}{{\cal V}}
\newcommand{\tW}{\widetilde{W}}
\newcommand{\cD}{{\cal D}}
\begin{document}

\title{The Confinement Problem in Lattice Gauge Theory}
\author{J.\ Greensite $^{1,2}$ \\ \\
$^1$Physics and Astronomy Department \\
San Francisco State University \\
San Francisco, CA 94132 USA \\ \\
$^2$Theory Group, Lawrence Berkeley National Laboratory \\
Berkeley, CA 94720 USA}

\maketitle
\begin{abstract}

  I review investigations of the quark confinement mechanism
that have been carried out in the framework of SU(N) lattice gauge theory.
The special role of $Z_N$ center symmetry is emphasized.

\end{abstract}

\section{Introduction}

   When the quark model of hadrons was first introduced by Gell-Mann and
Zweig in 1964, an obvious question was ``where are the quarks?''.
At the time, one could respond that the quark model was simply a
useful scheme for classifying hadrons, and the notion of quarks as
actual particles need not be taken seriously.  But the absence of
isolated quark states became a much more urgent issue with the
successes of the quark-parton model, and the introduction, in 1972, of
quantum chromodynamics as a fundamental theory of hadronic physics \cite{QCD}.
It was then necessary to suppose that, somehow or other, the dynamics
of the gluon sector of QCD contrives to eliminate free quark states
from the spectrum.

   Today almost no one seriously doubts that quantum chromodynamics
confines quarks.  Following many theoretical suggestions in the late
1970's about how quark confinement might come about, it was finally
the computer simulations of QCD, initiated by Creutz \cite{Creutz}
in 1980, that persuaded most skeptics.
Quark confinement is now an old and familiar
idea, routinely incorporated into the standard model and all its
proposed extensions, and the focus of particle phenomenology shifted
long ago to other issues.

   But familiarity is not the same thing as
understanding.  Despite efforts stretching over thirty years, there
exists no derivation of quark confinement starting from first
principles, nor is there a totally convincing explanation of the
effect.  It is fair to say that no theory of quark confinement is
generally accepted, and every proposal remains controversial.

   Nevertheless, there have been some very interesting developments
over the last few years, coming both from string/M-theory and from
lattice investigations.  On the string/M-theory side there is an
emphasis, motivated by Maldacena's AdS/CFT conjecture \cite{AdS},
on discovering
supergravity configurations which may provide a dual description of
supersymmetric Yang-Mills theories.  While the original AdS/CFT work
on confinement was limited to certain supersymmetric theories at
strong couplings, it is hoped that these (or related) ideas about duality
may eventually transcend those limitations.
Lattice investigations, from the point of view
of hadronic physics, have the advantage of being directly aimed
at the theory of interest, namely QCD.  On the lattice theory side, the
prevailing view is that quark confinement is the work of some special
class of gauge field configurations $-$ candidates have included instantons,
merons, abelian monopoles, and center vortices $-$ which for some
reason dominate the QCD vacuum on large distance scales.  What is new
in recent years is that algorithms have been invented which can
locate these types of objects in thermalized lattices, generated by
the lattice Monte Carlo technique.  This is an important development,
since it means that the underlying mechanism of quark confinement, via
definite classes of field configurations, is open to numerical
investigation.

   Careful lattice simulations have also confirmed, in recent years,
properties of the static quark potential which were only speculations
in the past.  These properties have to do with the color group
representation dependence of the string tension at intermediate and
asymptotic distance scales, and the ``fine structure'' (i.e.\
string-like behavior) of the QCD flux tube.  Taken together, such
features are very restrictive, and must be taken into account by any
proposed scenario for quark confinement.

   This article reviews some of the progress towards understanding
confinement that has been made over
the last several
years, in which the lattice formulation has played an important
role.  An underlying theme is the special significance of
the center of the gauge group, both in constructing relevant order
parameters, and in identifying the special class of field
configurations which are responsible for the confining force.
I will begin by focusing on what is actually known
about this force, either from numerical experiments or from
convincing theoretical arguments,  and
then concentrate on the two proposals which have received the most
attention in recent years: confinement by $Z_N$ center vortices, and
confinement by abelian monopoles.  Towards the end I will also
touch on some aspects of confinement in Coulomb gauge, and in
the large $N_c$ limit.  I regret that I do not have space here to cover
other interesting proposals, which may lie somewhat outside the mainstream.

\section{What is Confinement?}

   First of all, what \emph{is} quark confinement?

   The place to begin is with an experimental
result: the apparent absence of free quarks in Nature.  Free quark
searches are basically searches for particles with fractional
electric charge \cite{Lyons,Roger}, and
the term ``quark confinement'' in this context
is sometimes equated with the absence of free (hadronic) particles with
electric charge $\pm {1\over 3}e$ and $\pm {2 \over 3}e$.
But this experimental fact,
from a modern perspective, is not so very fundamental.
It may not even be true.  Suppose that Nature had
supplied, in addition to the usual quarks, a massive scalar field
in the {\bf 3} representation of color SU(3), having otherwise the
quantum numbers of the vacuum.  In that case there
would exist bound states of a quark and a massive scalar, which
together would have the flavor quantum numbers of the quark alone.
If the scalar were
not too massive, then fractionally charged particles
would have turned up
in particle detectors (and/or Millikan oil-drops)
many years ago.
It is conceivable (albeit unlikely) that such a scalar field
really exists, but that its mass is on the order of hundreds of GeV
or more.  If so, particles with fractional electric charge await
discovery at some future detector.

    Of course, despite the fractional
electric charge, bound state systems of this sort
hardly qualify as free quarks, and the discovery of such
heavy objects would not greatly change
prevailing theoretical ideas about non-perturbative QCD.
So the term ``quark confinement'' must mean
something more than just the absence of isolated objects with fractional
electric charge.  A popular definition is based on the fact that
all the low-lying hadrons fit nicely into a scheme in which the
constituent quarks combine in a color-singlet.  There is also no evidence
for the existence of isolated gluons, or any other particles in the spectrum,
in a color non-singlet state.  These facts suggest
identifying quark confinement with the more general concept of
color confinement, which means that
\begin{center}
   \textbf{\textit{There are no isolated particles in Nature
with non-vanishing color charge}}
\end{center}
\non i.e.\ all asymptotic particle states are color singlets.

   The problem with this definition of confinement is that it confuses
confinement with color screening, and applies
equally well to spontaneously broken gauge theories,
where there is not supposed to be any such thing
as confinement.  The definition even implies the ``confinement'' of electric
charge in a superconductor or a plasma.

   Imagine introducing some Higgs fields in the {\bf 3} representation
into the QCD Lagrangian, with couplings chosen such that all
gluons acquire a mass at tree level. In standard (but somewhat
inaccurate \cite{Elitzur}) terminology, this is characterized as
complete spontaneous symmetry breaking of the SU(3) gauge
invariance. As in the electroweak theory, where there exist
isolated leptons, W's, and Z bosons, in the broken SU(3) case the
spectrum will contain isolated quarks and massive gluons.  But
does this really mean that there is no color confinement,
according to the definition above? Note that whether the symmetry
is broken or not, the non-abelian Gauss law is \beq
    \vec{\nabla} \cdot \vec{E}^a = -c^{abc} A^b_k E^c_k -
       i {\pa {\cal L} \over \pa D^0 \p} t^a \p
\label{Gauss}
\eeq
where $c^{abc}$ are the structure constants, $\{ t^c \}$ the
SU(3) group generators, and $\phi$ represents the Higgs and any other
matter fields. The right-hand side of this equation happens to be the zero-th
component of a conserved Noether current
\beq
      j_\n^a =  -c^{abc} A^b_\m F^{c\m}_\n -
       i {\pa {\cal L} \over \pa D^\n \p} t^a \p
\label{Noether}
\eeq
and therefore the integral of $j_0^a$ in a volume $V$ can be identified
as the charge in that volume.  Just as in the abelian theory, the
integral of the electric field at the surface of $V$ measures the
charge $Q^a$ inside:
\beq
      \int_{\pa V} \vec{E}^a \cdot d\vec{S} = Q^a
\label{integralgauss}
\eeq
Now suppose that there is a quark in volume $V$, far from the
surface.  Since the gluons are massive, the color electric field due to the
quark falls
off exponentially from the source, and the charge in volume $V$,
as determined by the Gauss law, is essentially zero.  So the isolated
quark state, if viewed from a distance
much greater than the inverse gluon mass,
appears to be a color singlet.

   What is going on here is that the color charge of the quark source is
completely cancelled out by contributions due to the Higgs fields
and the gauge field surrounding the source.  The very same effect is found
in the abelian Higgs model, which is a relativistic generalization
of the Landau-Ginzburg superconductor.  Here too, the Higgs condensate
rearranges itself to screen the electric charge of any source.
Because of this screening effect there are no isolated particles,
in the spectrum of a spontaneously broken gauge theory,
which are charged with respect to a generator of the broken symmetry.

\begin{figure}[t!]
\begin{center}
\begin{minipage}[t]{8 cm}
\centerline{\scalebox{0.5}{\includegraphics{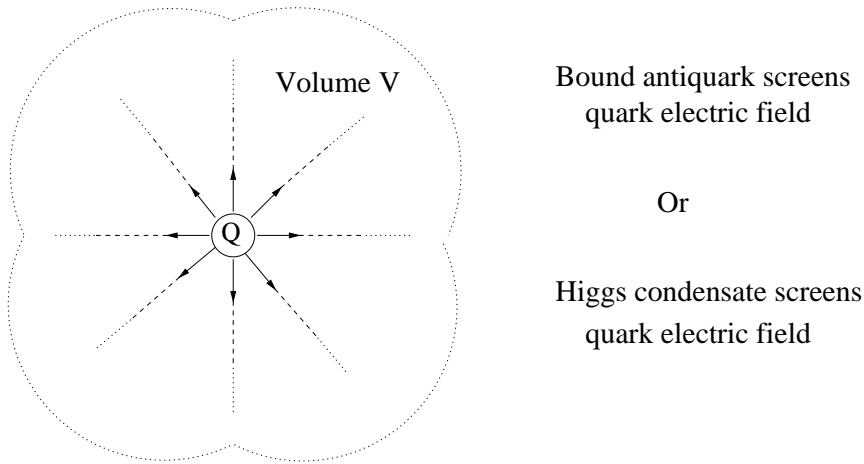}}}
\end{minipage}
\begin{minipage}[t]{16.5 cm}
\caption{In both QCD with dynamical quarks, and in a theory with
 complete spontaneous symmetry breaking, the field of a static color
 source $Q$ in the fundamental representation is completely screened.}
\label{Qscreen}
\end{minipage}
\end{center}
\end{figure}

  For QCD, the situation is as depicted in Fig.\ \ref{Qscreen}.  We consider
a heavy quark $Q$ sitting at the origin.  In ordinary, ``unbroken''
QCD, the color electric field of the heavy quark is absorbed by
a light antiquark,  or any other set of quarks or colored-charged
scalars forming a
color singlet bound state with the heavy quark.  In the
spontaneously broken case, where the gauge group is completely
broken, the charge of the heavy quark is shielded by
the compensating charge of the gauge and condensate fields.  In
either case the total color charge in volume $V$ is zero;
there are no asymptotic particles in the spectrum with non-vanishing
color charge.

   The similarity of the broken and unbroken gauge theories is not an accident.
Fradkin and Shenker \cite{FS}, in 1979, considered a lattice model
\bea
   -S_{FS} &=& \b_G \sum_x \sum_{\m>\n} \Bigl\{\tr[U_\m(x) U_\n(x+\widehat{\m})
              U^\dg_\m(x+\widehat{\n}) U^\dg_\n(x)] + \mbox{~c.c.} \Bigr\}
\non \\
   &+& \b_H \sum_x \sum_\m \Bigl\{ \p^\dg(x) U_\m(x) \p(x+\widehat{\m})
   + \mbox{~c.c.} \Bigr\}  ~~~,~~~ |\p|=1
\label{Fradkin}
\eea
which interpolates between the Higgs ($\b_H,\b_G \ra \infty$)
and the confinement ($\b_H,\b_G \ra 0$) limits.
They were able to show that for a Higgs field in the fundamental
representation of the gauge group, the two coupling regions are continuously
connected, rather than being separated everywhere in coupling space
by a phase boundary.  That is consistent with the fact that there
are only color singlet asymptotic states in both regimes.

    The absence of a phase separation may seem paradoxical, if we imagine
that the gauge symmetry is spontaneously broken at large
$\b_H,\b_G$ and restored at small $\b_H,\b_G$.  There is no real
paradox, of course, but unfortunately the phrase ``spontaneous
breaking of gauge symmetry'', although deeply embedded in the
lexicon of modern particle physics, is a little misleading.  In
fact there is no such thing as the spontaneous breaking of a
\emph{local} gauge symmetry in quantum field theory, according to
a celebrated theorem by Elitzur \cite{Elitzur}.  However, after
removing the redundant degrees of freedom by some choice of gauge,
there typically remain some \emph{global} symmetries, such as a
global center symmetry, which may or may not be spontaneously
broken. In the case that there are matter fields in the
fundamental representation of the gauge group, there is no
residual center symmetry, and the Fradkin-Shenker result assures
us that there is no symmetry breaking transition of any kind which
would serve to separate a Higgs from a confining phase.   The
notion that confining and Higgs physics, in theories with
fundamental matter fields, are separated by a symmetry breaking
transition serves only as a ``convenient fiction'' \cite{Wilcek}.

   The fiction is convenient because the spectrum of theories with
``broken'' gauge symmetry is
qualitatively very different from the QCD spectrum.  This qualitative
difference is not captured very well by the notion of color confinement
as it was defined above, because that property is found in both the Higgs
and the ``confining'' coupling regions.  What really distinguishes
QCD from a Higgs
theory with light quarks is the fact that meson states in QCD fall
on linear, nearly parallel, Regge trajectories.  This is a truly
striking feature of the hadron spectrum, it is not found in bound-state
systems with either Coulombic or Yukawa attractive forces, and it needs
to be explained.

\subsection{Regge Trajectories and Color Fields}

\begin{figure}[tb]
\begin{center}
\begin{minipage}[t]{8 cm}
\epsfig{file=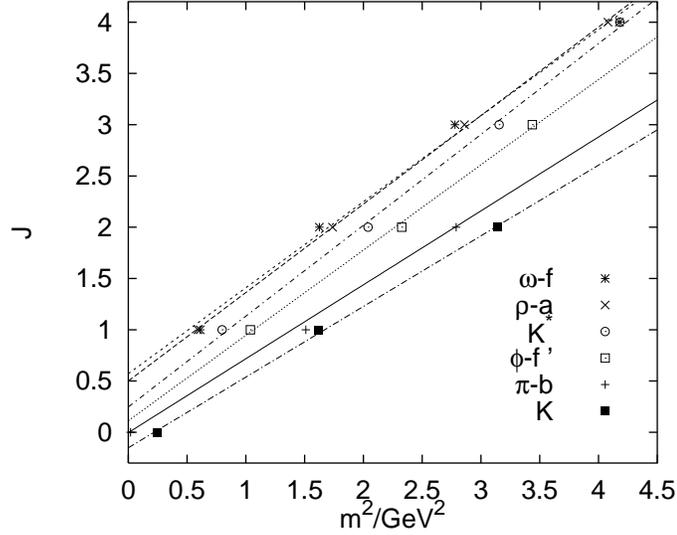,scale=0.5}
\end{minipage}
\begin{minipage}[t]{16.5 cm}
\caption{Regge trajectories for the low-lying mesons (figure from
 Bali, ref.\ \cite{Bali0}).}
\label{regge}
\end{minipage}
\end{center}
\end{figure}

    When the spin $J$ of mesons is plotted as a function of squared meson
mass $m^2$, it turns out that the resulting points can be sorted into
groups which lie on straight lines, and that the slopes of these lines are
nearly the same, as shown in Fig.\ \ref{regge}.
These lines are known as ``linear Regge trajectories,'' and the particles
associated with a given line all have the same flavor quantum numbers.
Similar linear trajectories are found for the baryons, out as far
as $J=11/2$.

   This remarkable feature of hadron phenomenology can
be reproduced by a very simple model.  Suppose that a meson
consists of a straight line-like object with a constant energy density $\s$
per unit length, having a nearly massless quark at one end of the
line, and a nearly massless antiquark at the other.  The quark and
antiquark carry the flavor quantum numbers of the system, and move
at nearly the speed of light.  For a straight line of length $L=2R$,
whose ends rotate at the speed of light, the energy of the system is
\bea
       m &=& E = 2 \int_0^R {\s dr \over \sqrt{1 - v^2(r)}}
\non \\
         &=&  2 \int_0^R {\s dr \over \sqrt{1 - r^2/R^2}}
\non \\
         &=& \pi \s R
\eea
while the angular momentum is
\bea
       J &=& 2 \int_0^R {\s r v(r)dr \over \sqrt{1 - v^2(r)}}
\non \\
         &=& {2\over R} \int_0^R {\s r^2 dr \over \sqrt{1 - r^2/R^2}}
\non \\
         &=& \oh \pi \s R^2
\eea
Comparing $m$ and $J$, we find that
\beq
       J = {m^2 \over 2 \pi \s}
\eeq
which means that this very simple model has caught the essential feature,
namely, a linear relationship between $m^2$ and $J$.
From the particle data, the slope of the Regge trajectories
is approximately
\beq
        \a' = {1\over 2\pi \s} \approx 0.9 ~ \mbox{GeV}^{-2}
\eeq
implying an energy/unit length of the line between the quarks, which
is known as the ``string tension'', of magnitude
\beq
         \s \approx 0.18 ~ \mbox{GeV}^2 \approx 0.9 ~ \mbox{GeV/fm}
\eeq

  Of course, the actual Regge trajectories don't intercept the x-axis
at $m^2=0$, and the slopes of the different trajectories
are slightly different, as can be seen
from Fig.\ \ref{regge}.  But the model can also be modified
by allowing for finite quark masses.  Note that since a crucial aspect of
the model is that the quarks  move at (nearly) the speed of light, the
low-lying heavy quark states (charmonium,
``toponium'', etc.), composed of the $c,t,b$ quarks, would not be expected
to lie on linear Regge trajectories.  Another way of making the model
more realistic would be to allow for quantum fluctuations of
the line-like object in directions transverse to the line. Those
considerations lead to
(and in fact inspired) the formidable subject of string theory \cite{string}.

   QCD can be made agree with the simple phenomenological
model if, for some reason,
the electric field diverging from a quark is collimated into a
flux tube of fixed cross-sectional area.  In that case
the string tension is simply
\beq
      \s = \int d^2 x_\perp ~ \oh \vec{E}^a \cdot \vec{E}^a
\eeq
where the integration is in a plane between the quarks,
perpendicular to the axis of the flux tube.
The problem is to explain why the electric field between a quark and
antiquark pair should be collimated in this way, instead of spreading
out into a dipole field, as in electrodynamics, or simply petering
out, as in a spontaneously broken theory.

\begin{figure}[htb]
\begin{center}
\begin{minipage}[t]{8 cm}
\centerline{\scalebox{0.4}{\includegraphics{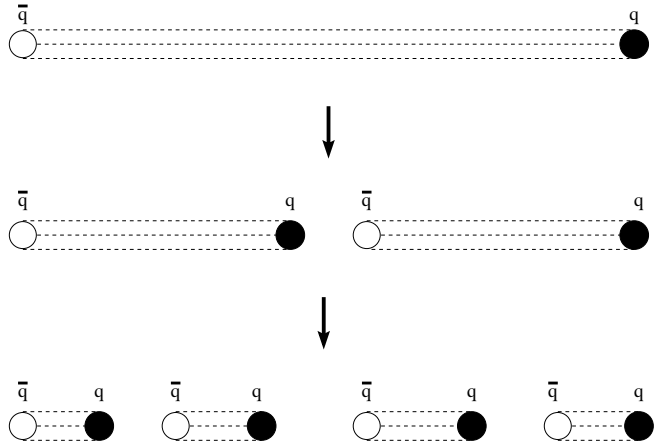}}}
\end{minipage}
\begin{minipage}[t]{16.5 cm}
\caption{String breaking by quark-antiquark pair production.}
\label{break}
\end{minipage}
\end{center}
\end{figure}

   In fact, as already emphasized, the color electric field of a quark
or any other color charge source \emph{does} peter out,
eventually. If a heavy quark and antiquark were suddenly separated
by a large distance (compared to usual hadronic scales), the
collimated electric field between the quarks would not last for
long. Instead the color electric flux tube will decay into states
of lesser energies by a process of ``string breaking'' (Fig.\
\ref{break}), which can be visualized as production of light
quark-antiquark pairs in the middle of the flux tube, producing
two or more meson states. The color field of each of the heavy
quarks is finally screened by a bound light quark, as indicated in
Fig.\ \ref{Qscreen}. This process also accounts for the
instability of excited particle states along Regge trajectories.

   Pair production, however, is suppressed if all quarks are very massive.
Suppose the lightest quark has mass $m_q$.  Then the energy of a flux
tube state between nearly static quarks will be approximately $\s L$,
while the mass of the pair-produced quarks associated with string-breaking
will be at least $2m_q$.  This means that the flux tube states will be
stable against string breaking up to quark separations of approximately
\beq
        L = {2 m_q \over \s}
\eeq
This brings us to the following statement of the problem we are
interested in: \\

\ni $\underline{\mbox{\bf The Confinement Problem (I)}}$ \\

  {\sl Show that in the limit that the masses of all quarks go to
infinity, the work required to increase the quark separation in
a quark-antiquark system by a distance $L$
approaches $\s L$ asymptotically, where $\s$ is a constant.} \\

\subsection{A First Encounter with the Center}

    It seems that in order to define ``quark confinement,''
we must effectively remove quarks as dynamical objects from the gauge theory.
In the infinite mass limit, of course, quarks do not contribute to
any virtual process, including string-breaking processes.
But the exclusion of quarks doesn't mean that
\emph{all} matter fields must be removed, or taken to the infinite mass
limit; we have only to exclude those
matter fields which can give rise to string breaking.  The criterion
here is group-theoretical:  If it is not possible for an individual quark
and some number of matter field quanta to form a color singlet, then
it is also not possible for the matter field to give rise to string-breaking.
Suppose, for example, that QCD included some scalar fields in the color
{\bf 8} representation.  It is not possible for a particle or particles in
the {\bf 8} representation to combine with a quark in the {\bf 3} or
antiquark in the $\overline{\mbox{\bf 3}}$
 representation to form a color singlet,
and therefore there can be no string breaking of the type shown in
Fig.\ \ref{break}.

   On the other hand, a set of Higgs field in the {\bf 8}
representation of color SU(3) can still break the symmetry in such
a way that all the gluons acquire a mass.  In that case only a
finite amount of work is required to separate two massive quarks
by an arbitrary distance, even in the $m_q \ra \infty$ limit.  QCD
with finite-mass matter fields in the {\bf 3} representation is
therefore quite different from QCD with finite-mass matter fields
in the {\bf 8} representation.  In the former case, according to
the Fradkin-Shenker result, there is no phase transition from the
Higgs phase to a confining phase, and the work required to
separate quarks by an infinite distance always has a finite limit.
In the latter situation, there can exist a true phase transition
between a confining phase, and a non-confining Higgs phase.  Which
phase is actually realized is a dynamical issue, and will depend
on the shape of the Higgs potential.

   Matter fields in color representations which cannot give rise to
string-breaking, such as the color {\bf 8} representation of SU(3),
are known as fields of zero {\bf N-ality}.  N-ality, as it is referred
to in the physics literature, or the representation ``class'', as it
is known in the mathematical literature, refers to the transformation properties
of a Lie group representation with respect to gauge group center.  We recall
that the {\bf center} of a group refers to that set of group elements which
commute with all other elements of the group.  For an SU(N) gauge group,
the center elements consist of all $g\in SU(N)$
proportional to the $N\times N$ unit matrix, subject to the condition that
$\det(g)=1$.  This is the set of $N$ SU(N)elements group elements $\{z_n\}$
\beq
       z_n  = \exp\left({2\pi i n \over N}\right)
             \left[ \begin{array}{cccccc}
             1 &   &   &   &   &  \cr
               & 1 &   &   &   &  \cr
               &   & . &   &   &  \cr
               &   &   & . &   &  \cr
               &   &   &   & . &  \cr
               &   &   &   &   & 1 \end{array} \right]
       ~~~~ (n=0,1,2,...,N-1)
\eeq
These center elements form a discrete abelian subgroup known as $Z_N$.

   Although there are an infinite number of representations of SU(N),
there are only a finite number of representations of $Z_N$, and every
representation of $SU(N)$ falls into one of $N$ subsets, depending on
the representation of the $Z_N$ subgroup in the given representation.
Each representation in
a given subset has the same N-ality, which is an integer $k$ defined
as the number of boxes in the corresponding Young tableau, mod N.
Transformation by $z_n \in Z_N$, for each representation of N-ality $k$,
corresponds to multiplication by a factor $\exp({2\pi i k n \over N})$.
Group representations of N-ality $k=0$ are special, in that all center
transformations are mapped to the identity.

   It is easy to see that a quark of non-zero N-ality can never form a
color singlet by binding to one or more particles of zero N-ality.
According to the usual rules of group theory, the possible
irreducible color representations of the bound states are formed
by combining the boxes in the Young tableaux of the constituents.
If only the quark has non-zero N-ality, then the N-ality of the
bound state is identical to that of the quark.

  If follows that matter fields in N-ality=0 color representations
cannot cause string breaking, and it is therefore unnecessary to
take their masses to infinity, in order to properly define quark
confinement.  We can therefore restate the quark confinement problem
a little more generally as follows: \\

\ni $\underline{\mbox{\bf The Confinement Problem (II)}}$ \\

  {\sl Consider an SU(N) gauge theory with matter fields in various
representations of the gauge group, and take the limit that the
masses of the non-zero N-ality matter fields go to infinity.
Show that in this limit there exists a confining phase,
in which the work required to increase the separation of a
non-zero N-ality particle-antiparticle pair by a distance $L$
approaches $\s L$ asymptotically, where $\s$ is a (representation-dependent)
constant.} \\

  The next task is to formulate order parameters which can
distinguish the confining phase of an SU(N) gauge theory from other
possible phases

\section{Signals of the Confinement Phase}

\subsection{The Wilson Loop}

    We begin with the lattice SU(N) gauge theory
Lagrangian containing a single, very massive (Wilson) quark field in a color
representation $r$ of SU(N)
\bea
      S &=& \b \sum_p \left(1 - {1\over N} \mbox{ReTr}[U(p)] \right)
\non \\
        &+& \sum_{x} \left\{ (m_q + 4\a) \overline{\psi}(x) \psi(x)
             - \oh \sum_{\m=\pm 1}^{\pm 4} \overline{\psi}(x)
         (\a+\gamma_\m) U^{(r)}_\m(x)\psi(x+\widehat{\m}) \right\}
\eea
where $p$ denotes plaquettes, $\gamma_{-\m}=-\gamma_\m, ~ 0\le \a \le 1$,
and $U^{(r)}_\m$ is the link variable in representation $r$, with
$U_{-\m}(x)=U^\dg_\m(x-\widehat{\m})$.
A Wick rotation to imaginary time is understood.
Suppose we create a quark-antiquark pair at time $t=0$ separated by
a distance $R$ along, say, the x-axis, and let this system
propagate for a time interval $T$.
In the absence of gauge fixing, the expectation value of a color
non-singlet state will average to zero, so it is necessary to include a product
of link variables (a ``Wilson line'') between the quarks in order
to form a gauge-invariant creation operator:
\beq
      Q(t) = \overline{\psi}(0,t) \Gamma
               \prod_{n=0}^{R-1} U^{(r)}_x(n\hat{i},t) \psi(R\widehat{i},t)
\label{Qt}
\eeq
where $\Gamma$ is some $4\times 4$ matrix, constructed from Dirac $\gamma$
matrices, acting on the spinor indices.
Then, by the usual rules of quantum mechanics in imaginary time,
\bea
      \langle Q^\dg(T) Q(0) \rangle &=&
   { \sum_{nm} \langle 0 |Q^\dg |n\rangle \langle n| e^{-H T} |m \rangle
     \langle m | Q | 0 \rangle \over \langle 0| e^{-HT} |0 \rangle}
\non \\
         &=&  \sum_n |c_n|^2 e^{-\D E_n T}
\label{sos}
\eea
where $\lla \rra$ indicates the Euclidean vacuum expectation value,
and $\D E_n = E_n-E_0$.
Now integrate out the quark fields in the path integral.
The leading contribution, at large $m_q$, is obtained
by bringing down from the action a set of terms
$\overline{\psi}(\a+\gamma_4) U\psi$ along the shortest lines joining the
quark operators in $Q$ and $Q^\dg$, and these shortest lines form
the timelike sides of an $R\times T$ rectangle.  This contribution
corresponds to the two quarks propagating from $t=0$ to $t=T$ without
changing their spatial positions, and in the $m_q \ra \infty$
limit all other quark contributions are negligible by comparison.
We then have
\bea
     \langle Q^\dg(T) Q(0) \rangle &=& {1\over Z}
      \int DU D\psi D\overline{\psi} ~ Q^\dg(T) Q(0) e^{-S}
\non \\
     &\sim&  C (m_q+4\a)^{-2T} {1\over Z_U} \int DU ~ \chi_r[U(R,T)] e^{-S_U}
\non \\
     &\sim&  C (m_q+4\a)^{-2T} W_r(R,T)
\eea
where $U(R,T)$ is the path-ordered
product of links along the rectangular
contour with opposite sides of lengths $R$ separated by time $T$,
$\chi_r(g)$ is the group character (trace) of group element
$g$ in representation $r$, and $C$ is a constant arising from a trace
over spinor indices.
The Wilson loop $W_r(R,T)$ is defined as the expectation value of
$\chi_r[U(R,T)]$, and $S_U$ is the Wilson action of the pure gauge
theory. It follows that
\beq
        \sum_n |c_n|^2 e^{-\D E_n T} \sim  C (m_q+4\a)^{-2T} W_r(R,T)
\eeq
In this relation, $\D E_n$ refers to the energy, above the vacuum
energy, of the n-th energy eigenstate having a non-vanishing overlap
with the state created by $Q$.  These can be understood as the flux tube
eigenstates, and as $T \ra \infty$, only the lowest
energy contribution $\D E_{min}$ contributes.  Subtracting the
$\ln(m_q+4\a)$
self-energy terms, which are independent of $R$ and therefore irrelevant
for our purposes, the $R$-dependent part $\D E_{min}$ is contained in
the quantity
\beq
     V_r(R) = -\lim_{T \ra \infty} \log \left[ {W_r(R,T+1)
                           \over W_r(R,T)} \right]
\eeq
which will be referred to from here on as the static quark potential.

   The confinement problem, then, is to show that $V_r(R)$ has the
asymptotic behavior
\beq
      V_r(R) \sim \s_r R
\eeq
at large $R$, for non-zero N-ality representations $r$.
Note that in the $m_q \ra \infty$ limit, this is an observable of the
pure gauge theory.  Put another way, confinement is a property of the
gauge theory vacuum in the absence of matter fields.

   In general, though, the confinement criterion
for non-zero N-ality allows
there to be finite mass matter fields in zero N-ality representations,
as discussed in section 2.
In fact, gluons themselves are zero N-ality particles.  This means
that gluons can cause string-breaking for heavy quark sources in
zero N-ality representations.  This is an important point, which we
will return to in the next section.  It means that the confinement
criterion can only refer to quarks in non-zero N-ality representations;
the color field of a zero N-ality source will be screened in both the
Higgs and confined phases, rather than collimated into a flux tube.
Only the string tension $\s_r$ of non-zero N-ality sources can serve
as an order parameter for the confined phase, in which
\beq
       \s_r \ne 0 ~~~\mbox{for all color charge sources of non-zero N-ality}
\eeq

\subsection{The Polyakov Line}

   As already noted, a gauge theory with Higgs fields in the adjoint
(or other zero N-ality) representation  can have distinct
confining and Higgs phases.  Deconfining phase transitions can also
occur in a pure gauge theory at finite temperature.  Any lattice Monte
Carlo simulation is a simulation at finite temperature, with the inverse
temperature  equal to the extent $L_t$
of the lattice in the time direction.
Numerical studies of finite temperature phase transitions are carried out on
$L_s^3\times L_t$ lattices, with $L_t < L_s$, and
deconfining phase transitions are found, at fixed gauge coupling, as
$L_t$ is reduced below some critical value.

\begin{figure}[t!]
\begin{center}
\begin{minipage}[t]{8 cm}
\centerline{\scalebox{0.5}{\includegraphics{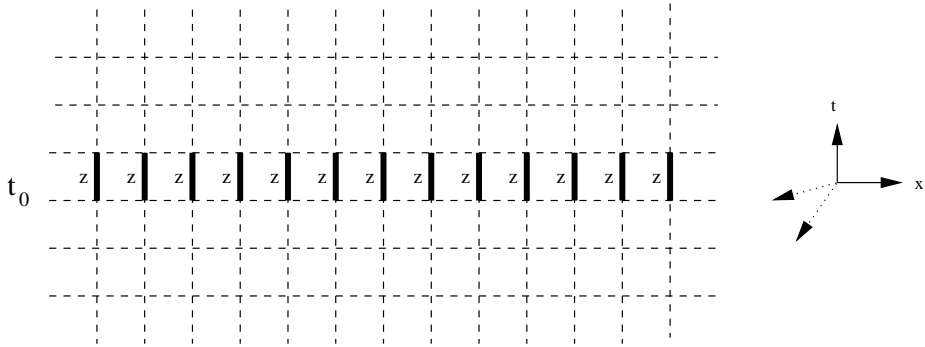}}}
\end{minipage}
\begin{minipage}[t]{16.5 cm}
\caption{The global center transformation.  Each of the indicated links in the
$t$-direction, at $t=t_0$, is multiplied by a center element $z$.  The lattice
action is left unchanged by this operation.}
\label{global}
\end{minipage}
\end{center}
\end{figure}

   Different phases of a statistical system
are often characterized by the broken or unbroken
realization of some global symmetry (it is impossible to break
a \emph{local} gauge symmetry, as we know from Elitzur's theorem).
In an SU(N) gauge theory, with only zero N-ality matter fields, there
exists the following global $Z_N$ symmetry transformation on a finite periodic
lattice (Fig.\ \ref{global}):
\beq
       U_0(\vec{x},t_0) \ra z U_0(\vec{x},t_0) ~~,~~z \in Z_N
             ~~,~~ \mbox{all~} \vec{x}
\label{globalz}
\eeq
with all other link variables unchanged.
The action is obviously invariant under this transformation, with
factors $z$ and $z^{-1}$ cancelling in every timelike plaquette at $t=t_0$
(a property which is only true in general if $z$ is a center element).  For the
same reason, any ordinary Wilson loop is invariant under this transformation.

   However, not every gauge-invariant observable is unchanged by
the transformation \rf{globalz}.  An important observable which is
sensitive to this transformation is a Wilson line which winds once through
the lattice in the periodic time direction
\beq
      P(\vec{x}) =
    \mbox{Tr}\left[ U_0(\vec{x},1)U_0(\vec{x},2)...U_0(\vec{x},L_t) \right]
\eeq
and which transforms under \rf{globalz} as
\beq
      P(\vec{x}) \ra z P(\vec{x})
\eeq
This observable, which is simply a Wilson loop with non-zero winding
number in the time direction, is known as a {\bf Polyakov Loop}.  The
global symmetry \rf{globalz} can then be realized on the lattice in one
of two ways:
\beq
       \langle P(\vec{x}) \rangle = \left\{
       \begin{array}{cl}
         0      &  \mbox{~~unbroken $Z_N$ symmetry phase} \cr
\mbox{non-zero} &  \mbox{~~broken $Z_N$ symmetry phase}
       \end{array} \right.
\eeq
The relation of the Polyakov loop VEV to confinement
is quite direct.  A Polyakov loop
can be thought of as the world-line of a massive static quark at
spatial position $\vec{x}$,
propagating only in the periodic time direction.
Then
\beq
        \lla P(\vec{x}) \rra = e^{-F_q L_t}
\eeq
where $F_q$ is the free energy of the isolated quark.  In the confinement
phase, the free energy of an isolated quark is infinite, while it is
finite in a non-confined phase.  Therefore, on a lattice with finite time
extension
\begin{center}
   {\bf unbroken $Z_N$ symmetry $\Leftrightarrow$ confinement phase}
\end{center}
In other words, confinement can be identified as the phase in which global
center symmetry is also a symmetry of the vacuum \cite{book}.
The Polyakov loop is
a true order parameter: zero in one phase, non-zero in another,
which associates the breaking of a global symmetry with the transition
from one phase to another.

    This fact provides further insight into the Fradkin-Shenker result.
A gauge theory with matter fields in the fundamental representation,
such as the action in eq.\ \rf{Fradkin}, is not invariant under global
center symmetry.  Since the symmetry itself doesn't exist at the
level of the Lagrangian, there can clearly be no phase transition between
its broken and unbroken realization, and therefore no
transition from the Higgs to a genuine confining phase.
On the other hand, for matter fields in the adjoint (or any zero N-ality)
representation, the Lagrangian does have global center symmetry, and
a distinct confinement phase can exist.

   The transformation \rf{globalz} can be generalized: It is an example
of a wider class of {\bf singular gauge transformations}.
We consider gauge transformations on a periodic lattice
\beq
     U_0(x,t) \ra g(x,t) U_0(x,t) g^\dg(x,t+1)
\eeq
which are periodic in the time direction only up to a $Z_N$
transformation:
\beq
       g(x,L_t+1) = z^* g(x,1)
\label{aperiodic} \eeq which affects Polyakov loops in the same
way as \rf{globalz}. The particular transformation \rf{globalz} is
generated by the singular transformation \beq
       g(x,t) = \left\{ \begin{array}{cc}
                  I & t \le L_t \cr
                z^* & t = L_t+1 \end{array} \right.
\eeq

    In the continuum gauge theory in a finite
volume, the singular gauge transformation is again periodic only up
to a center transformation
\beq
     g(x,L_t) = z^* g(x,0)
\eeq
and the spatial gauge potentials $A_k(x)$ transform in the usual way.
The $A_0(x,t)$ potential, however, transforms in the following way:
At $t\ne 0,L_t$,
\beq
      A'_0(x,t) = g(x,t) A_0 g^\dg(x,t) - {i\over g_s}g(x,t) \pa_0 g^\dg(x,t)
\label{usual}
\eeq
as usual, where $g_s$ is the gauge coupling.  However, at $t=0,L_t$,
we define
\bea
      A'_0(x,0) &\equiv& g(x,0) A_0(x,0) g^\dg(x,0) -
        \lim_{\e \ra 0} {i\over g_s}g(x,\e) \pa_0 g^\dg(x,\e)
\non \\
                &=& A'_0(x,L_t)
\non \\
                &\equiv& g(x,L_t) A_0(x,L_t) g^\dg(x,L_t) -
        \lim_{\e \ra 0} {i\over g_s}g(x,L_t-\e) \pa_0 g^\dg(x,L_t-\e)
\eea
What this definition does is to drop the delta function at $t=0,L_t$
which would normally be present in \rf{usual} due to the discontinuity
in the transformation $g(x,t)$.  That means that a ``singular''
gauge transformation is not really a gauge transformation, and this should
not be a surprise.  If singular gauge transformations were true gauge
transformations, then all gauge-invariant observables, including Polyakov
loops, would be unaffected by the transformation.  The term ``singular
gauge transformation'' is therefore slightly misleading, but because of
common usage it will be retained here.

\subsection{The 't Hooft Operator}

   In ordinary electrodynamics, the Wilson loop holonomy
\beq
U(C) =\exp\left[ie\oint_C dx^\m A_\m(x) \right] = e^{i\Phi_B}
\eeq
of a space-like loop $C$ is simply the exponential of the magnetic flux
through a surface bounded by the loop
$C$.  Although $\Phi_B \ne 0$ implies that there must be non-zero
field strength somewhere on any surface bounded by $C$, it is possible that
for a large loop this field strength is localized far from loop $C$
itself, and that the field strength in the neighborhood of the loop is
actually zero.  A familiar example is the vector potential in the
exterior region of a solenoid, of radius $R$, oriented along the $z$-axis
\beq
     A_\th = {\Phi_B \over 2\pi e r} \widehat{\th} ~~~~(r>R)
\label{solenoid}
\eeq
where the field strength vanishes outside the coil at $r>R$. Although
the field strength vanishes, we know that the vector potential in this
region can still affect the motion of electrons.  This is the well-known
Bohm-Aharonov effect, which makes use of the fact that for a loop $C$
winding around the exterior of the solenoid, we have $U(C) \ne 1$.

   The exterior solenoid field can be expressed as a singular gauge
transformation of the classical $A_\m=0$ vacuum state, with the
transformation given by
\beq
      g(r,\th,z,t) = \exp\left[-i\Phi_B {\th \over 2\pi}\right]
\label{g1}
\eeq
This transformation is obviously aperiodic around the loop $C$
\beq
      g(r,\th=2\pi,z,t) = e^{-i\Phi_B} g(r,\th=0,z,t)
\label{aperiod}
\eeq
with the aperiodicity associated with an element $e^{-i\Phi_B} \in U(1)$ of
the center of the gauge group (for an abelian gauge group such as U(1), the
center of the group is the group itself, but we are anticipating the
SU(N) case).  The solenoid field in the exterior $r>R$ region, at
$\th \ne 0,2\pi$ is given by
\beq
    A_\mu(r>R,\th,z,t) = - {i\over e} g(r,\th,z,t) \pa_\m g^\dg(r,\th,z,t)
\eeq
while at $\th=0,2\pi$ we define
\bea
      A_\m(r,0,z,t) &\equiv&
       - \lim_{\e \ra 0} {i\over e}g(r,\e,z,t) \pa_\m g^\dg(r,\e,z,t)
\non \\
                &=& A_\m(r,2\pi,z,t)
\non \\
                &\equiv&
  - \lim_{\e \ra 0} {i\over e}g(r,2\pi-\e,z,t) \pa_\m g^\dg(r,2\pi-\e,z,t)
\eea
Once again, this definition has the effect of dropping the
delta function which would normally arise from the derivative of
$g(r,\th,z,t)$ along the hypersurface $\th=0$
of discontinuity.  As pointed out in the
last section, this means that the singular
gauge transformation is not a true gauge transformation (otherwise it could
not possibly affect a gauge-invariant operator such as a Wilson loop).
In the $R \ra 0$ limit, the singular gauge transformation
creates a line of infinite
field strength along the z-axis.  This is the $R=0$ limit of the interior field
of the solenoid.

   The vector potential \rf{solenoid}, resulting from the singular
gauge transformation \rf{g1}, is certainly not the unique form for the
exterior field of a solenoid, and can be altered by any non-singular
gauge transformation.  What is essential is the aperiodicity
\rf{aperiod} along some hypersurface of dimension $D-1$ (in this case
the hypersurface at $\th=0$).  The hypersurface itself carries no
action, and its position, apart from its boundary, is not a physical
observable.  The boundary of
the hypersurface, however, is the solenoid, which carries
non-vanishing field strength.  In this example, in the $R=0$ limit, the
boundary of the hypersurface is the z-axis, which carries an infinite
field strength. Any loop $C$ which winds $n$ times around the z-axis,
i.e.\ which has {\bf linking number} $n$ with the (infinite or
periodic) z-axis, has a holonomy which is altered by the singular
gauge transformation in this way:
\beq
        U(C) \ra e^{\pm in \Phi_B} U(C)
\eeq
with the sign in the exponent depending on the orientation of the loop
relative to the direction of the B-field.

\begin{figure}[htb]
\begin{center}
\begin{minipage}[t]{8 cm}
\centerline{\scalebox{0.4}{\includegraphics{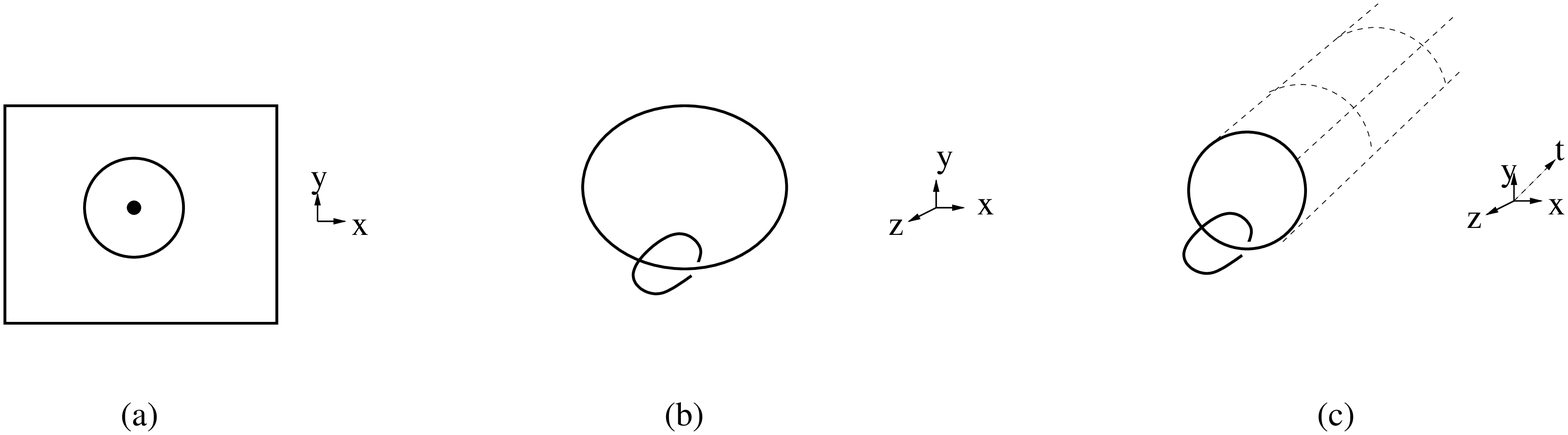}}}
\end{minipage}
\begin{minipage}[t]{16.5 cm}
\caption{A loop topologically linked to: a) a point, $D=2$;
b) another loop, $D=3$; 4) a surface, $D=4$.}
\label{link}
\end{minipage}
\end{center}
\end{figure}

   The linking of Wilson loops and vortices is a crucial concept, and deserves
further elaboration.   In
D=2 dimensions, a loop can be topologically linked to a point, in the
sense that such a loop cannot be moved far from the point without actually
crossing the point (Fig.\ \ref{link}).  Obviously this doesn't make sense
in D=3 dimensions, where a loop and point can be moved apart without crossing.
Likewise, in D=3 dimensions, a closed
loop can be topologically linked to another closed loop, but there
is no such linking in D=4 dimensions where any two loops can be separated
without crossing.  In D=4 dimensions, closed loops can be topologically
linked to surfaces.  To visualize this last statement, suppose that a
loop $C$, embedded in a 3-dimensional subspace, is
topologically linked in D=4 dimensions to a closed surface $S$.  In the
3D subspace, $S$ appears as a closed curve $C'$.  If $S$ and $C$ are
topologically linked in D=4 dimensions, then $C$ and $C'$ are topologically
linked in the D=3 subspace (otherwise $C$ could be moved arbitrarily
far away from $S$, without crossing $S$, in the D=3 subspace).
The general statement, in any number of dimensions,
is that a closed loop can be topologically linked to a hypersurface
of co-dimension 2.
In our example, the singular gauge transformation \rf{aperiod}
actually creates field strength along the z-axis at every time $t$, i.e.\
a \emph{surface} of field strength in the z-t plane, and a loop winding
once around the z-axis at a fixed time $t$ is linked topologically
to this surface.

   Now we generalize to SU(N), and again consider transformations of
the gauge field which can alter
loop holonomies, without changing the action in the neighborhood of the loop.
In D=4 dimensions, let $V_3$ denote a compact,
simply-connected {\bf Dirac 3-volume}, whose closed boundary is the
surface $S$.  Consider any loop $C$ topologically linked to $S$,
where the loop is parametrized by
$\vec{x}(\t),~\t\in [0,1]$, and
$\vec{x}(1)=\vec{x}(0) \in V_3$.
Let the gauge field along the curve $C$ be transformed
by a singular gauge transformation with the discontinuity
\beq
      g(\vec{x}(1)) = z g(\vec{x}(0)) ~~~~,~~~ z\in Z_N
\label{ap1}
\eeq
which means that
\beq
        U(C) \ra z^* U(C)
\eeq
and the transformed gauge potential $A_\m(x)$ is defined in the
usual way, except that a delta function
arising from the discontinuity of $g(\vec{x}(\t))$ on the hypersurface
$V_3$ is dropped.  As in the abelian
case, the singular transformation creates a surface of infinite field
strength along $S$ in the continuum gauge theory,
which is known as a {\bf thin center
vortex}.  For an SU(N) gauge group there are $N-1$ possible vortices,
corresponding to the number of elements in $Z_N$ different from the
identity.  As in the abelian case, it is possible to modify the
transformation \rf{ap1} near $S$, and smear out the thin vortex
into a surface-like region of finite thickness, and finite field
strength.  This is a ``thick'' center vortex.

\begin{figure}[t!]
\begin{center}
\begin{minipage}[t]{8 cm}
\centerline{\scalebox{0.5}{\includegraphics{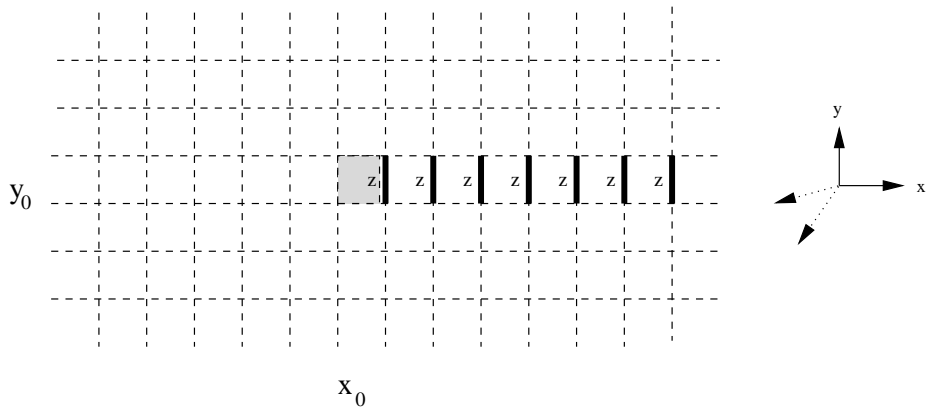}}}
\end{minipage}
\begin{minipage}[t]{16.5 cm}
\caption{Creation of a thin center vortex.  The shaded plaquette, and
all other x-y plaquettes at sites $(x_0,y_0,\xp)$ form the center vortex.
The stack of vortex plaquettes lie along a line in $D=3$ dimensions,
or a surface in $D=4$ dimensions.}
\label{cvfig}
\end{minipage}
\end{center}
\end{figure}

   A particular example of a singular gauge transformation on the lattice is
the transformation
\beq
      U_y(x,y_0,\xp) \rightarrow z U_y(x,y_0,\xp) ~~~\mbox{for~~} x > x_0
\label{cv}
\eeq
with all other links unchanged, as indicated in Fig.\ \ref{cvfig}.
The (half-infinite) Dirac volume in this case is all sites with
$x>x_0,~y=y_0$.  All $x-y$ plaquettes on the surface $S$, parallel
to the $z-t$ plane at $x=x_0,~y=y_0$, are transformed by a center element
\beq
      U(P) \ra z U(P)
\eeq
The surface $S$ is a thin center vortex.

   Following 't Hooft, we now go to a Hamiltonian formulation,
and consider an operator
$B(C)$ which creates a thin center vortex at a fixed time $t$ on curve $C$
(cf.\ ref.\ \cite{Bop} for an
explicit construction of this operator).
This means that the gauge fields at time $t$ are transformed by
a singular gauge transformation satisfying \rf{ap1} on any curve
$C'$ at time $t$, parametrized by $\vec{x}(\t)$,
which has linking number one with loop $C$ .  Then
\beq
      U(C') B(C) = z B(C) U(C') ~~~,~~~~ z\in Z_N
\eeq

  Using only this commutation relation, 't Hooft argued in ref.\ \cite{thooft1}
that only area-law or perimeter-law falloff for the $U(C),~B(C)$ expectation
values is possible, and that in the
absence of massless excitations, the simultaneous behavior
\beq
     \lla U(C) \rra \sim e^{-a P(C)} ~~\mbox{and}~~~ \lla B(C) \rra \sim
                                                       e^{-b P(C)}
\eeq
is ruled out, where $P(C)$ is the loop perimeter.
From this it follows that if there are no massless excitations,
then $B(C)$ is an order parameter for confinement, in the sense that
a perimeter-law falloff $\lla B(C) \rra \sim \exp[-b P(C)]$ implies that
the theory is in the confinement phase, while an area-law falloff for
$\lla B(C) \rra$ implies spontaneous breaking of at least part of the gauge symmetry.

\subsection{The Vortex Free Energy}

   On a finite lattice, it is impossible to create a single vortex
sheet winding through the periodic lattice along, say, the z-t plane,
because it requires a half-infinite Dirac volume.  Instead, vortices
which are closed by lattice periodicity can only be created or destroyed
in pairs.

   There is however, a trick (due to 't Hooft \cite{thooft2}) known as
``twisted boundary conditions,'' which forces the number of vortices
winding through the periodic lattice to be odd, rather than even.  Consider an
SU(2) lattice gauge theory in D=4 dimensions, and imagine changing the
sign $\b \ra -\b$ of the coupling on the set of x-y plaquettes at
$x_0,y_0$ and all $z,t$.  It is not hard to see that the minimal
action configuration of such a modified action, on a lattice which
is infinite in the x-direction, is gauge equivalent to
\beq
       U_y(x,y_0,\xp) = -I ~~~\mbox{for} ~~ x>x_0
\eeq
with all other links equal to the identity matrix $I$ ($\xp$ denotes axes
perpendicular to the $x,y$ directions).  In this
configuration the plaquettes at negative coupling are equal to $-\tr[I]$,
with all other plaquettes equal to $+\tr[I]$.  The negative $U_y$ links
are located in a half-infinite Dirac volume.  On a finite lattice,
however, the Dirac volume cannot extend indefinitely in the positive
x-direction, and must end on a center vortex sheet, which winds
through the lattice in the $z$ and $t$ directions.  There
is no need for this vortex sheet to be thin, which costs a great deal
of action at large $\b$.
The action can be lowered if the vortex sheet has some
finite thickness (just \emph{how} thick is a dynamical issue at the
quantum level, see section 6.6 below).  The upshot is that twisted
boundary conditions, implemented by setting $\b \ra -\b$ on a
``co-closed'' set of plaquettes (x-y plaquettes on a particular
z-t plane closed by lattice periodicity), requires that there are an
odd number (at least one) of thick center vortices in the periodic
lattice.

    The magnetic free energy of a $Z_2$ vortex in a finite volume
$V=L_x L_y L_z L_t$ is
defined as the excess free energy of a periodic lattice with twisted
boundary conditions, as compared to a lattice without twist, i.e.
\beq
      e^{-F_{mg}} = {Z_- \over Z_+}
\eeq
where $Z_{\pm}$ indicates the partition function with normal ($+$) and twisted
($-$) boundary conditions.  The ``electric'' field energy is defined by
a $Z_2$ Fourier transform
\bea
      e^{-F_{el}} &=& \sum_{z=\pm } z {Z_z \over Z_+}
\non \\
                  &=& 1 - e^{-F_{mg}}
\eea
Let $C$ be a rectangular loop of area $\A(C)$.  The following inequality
was proven by Tomboulis and Yaffe \cite{TY}:
\beq
       \lla \tr[U(C)] \rra \le \left\{ \exp[-F_{el}]
                \right\}^{\A(C) \over L_x L_y}
\eeq
Therefore, if the free energy of a thick vortex surface running in the periodic
z-t directions has an area-law falloff with respect to the cross-sectional
x-y area
\beq
         F_{mg} = c L_z L_t e^{-\r L_x L_y}
\label{vfe}
\eeq
then this is a sufficient condition for confinement, because it implies
an area-law bound for Wilson loops.

\begin{figure}[tb]
\begin{center}
\begin{minipage}[t]{8 cm}
\centerline{\scalebox{0.5}{\includegraphics{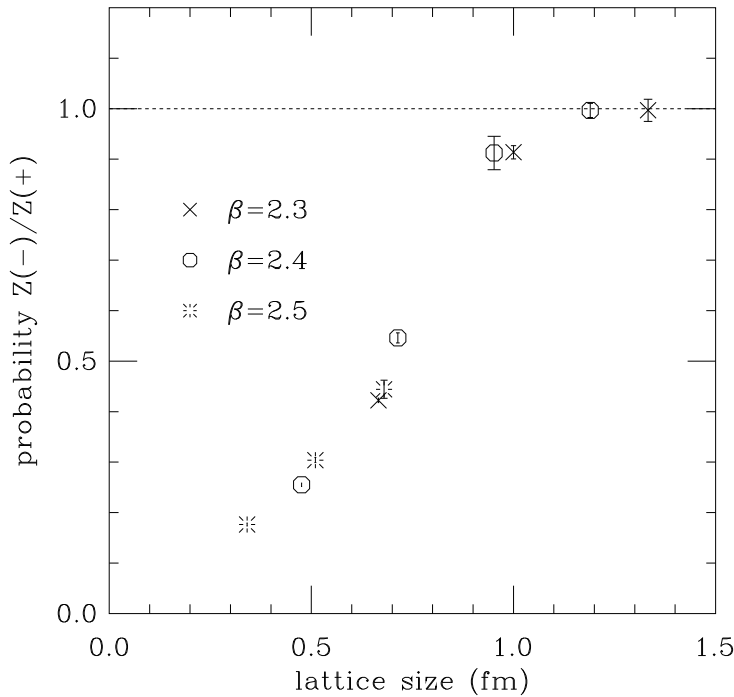}}}
\end{minipage}
\begin{minipage}[t]{16.5 cm}
\caption{The behavior of the vortex free energy vs.\ lattice extension
(same in all directions).  From
Kov\'acs and Tomboulis, ref.\ \cite{KT}.}
\label{tkfig}
\end{minipage}
\end{center}
\end{figure}

   The ratio $Z_-/Z_+$ has been calculated numerically in SU(2)
lattice gauge theory via lattice Monte Carlo, by Kov\'acs and Tomboulis
in ref.\ \cite{KT}.  Their result is shown in Fig.\ \ref{tkfig}.
The rapid rise of the ratio to unity, with increasing
lattice extension in the $x,y$ directions,
is consistent with confining behavior
\rf{vfe} of the vortex free energy.
The twist trick can also be used to insert center vortex flux
through any plane, or several planes simultaneously \cite{dF1}.

   The drop in vortex free energy with lattice size is naturally associated
with the vortex ``spreading out'' to its
natural thickness as lattice size increases.  It is possible to insert
constraints which prevent this spreading out \cite{Yaffe}, and then the
behavior \rf{vfe} is lost.\footnote{A related constraint can be imposed
on SU(2) Wilson loop operators, such that Tr$[U(C)]$ is effectively positive
at weak couplings, and this very strong constraint on the loop removes the area
law falloff \cite{KTref}.  That result is not unexpected, since the rapid
falloff of the Wilson loop is due to strong cancellations between positive
and negative contributions to the loop expectation value.}
Finite temperature studies involving the vortex free energy
are also relevant \cite{vS,Adriano1,Hart} (see section 6.4 below), because
they involve ``squeezing''
vortices in one direction.

\subsection{Confinement Without Center Symmetry?}

   All four of the confinement criteria listed above
\begin{enumerate}
\item area law for Wilson loops,
\item vanishing Polyakov lines,
\item perimeter-law behavior for 't Hooft loops,
\item exponential area falloff of the vortex free energy,
\end{enumerate}
can only be satisfied if the Lagrangian is invariant
under the global center symmetry of eq.\ \rf{globalz}.  If, on
the other hand, the theory contains matter fields of non-zero N-ality
which completely break the center symmetry, then all Wilson loops have
perimeter law falloff, all Polyakov loops are finite, and the Dirac
volumes of center vortices carry an action density which is infinite
in the continuum limit.  We may ask whether it is possible to
find some other observable or order parameter, not listed above, which
would distinguish a ``confinement'' phase from a Higgs phase even
when the global center symmetry is explicitly broken
by matter fields in the fundamental representation of the gauge group.

   The Fradkin-Shenker result argues very strongly against this possibility.
If there were some order parameter, call it Q, whose expectation value could
distinguish between the confined phase and the Higgs phase of the theory,
then there would have to be a line of transitions completely isolating
these regions
from one another in the coupling-constant
phase diagram.  But there is no such line of
transitions and, in consequence, no such operator Q, at least in
a gauge theory with scalar fields in the fundamental representation of
the gauge group.  An order parameter distinguishing
between the Higgs and Coulomb phases is a possibility, but an order
parameter distinguishing between the Higgs and confinement phase
appears to be ruled out.

\begin{figure}[t!]
\begin{center}
\begin{minipage}[t]{8 cm}
\centerline{\scalebox{0.5}{\includegraphics{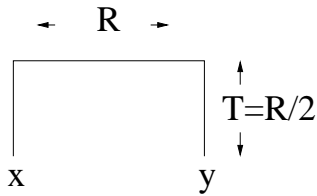}}}
\end{minipage}
\begin{minipage}[t]{16.5 cm}
\caption{Contour for the Fredenhagen-Marcu operator.}
\label{fmfig}
\end{minipage}
\end{center}
\end{figure}

   A case in point is the operator suggested by Fredenhagen
and Marcu (FM) \cite{FM}.  The idea behind their proposal is that if
there exist dynamical matter fields, then one might redefine ``confinement''
to mean string breaking, rather than a linearly rising static potential.
Consider the state
\beq
      |\Psi_{xy} \rangle =  {1 \over \sqrt{W(R,R)} }
           \sum_i \phi^\dg_i(x) U(C_{xy}) \phi_i(y) |\Psi_0 \rangle
\eeq
where the sum runs over all matter fields in the fundamental
representation, $C_{xy}$ is the contour shown in Fig.\ \ref{fmfig}, and
$\Psi_0$ is the vacuum.
Sites $x$ and $y$ are at equal times, and we define $R = |\vec{x}-\vec{y}|$
as the spatial separation.  The
Wilson loop $W(R,R)$ runs over a rectangular $R \times R$ contour, and
the factor $1/\sqrt{W}$ is introduced so that the norm of $|\Psi_{xy} \rangle$
is of order one.

   The FM order parameter is defined as
\bea
      \rho &=&
       \lim_{R\ra \infty} \Bigl| \langle \Psi_0 | \Psi_{xy} \rangle \Bigr|^2
\non \\
  &=& \lim_{R\ra \infty} { \Bigl| \sum_i \langle \phi^\dg_i(x)
       U(C_{xy}) \phi_i(y) \rangle \Bigr|^2 \over W(R,R)}
\label{FM}
\eea
This order parameter is sensitive to the origin of the perimeter
law behavior of Wilson loops.  If the perimeter-law falloff of a Wilson loop is
dominated by charge screening due to matter fields,
then the numerator and denominator
in eq.\ \rf{FM} are comparable, and the ratio $\rho$ is non-zero in the
large $R$ limit.  This is the FM criterion for confinement.  On the other
hand, if the perimeter-law falloff of the Wilson loop is independent
of charge screening, then the denominator may be
much larger than the numerator,
and $\r \ra 0$ in the limit.

   It should be clear that this is really a color screening criterion,
rather than a confinement criterion.  In the Fradkin-Shenker model, the
FM criterion is certainly satisfied in the $\b_G,\b_H \ll 1$ regime, where the
dynamics is ``confinement-like'', but it is also satisfied deep in the
Higgs regime $\b_G,\b_H \gg 1$, where screening also takes place.
The FM criterion has, in fact, been applied numerically to
the adjoint Higgs model (which actually has a transition between the
confining and Higgs phases), with the matter field $\phi$ at points $x$
and $y$ in the adjoint representation.  It was found numerically in ref.\
\cite{Azcoiti} that the criterion $\rho \ne 0$ was satisfied in both
the Higgs and the confinement phases.  

   The term "confinement" is sometimes taken to be
synonymous with the absence of color-charged states in the spectrum,
whether or not there exists a confining static potential. With
this usage, a theory based on, e.g., the G(2) gauge group, which
has a trivial center, a trivial first homotopy group (see
below), and an asymptotically flat static potential, 
belongs on the list of confining theories \cite{G2}.
However, terminology which essentially identifies confinement
with color screening has drawbacks that have already been pointed
out. It implies, for example, that electric charge is "confined"
in an ordinary superconductor.  In a gauge theory with scalars in
the fundamental representation, it implies that there is
confinement deep in the Higgs regime.  In a gauge theory with
scalars (and other matter fields) only in the adjoint
representation, which really does have a transition between the
Higgs and confinement phases, this use of the word "confinement"
means that \emph{both} phases are confining. In theories without a
high-temperature deconfinement transition, the theory at high
temperature would have to be taken as "confining" if the low
temperature phase, which fulfils the FM criterion, is deemed to be
in a confining phase.

   We conclude that while the FM operator is a good order parameter for
color screening, there is a legitimate distinction to be made 
between the screening of charge by, e.g., a condensate of some kind, and 
confinement by a linear potential. This distinction means that
confinement and color screening are not truly synonymous.  The transition
from a confining to a screened potential is always associated with the
breaking of a global center symmetry, and in the absence of a non-trivial
center symmetry, only screened or Coulombic potentials can be realized.  
One would therefore expect that this symmetry is also important in 
understanding the origin of the confining force.

\subsubsection{The Case of SO(3) Gauge Theory}

   There is still one puzzle: gluons actually belong to the adjoint
representation of the gauge group, and it would seem to make no difference
at all, in the continuum limit, whether the gauge group is SU(N) or
$SU(N)/Z_N$.  Both groups have the same Lie algebra, but in the latter case
the center of the gauge group is trivial.  Strictly speaking, the
$SU(N)/Z_N$ theory is non-confining, simply because it is impossible to
introduce color sources of non-zero N-ality.  The static potential therefore
tends to a constant at large distances.   Still, as noted in ref.\
\cite{deF-J}, the presumed
universality of the continuum fixed point suggests that
the SU(N) and $SU(N)/Z_N$ theories should be in some sense equivalent in
the continuum.  For example, one might expect a finite temperature
``deconfinement'' transition in both cases,
and whatever extended objects dominate the vacuum
of the SU(N) gauge theory, in the continuum limit,
should also dominate the vacuum of the $SU(N)/Z_N$ theory.

   This puzzle has recently been addressed by de Forcrand and Jahn in
ref.\ \cite{deF-J}, who find that the center vortex free energy remains
a good order parameter for, e.g., the confinement-deconfinement transition
both in SU(2) and in SO(3)=SU(2)$/Z_2$ gauge theories.
The key point is that center vortex creation is well defined in both
SU(N) and $SU(N)/Z_N$ theories.  Consider a path
in the SU(N) group manifold, parametrized by $\t\in [0,1]$,
which is traced by a singular gauge transformation $g(\t)$ around a closed
loop $x(\t)$ in Euclidean space .  On the SU(N) manifold, a vortex-creating
transformation is discontinuous, i.e.\ $g(1) = z g(0)$.
The same transformation, mapped to the SU(N)$/Z_N$ group, traces a continuous
but topologically non-contractible path on group manifold.  In this way,
the vortex-creating transformations have an unambiguous topological
signature both in SU(N) and SU(N)/$Z_N$.  Formally, the definition
of center vortices rests on the existence of a non-trivial first homotopy
group of the gauge group modulo its center.
This homotopy group, $\pi_1(SU(N)/Z_N)=Z_N$, is the same for both
the SU(N) and SU(N)/$Z_N$ groups.

   In SU(2) lattice gauge theory, it is possible to specify the number of
center vortices piercing a plane, mod 2, by imposing twisted boundary
conditions as discussed above.  De Forcrand and Jahn point out that
SO(3) lattice gauge theory can be reformulated as an SU(2) lattice gauge theory
in which all possible twisted boundary conditions are summed over. Their
construction is motivated by the formulation of lattice SU(2) gauge theory
in terms of SO(3) and $Z_2$ variables, which was developed by the authors
of refs.\ \cite{KTref,MP,TK1,AH}.
Instead of being imposed boundary conditions, the ``twists''
$z_{\m\n}=\mbox{sgn}(\eta_{\m\n})$ are observables
in SO(3) lattice theory, and are extracted in the $\m\n$ plane from
the quantity
\beq
       \eta_{\m\n} = {1\over L_\r L_\s}\sum_{x_\r,x_\s}
            \prod_{x_\m x_\n} \mbox{sgn} \tr_F \Bigl[U(p_{x,\m\n})\Bigl]
\eeq
where $p_{x,\m\n}$ is a plaquette at point $x$ in the $\m\n$ plane.
With the help of this observable, it is possible to define and calculate
the vortex (or ``twist'') free energies
\beq
      F(z) = -\log {Z(z)\over Z(1)}
\eeq
in SO(3) gauge theory as well as SU(2) gauge theory.

\section{Properties of the Confining Force}

   Every theory of confinement aims at explaining the linear rise of
the static quark potential, which is suggested by the linearity of
meson Regge trajectories.  The linear potential, however, is only one
of a number of properties of the confining force that a satisfactory
theory of confinement is obligated to explain.  A complete list would
include at least the following:
\begin{itemize}
\item Linearity of the Static Potential
\item Casimir Scaling
\item N-ality Dependence
\item String Behavior: Roughening and the L\"{u}scher term
\end{itemize}
Each item on the list is supported by strong evidence from lattice
Monte Carlo simulations; the last two items are also bolstered
by fairly persuasive theoretical arguments.  We will consider
each item in turn.

\subsection{Linearity}

    There is a theorem which can be proven in lattice gauge theory, which
says that the force between a static quark and antiquark is everywhere
attractive but cannot increase with distance; i.e.\
\beq
       {dV \over dR} > 0 ~~~\mbox{and}~~~ {d^2 V \over dR^2} \le 0
\eeq
The proof of this statement, which holds in any number of spacetime dimensions,
requires nothing more than the reflection positivity of
the lattice action, and a clever application of Schwarz-type inequalities
to certain Wilson loops used in potential calculations \cite{Bachas}.
Therefore the linearity of the static quark potential is in fact the
limiting behavior; the potential is constrained to be increasing
but concave downward.  If the quark potential grew at a rate faster than
linear, it would violate the above bound on the second derivative.

   It has been clear for a long time, from lattice Monte Carlo simulations,
that the static quark potential is asymptotically linear at long distances.
In practice, it is useful to replace the Wilson line operator $Q(t)$ in
eq.\ \rf{Qt} with an operator which has a larger overlap with the ground
state of the QCD flux tube.  A number of different operators have been
used, but a typical method \cite{APE}
(``APE smearing'', see also refs.\ \cite{smear})
is to first compute for each spatial link
\bea
\lefteqn{U'_i({\bf x},t) =  U_i({\bf x},t) }
\non \\
  & & + c \sum_{k\ne i}
\left[ U_k({\bf x},t) U_i({\bf x} +\widehat{k},t)
 U_k^\dg({\bf x}+\widehat{i},t)
 + U^\dg_k({\bf x}-\widehat{k},t) U_i({\bf x} -\widehat{k},t)
 U_k({\bf x}-\widehat{k}+\widehat{i},t) \right]
\eea
and then project the ``thick'' link
variables $U'_i$ back to the SU(N) group manifold.  This procedure can be
iterated, with $c$ a fixed parameter.
If parameter $c$ and the number
of smearing iterations are chosen optimally, then the overlap of
$Q'(t)$, constructed from thick links,
onto the flux-tube ground state is large.  Defining $\tW(R,T)$ to
be timelike Wilson loops with thick spacelike links, the quantity
\beq
        \V(R,T) = -\log\left[ \tW(R,T+1) \over \tW(R,T)\right]
\eeq
converges rapidly to its $T\ra \infty$ limit, which is the static quark
potential $V(R)$.

\begin{figure}[tb]
\begin{center}
\begin{minipage}[t]{8 cm}
\centerline{\scalebox{0.5}{\includegraphics{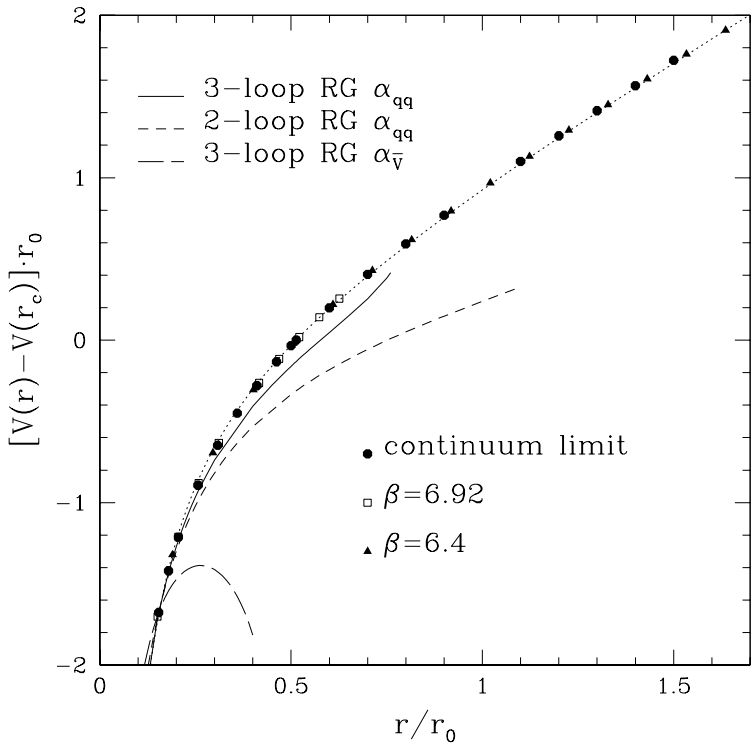}}}
\end{minipage}
\begin{minipage}[t]{16.5 cm}
\caption{The static quark potential in SU(3) lattice gauge theory.
The data points are compared
to perturbative calculations with different renormalization schemes; the dotted
line is a fit to the string-inspired potential $a - \pi/(12 R) + \s R$.
From Necco and Sommer, ref.\ \cite{NS}.}
\label{nspot}
\end{minipage}
\end{center}
\end{figure}

      Figure \rf{nspot}, taken from ref.\ \cite{NS}, displays some typical
results for the static potential $V(R)$ in SU(3) lattice gauge theory,
obtained from the lattice
Monte Carlo procedure.  In this figure the Monte Carlo data is
compared to perturbative calculations of the potential, with differing
renormalization schemes.  The linear growth of the potential beyond $0.7 r_0$
is evident from the figure.

   The x-axis of the figure shows quark separation
in units of the Sommer scale $r_0 \approx 0.5$ fm,
which is often used in studies
of the static potential. The Sommer scale is defined as the separation at which
\cite{Sommer}
\beq
        F(r_0) r_0^2 = 1.65
\eeq
where $F(r)$ is the force between static quarks in the fundamental
representation.

\subsection{Casimir Scaling}

   ``Casimir scaling'' \cite{Us1}
refers to the fact that there is an intermediate
range of distances where the string tension of static sources in
color representation $r$ is approximately proportional to the
quadratic Casimir of the representation; i.e.
\beq
        \s_r = {C_r \over C_F} \s_F
\eeq
where the subscript $F$ refers to the fundamental representation.
This behavior was first suggested in ref.\ \cite{Poul}.
The term `''Casimir scaling'' was introduced much later, in ref.\ \cite{Us1},
where it was emphasized that this behavior poses a serious challenge to
some prevailing ideas about confinement (see also the earlier comments
in ref.\ \cite{Marty}).

\begin{figure}[t!]
\begin{center}
\begin{minipage}[t]{8 cm}
\centerline{\scalebox{0.5}{\includegraphics{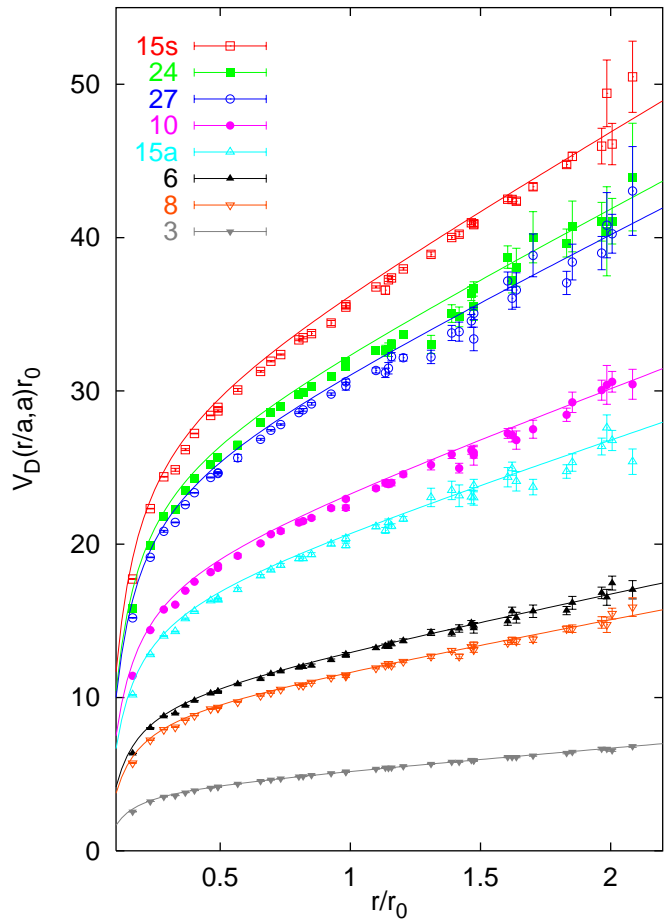}}}
\end{minipage}
\begin{minipage}[t]{16.5 cm}
\caption{Numerical evidence for Casimir scaling in SU(3) lattice gauge
theory. The solid lines are obtained
from a fit of the potential in fundamental representation,
multiplied by a ratio of
quadratic Casimirs $C_r/C_F$. From Bali, ref.\ \cite{Bali}.}
\label{Cas}
\end{minipage}
\end{center}
\end{figure}

\begin{figure}[t]
\begin{center}
\begin{minipage}[t]{8 cm}
\centerline{\scalebox{0.6}{\includegraphics{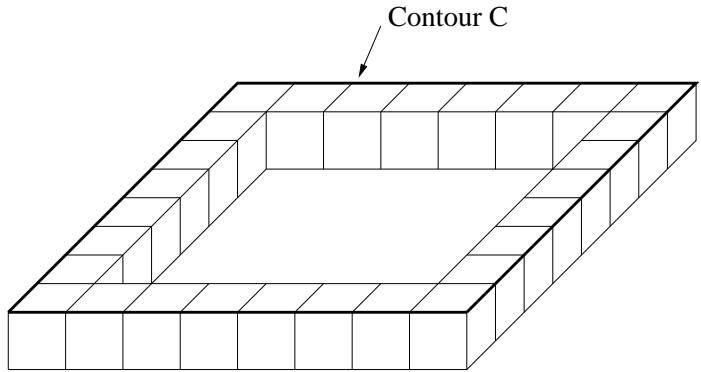}}}
\end{minipage}
\begin{minipage}[t]{16.5 cm}
\caption{Strong-coupling diagram contributing to the perimeter-law
contribution to a Wilson loop in the adjoint representation.}
\label{scup}
\end{minipage}
\end{center}
\end{figure}

   There are two theoretical arguments for Casimir scaling:
dimensional reduction \cite{Me1,Poul2} and factorization in the
large-N limit \cite{Witten}.  We begin with factorization.  The trace
of a Wilson loop holonomy $U(C)$, in a representation $r$ of SU(N)
gauge theory, is the group character $\chi_r[U(C)]$, and this can
always be expressed in terms of a sum of products of the group
character in the defining representation $F$ and its conjugate.  For a
large number $N$ of colors, only the leading term
\beq
      \chi_r(g) \sim \chi_F^n(g) \chi^{* m}_F(g) + O(N^{-1})
\eeq
is important.  Here $n+m \ll N$ is the smallest integer such that the
irreducible representation
$r$ is obtained from the reduction of a product of $n$ defining (``quark'')
representations, and $m$ conjugate (``antiquark'') representations.
Large-N factorization tells us that if $A$ and $B$ are
two gauge-invariant operators, then in the $N \ra \infty$ limit
\beq
       \lla A B \rra  = \lla A \rra \lla B \rra
\eeq
Applying this property to Wilson loops in representation $r$,
we have
\bea
       W_r(C) &=& \lla \chi_r[U(C)] \rra
\non \\
              & \stackrel{N\ra \infty}{\sim} &
          \lla \chi_F^n[U(C) \chi^{* m}_F[U(C)] \rra
\non \\
              &\sim& W_F^{n+m}
\eea
From this it follows that at large N
\beq
       \s_r = (n+m) \s_F
\label{casN}
\eeq
which is precisely Casimir scaling in this limit.
Therefore Casimir scaling is
\emph{exact}, out to infinite quark separation, at $N=\infty$.
In particular, as first noted in ref.\ \cite{Marty},
the string tension of the adjoint representation must be twice the
string tension of the fundamental representation in the large N limit.
Assuming that there is nothing singular or pathological about the
large-N limit, it follows that some approximate Casimir scaling, at least
up to some finite range of distances, should be observed at finite $N$.
In fact, from the strong-coupling expansion in lattice gauge theory,
one finds for square $L\times L$ Wilson loops in the adjoint representation
that
\beq
       W_A[C] = N^2 e^{-2 \s_F L^2} + e^{-16 \s_F L}
\label{AandP}
\eeq
where the strong-coupling diagram contributing to the perimeter-law
falloff is shown in Fig.\ \ref{scup}.
For sufficiently large $L$ the second term dominates, and therefore
the asymptotic string tension is zero.  At $N\ra \infty$ only the
first term is important, and the string tension in the adjoint
representation is $\s_A = 2 \s_F$, in agreement with Casimir scaling.
At finite values of $N$, and strong-couplings, we would have
$\s_A \approx 2 \s_F$ up to the distance
\beq
       L = \sqrt{{1\over \s_F}\log(N) + 16} + 4
\eeq
which increases logarithmically with $N$ at large $N$.

    The second argument for Casimir scaling is the (supposed) property
of dimensional reduction at large distance scales.  One argument for
this property goes as follows \cite{Me1}:  The expectation value
of a planar, spacelike Wilson loop in $D=4$ dimensions
can be expressed in terms of the
vacuum wavefunctional
\beq
     W^{D=4}_r(C) = \lla \Psi_0 | \chi_r[U(C)] | \Psi_0 \rra
\eeq
with
\beq
     \Psi_0 = \exp[-R[A]]
\eeq
where $R[A]$ is some gauge-invariant functional of the spatial components
$A_k(x)$ of the gauge field at a fixed time.  Suppose that $R[A]$
can be expanded in a power
series of the field strength and its covariant derivatives
\beq
      R[A] = \int d^3x \Bigl[ \a \tr[F^2] + \b \tr[F^4] + \gamma
                \tr[DF DF] + ...\Bigr]
\eeq
For small amplitude, long-wavelength configurations, which are assumed
to dominate the functional integral at long range, we then have
\beq
     \Psi_0[A] \sim \exp\left[-\a \int d^3x \tr[F^2] \right]
\label{vwf} \eeq Lattice Monte Carlo simulations have verified
this form of the ground-state wavefunctional for two instances of
Yang-Mills field configurations:  non-abelian constant fields, and
``abelian'' ($[A_i,A_j]=0$) plane-wave configurations
\cite{Junichi}.

   Assuming the form \rf{vwf} for the vacuum wavefunctional,  we have
that for sufficiently large Wilson loops
\bea
       W_r^{D=4}(C) &\sim& \int DA_k ~ \chi_r[U(C)]
         e^{-2\a \int d^3x \tr[F^2]}
\non \\
              &=& W_r^{D=3}(C)
\eea
and the calculation of the planar Wilson loop in a D=4 dimensional theory
can be reduced to D=3.  The reasoning can be repeated to express the
expectation value of large planar
Wilson loops in $D=3$ dimensions to the corresponding $D=2$ dimensional
calculation, so we have
\beq
       W_r^{D=4}(C) \sim W_r^{D=2}(C)
\eeq
But Wilson loops in D=2 dimensions can be calculated directly from
perturbation theory (the answer is obtained, in an appropriate gauge,
just by exponentiating one-gluon exchange), and the resulting string
tensions satisfy Casimir scaling exactly.  On these grounds,
one argues that string tensions in the $D=4$ dimensional theory should
also satisfy Casimir scaling.

   After the significance of Casimir scaling was re-emphasized in
\cite{Us1}, numerical studies were undertaken by Bali \cite{Bali}
and by Deldar \cite{Deldar} for the SU(3) group, to check just how
accurately this law is satisfied.  It appears that string tensions
satisfy Casimir scaling to quite a high degree of accuracy, as
shown in Fig.\ \ref{Cas}, taken from ref.\ \cite{Bali}.

\subsection{N-ality Dependence}

   For finite $N$ the Casimir scaling law must break down at
some point, to be replaced by a dependence on the N-ality $k$
of the representation
\beq
       \s_r = f(k) \s_F
\label{fk}
\eeq
The reason for this N-ality dependence is color screening by gluons.
Quarks in the adjoint representation, for example, have zero N-ality,
and according to Casimir scaling
\beq
       \s_A = {2 N^2 \over N^2 - 1} \s_F
\eeq
The static quark potential for adjoint quarks separated by a large distance
$R$ would then be roughly $V(R) = \s_A R$.  At some point it becomes
energetically favorable for a pair of gluons to pop out of the vacuum
and bind to each of the quarks, forming two ``gluelumps'' of mass $M_{GL}$.
This is a string-breaking process, and if we only look at energetics it
should be favorable at
\beq
       R_c = {2 M_{GL} \over \s_A}
\eeq
However, expressed in terms of either Feynman or lattice strong-coupling
diagrams, string-breaking is a non-planar process, and its amplitude
is suppressed by a factor of $1/N^2$, as already seen in eq.\ \rf{AandP}
and Fig.\ \ref{scup}.  This explains why Casimir scaling
is exact out to infinite $R$ in the $N=\infty$ limit, even if $M_{GL}$ is
finite in the same limit.  But allowing for non-planar contributions
at finite $N$, we expect that Casimir scaling must break down for, e.g.,
square $L\times L$ Wilson loops at a length scale which grows only
logarithmically with $N$.

   In general, consider a static quark-antiquark pair with the quark
in representation $r$ of N-ality $k$.  As the separation $R$ is increased,
it can become energetically favorable to pop some number of gluons out
of the vacuum to bind with the quark and antiquark charges.  The
energetically most favorable representation of the quark-gluon bound state
will be the lowest dimensional representation $r'$ with same N-ality as
$r$, since gluons (themselves of zero N-ality) cannot change the N-ality
of a source.  But this means that the \emph{asymptotic} string tension
of the quark-antiquark pair is the same as that of a quark-antiquark pair
in the lowest dimensional representations of the same N-ality, which implies
that asymptotic string tension depends only on N-ality $k$, as indicated
in eq.\ \rf{fk}.  This asymptotic string tension, depending only on
N-ality, is known as the ``k-string tension'' $\s(k)$.  The k-representations
are the representations of lowest dimensionality of N-ality $k$.  In terms
of Young tableaux, these are denoted by a single column of $k$ boxes.
Since a charge in a $k$ representation cannot be screened to some lower
representation by binding to gluons, it is reasonable to suppose that
the $k$-string tension is the same as the Casimir string tension, in which
case, asymptotically,
\beq
      \s_r \ra \s(k) = {k(N-k)\over N-1} \s_F
\eeq
for representation $r$ of N-ality $k$.
An alternative behavior is suggested by MQCD and softly broken
${\cal N}=2$ super Yang-Mills theory,
\beq
 \s_r \ra \s(k) = {\sin{\pi k\over N} \over \sin{\pi \over N}}  \s_F
\eeq
which is known as ``Sine-Law scaling'' \cite{DS,Strassler1}.  We note that
for fixed $k$, Casimir and Sine-Law scaling are identical in the
$N\ra \infty$ limit,
\beq
      \s(k) = k \s_F
\eeq

\begin{figure}[h!]
\begin{center}
\begin{minipage}[t]{8 cm}
\centerline{\scalebox{0.50}{\rotatebox{270}
 {\includegraphics{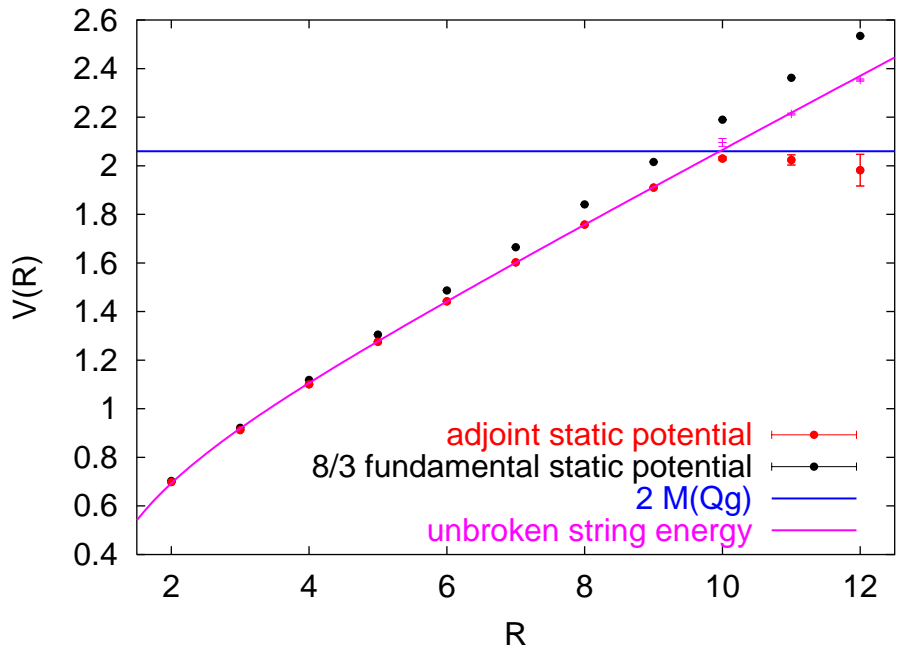}}}}
\end{minipage}
\begin{minipage}[t]{16.5 cm}
\caption{The adjoint and
$\frac{8}{3}\times$fundamental static potentials $V(R)$ vs $R$, in
$D=3$ dimensional SU(2) lattice gauge at $\b=6.0$.
The horizontal line at 2.06(1) represents
twice the energy of a gluelump. From de Forcrand and Kratochvila,
ref.\ \cite{deF-K}. }
\label{sbreak}
\end{minipage}
\end{center}
\end{figure}

   Although we can be fairly sure that the asymptotic string
tension depends only on N-ality, both from the color-screening argument
and from explicit lattice strong-coupling calculations, it is actually
rather difficult to see this dependence explicitly.  The problem is
clear from Fig.\ \ref{Cas}
above, where the color {\bf 8} (adjoint) representation of $SU(3)$ appears
to have a string tension $\s_A$
which is $9/4$ that of the fundamental representation
out to the largest separation calculated. There is no hint of a crossover
to $\s_A=0$, where the potential would go from linearly rising to flat.
The reasons for this are technical.  The potentials in Fig.\ \ref{Cas}
were calculated at $T \ll R$, using the smearing technique discussed in
section 4.1 to increase the overlap to the lowest energy flux tube
state.  This method works very well for the fundamental representation.
However, for the adjoint representation, it may be that the overlap of
the ``smeared'' Wilson line with the two gluelump state is very small,
and therefore one really has to go to time separations $T \sim R$
in order for this state to become dominant in the sum over states
\rf{sos}.  As a practical matter, it is difficult to compute loops
in the adjoint representation which are large enough to observe the
string-breaking effect.

   This problem was originally overcome by operator-mixing methods,
and crossover from Casimir-scaling to asymptotic behavior for
adjoint representation charges was observed numerically in ref.\
\cite{deF-P}. In the last year, however, a clever algorithm
invented by L\"{u}scher and Weisz \cite{lush1} has greatly
increased the attainable accuracy of Wilson loop calculations, and
allows numerical calculation of much larger loops than was
possible previously, even with no further increase in processor
power.  The algorithm was recently exploited by de Forcrand and
Kratochvila \cite{deF-K} to compute $V(R)$ from $\V(R,T)$ at up to
$T \approx 2R$, and at the largest values of $R$ the crossover
behavior to $\s_A=0$ from the (approximately) Casimir scaling
value is clearly seen. In Fig.\ \ref{sbreak} the adjoint
representation potential is shown, for SU(2) lattice gauge theory
at $\b=6.0$ in $D=3$ dimensions, compared to the fundamental
representation string tension multiplied by the Casimir ratio
$C_A/C_F = 8/3$.  The crossover from a linearly-rising,
Casimir-scaling potential to the asymptotic flat behavior is seen
at a separation of 10 lattice spacings.  Note also that, prior to
color-screening, there is a small but significant deviation of the
adjoint potential from exact Casimir scaling.

   Color-screening for representations beyond the adjoint have not yet
been studied numerically.  On the grounds of very general energetics
arguments, however, the dependence of string tensions on N-ality alone,
i.e.\ $\s_r = \s(k)$ is sure to hold.

\subsection{String Behavior}

    Linear Regge trajectories are explained, as we have seen, by a
model in which a meson is a rotating straight line, of constant energy
density, running between a massless quark and antiquark.
Allowing for quantum fluctuations of the line in the transverse direction
leads to the Nambu-Goto action
\beq
        S_N = {1\over 2\pi \a'} \int d^2\s ~ \sqrt{\det[g]}
\eeq
where $x^\m(\s^1,\s^2)$ are coordinates of the worldsheet swept out by
the line running between the
quark and antiquark, as it propagates through D dimensional spacetime.
Wick rotation to Euclidean spacetime is understood, and $g_{ab}$
is the induced metric on the worldsheet
\beq
        g_{ab} = {\pa x^\m \over \pa \s^a} {\pa x^\m \over \pa \s^b}
\eeq

    The Nambu-Goto action, although inspired from hadronic physics, is
not an adequate fundamental theory of mesons.  For one thing, quantization
is only consistent in D=26 dimensions, and for another, the lowest lying
state is tachyonic.  Nevertheless, if we view this action as an effective
theory of the QCD flux tube (with an implicit high-momentum cutoff of
some kind), then it is possible to explore the effect, on the static
quark potential, of quantum fluctuations of the electric flux tube in
the transverse directions.
This investigation was first carried out by L\"{u}scher
in ref.\ \cite{lush2} (see also ref.\ \cite{Orlando}).

   The first step is to introduce static quarks by demanding that the
string worldsheet is bounded by an $R\times T$ loop, with $T \gg R$.
Denote coordinates by $x^\m = (x_0,x_1,\xp)$, and use the invariance
of Nambu-Goto action under reparametrizations to set $\s^0 = x_0,~
\s^1 = x_1$, with the $R\times T$ loop lying in the $x_0-x_1$ plane at $\xp=0$.
Then the static potential, in this model, is given by
\bea
      e^{-V(R)T} &=& \int Dx^\m ~ e^{-S_N}
\non \\
                 &=& \int D\xp ~ \exp\left[-{1\over 2\pi \a'}
 \int_0^T dx_0 \int_0^R dx_1 \sqrt{1 + (\pa_0 \xp)^2 + (\pa_0 \xp)^2
           + O[(\pa \xp)^4]} \right]
\non \\
\eea
Expanding the action to second order in the transverse fluctuations
\bea
      e^{-V(R)T} &=& e^{-\s RT} \int D\xp ~ \exp\left[ - \s
   \int_0^T dx_0 \int_0^R dx_1 ~
    \left( \oh (\pa_0 \xp)^2 + \oh (\pa_1 \xp)^2 \right) \right]
\non \\
          &=&  e^{-\s RT} \Bigl[\det(-\pa^2_0 - \pa^2_1)\Bigr]^{(D-2)/2}
\eea
where the determinant of the two-dimensional Laplacian operator, subject
to Dirichlet boundary conditions on the $R\times T$ loop, is regulated
and evaluated via standard zeta-function methods.  The result (for $T\gg R$)
is that
\beq
     \Bigl[\det(-\pa^2_0 - \pa^2_1)\Bigr]^{(D-2)/2} = \exp\left[-{D-2\over 2}
         {\pi T \over 12 R} \right]
\eeq
which leads to the static potential
\beq
       V(R) = \s R - {\pi (D-2) \over 24} {1 \over R}
\label{lushterm}
\eeq
The term proportional to $1/R$ is known as the {\bf L\"{u}scher term},
and its origin is completely different from the Coulombic term in
3+1 dimensional QCD
due to one-gluon exchange at short distances.  The L\"{u}scher term
is due to quantum fluctuations of the string, rather than one-gluon
exchange, and can viewed as the Casimir energy of the string arising
from the Dirichlet boundary conditions.

     The bosonic string model also predicts how the thickness of a
QCD flux tube depends on the separation $R$ of the quarks.  In QCD,
the energy density due to the color electric field parallel to the
flux tube is determined by the connected
correlator of a timelike plaquette $P$ parallel to
the $R\times T$ loop $C$
\beq
      \E = {\lla \tr[U(C)] \tr[U(P)] \rra \over
               \lla \tr[U(C)] \rra } - \lla \tr[U(P)] \rra
\label{corr}
\eeq
and the thickness of the flux tube is determined from the falloff
of $\E$ in the directions $\xp$ transverse to the plane of the $R\times T$
loop.  The analogous calculation, in string theory, is to calculate
the loop-loop correlator $G(C_1,C_2,h)$  given by the functional integral
over string worldsheets with two parallel loops $C_1$ and $C_2$,
separated by a transverse distance $h$, as boundaries of the worldsheet.
Loop $C_1$ represents the large $R\times T$ loop, while $C_2$ represents
the probe plaquette.

   It is not essential for the loops to be rectangular, and it simplifies
the calculation if both loops are circular and concentric.  Providing
the transverse displacement $h$ is not too large compared to the radii
$R_1 > R_2$ of loops $C_1,C_2$, the leading contribution to the
loop-loop correlation function can be computed semiclassically
\beq
       G(C_1,C_2,h) = \exp\left[-\s \A_{min}(C_1,C_2,h)\right]
\label{GC}
\eeq
where $\A_{min}(C_1,C_2,h)$ is the area of the
minimal surface with boundaries at $C_1,C_2$.  The answer is found to
be \cite{lush2}
\beq
        \A_{min}(C_1,C_2,h) \approx \pi(R_1^2 - R_2^2)
          +{\pi \over \log(R_1/R_2)} h^2
\eeq
Plugging this into \rf{GC}, we see from the Gaussian falloff of
$G(C_1,C_2,h)$ with $h$
that the width of the flux tube grows logarithmically
with the quark-antiquark separation (which for this geometry is on the
order of $R_1$).  The semiclassical calculation breaks down at some point,
since at some $h$ the surface degenerates into a sheet spanning the minimal
area of $C_1$, a sheet spanning the minimal area of $C_2$, with a tube
of infinitesimal radius connecting to the two sheets. Beyond this point,
a quantum treatment is required.  One approach is to simply calculate,
for the quantized Nambu string, the expectation value
\beq
         d^2 = \lla \xp^2(x_0=T/2,x_1=R/2) \rra
\eeq
The answer can be expressed as a divergent sum over string modes.
Cutting off the sum at some $k=k_{max}$ gives the result
\beq
         d^2 \sim {D-2 \over \s} \log(k_{max}R)
\eeq
As in the semiclassical calculation, the width of the flux tube is found
to grow logarithmically with quark separation.

   Since the Nambu action is clearly not the correct description of
quark-antiquark forces, because of the pathologies already mentioned,
one might doubt that the L\"{u}scher
term or the logarithmic growth of the flux tube width, derived from
bosonic string theory, have anything to do with QCD.  In recent years,
however, developments in M-theory $-$ specifically the AdS/CFT correspondence
conjectured by Maldacena  \cite{AdS} $-$ have lent strong support to the string
picture of the QCD flux tube.  The string description of the
confining force, in the AdS/CFT approach, is not the old Nambu string.
Instead it is a 10-dimensional superstring in a peculiar metric, which
nonetheless describes the physics of $D=4$ dimensional QCD.
Both the L\"{u}scher term and the logarithmic growth of the flux tube
are natural in this framework \cite{Israelis,Me+Poul}.  The catch is
that the AdS/CFT correspondence for QCD is (so far) restricted to
strong-coupling calculations with a short-distance cutoff, much like
strong-coupling lattice gauge theory.  The  AdS/CFT
correspondence is beyond the scope of this review, and the reader is
referred to ref.\ \cite{AdS} for an exposition of the subject.

\begin{figure}[t!]
\begin{center}
\begin{minipage}[t]{8 cm}
\centerline{\includegraphics[scale=0.5]{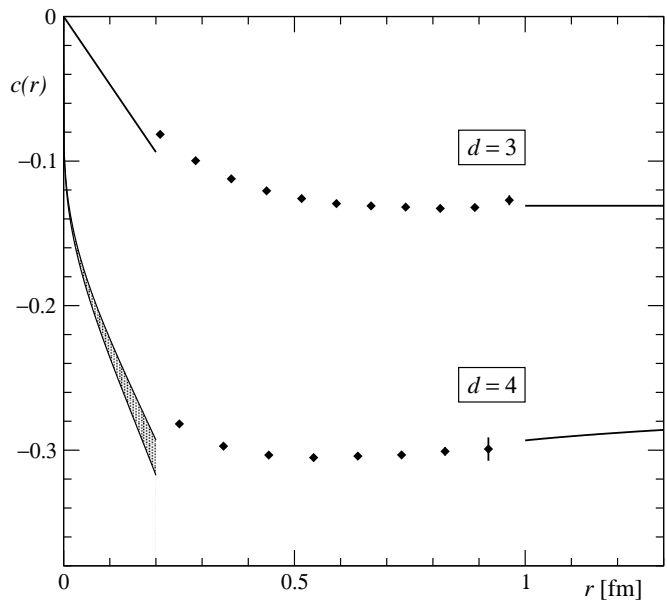}}
\end{minipage}
\begin{minipage}[t]{16.5 cm}
\caption{Coefficient of the L\"{u}scher term $c(r)$ in D=3 and D=4
dimensions.  The Nambu-Goto string action predicts $c=-0.262$ for D=4
and $c=-0.131$ for D=3. From L\"{u}scher and Weisz, ref.\ \cite{lush3}.}
\label{lush}
\end{minipage}
\end{center}
\end{figure}

   One of the implications of a logarithmic growth in the width of
the QCD flux tube with quark separation is that the width is infinite
at $R=\infty$.  This is a clear prediction of string models.
In contrast, in strong-coupling lattice gauge theory the loop-plaquette
correlation function \rf{corr} can be calculated to high orders in $\b$,
and the width of the flux tube $w_\infty$ is found to be finite at $R=\infty$
for small $\b$.  However, the power series has a finite radius of convergence,
and Pade and other extrapolation methods indicate that $w_\infty \ra \infty$
at a finite $\b=\b_c$ \cite{rough}.
The phase transition at this point is known as the ``roughening transition,''
which is known to occur in various models in statistical mechanics which
involve the thermal fluctuations of a boundary.  It therefore appears
likely that $w_\infty=\infty$ is a genuine feature of lattice QCD at
weak couplings, in accordance with the prediction of string theory.

   Verification of the L\"{u}scher term via lattice Monte Carlo simulations
is a difficult task, and until rather recently the numerical evidence favoring
this term was unconvincing.  It
requires a very precise measurement of the static potential, at distances
where the potential is almost linearly rising,
to accurately detect the small deviation from linearity predicted by
string theory.  Recently, using a powerful algorithm for computing
Polyakov line correlators \cite{lush1}, L\"{u}scher and Weisz \cite{lush3}
were able
to compute the static potential in SU(3) lattice gauge theory to
very high accuracy, and extracted
\beq
       c(R) = \oh R^3 {\pa^2 V \over \pa R^2}
\eeq
from the second lattice derivative of $V(R)$.
The results for $c(R)$ vs.\ $R$ in $D=3$ dimensions
at $\b=20$, and $D=4$ at $\b=6.0$,
are shown in Fig.\ \ref{lush}.  In both cases, the asymptotic value of
$c(R)$ is consistent with the Casimir energy of bosonic string theory,
i.e.\ the L\"{u}scher term in eq.\
\rf{lushterm}.  This result has been confirmed by the independent
investigation of Juge et al.\ \cite{Kuti} (see also Necco and Sommer
\cite{NS} and Teper \cite{Teper1}).  A common conclusion of all these
investigations is that the L\"{u}scher term (which is evidence of
string-like behavior) makes its appearance at a surprisingly short quark
separation, on the order of $0.5$ fm.

\section{The Center Vortex Confinement Mechanism}

   In this and the
following sections we explore some possible explanations of the
confining force, beginning with the proposal which is most closely
tied to the $Z_N$ center symmetry \cite{thooft1,cv1,cv2,cv3,cv4}.

   There are several motivations for the center vortex mechanism.
The first is the fact
that the asymptotic string tension $\s_r$, extracted from
Wilson loops in the $r$ representation of color SU(N), depends only on
the N-ality $k_r$ of the representation; i.e.
\beq
      \s_r = \s(k_r)
\label{sr}
\eeq
In particular, $\s_r=0$ if $k_r=0$.  A second motivation is the fact that
confinement is associated with the unbroken realization of a global
center symmetry, as explained in section 3.2.
Spontaneous or explicit breaking of this global
symmetry results in a non-confining theory.   Finally there is the
idea, which is common to a number of proposals about the confinement
mechanism, that vacuum configurations in pure SU(N) gauge theory
can be decomposed into high-frequency, non-confining perturbations $\tA_\m(x)$
around a smooth confining background $\A_\m(x)$.  The sub-leading,
perimeter-law falloff is mainly due to fluctuations in $\tA_\m$, while the
long-wavelength background is responsible for the area law.\footnote{Note
that $\tA_\m$ is not necessarily entirely perturbative. It might
include, e.g., short-range effects due to the finite thickness of
monopoles, vortices, instantons, or any other solitons of importance.}

   Let us focus in particular on large N-ality $=0$
Wilson loops which have asymptotic string tensions $\s_r=0$.
Somehow the long-wavelength
confining fluctuations $\A_\m$ don't affect the zero N-ality loops, which
implies
\beq
       \chi_r[U(C)] \approx \chi_r[\tU(C)]
\label{zeroN}
\eeq
where
\bea
    U(C) &=& P\exp\left[i\oint_C dx^\m ~
           \Bigl(\A_\m(x) + \tA_\m(x) \Bigr) \right]
\non \\
   \tU(C) &=& P\exp\left[i\oint_C dx^\m ~
                 \tA_\m(x) \right]
\non \\
   \U(C)  &=& P\exp\left[i\oint_C dx^\m ~
                 \A_\m(x) \right]
\eea
But how can eq.\ \rf{zeroN} be true, for any confining background
$\A_\m$?  The most straightforward possibility is that
\beq
       U(C) = Z(C) g\tU(C)g^{-1}
\eeq
where $Z(C) \in Z_N$ is a center element, and therefore $\U(C) = Z(C)$.
The lattice configurations $\U_\m(x)$ which satisfy $\U(C)=Z(C)$
on any loop $C$ can be identified uniquely:
\beq
      \U_\m(x) = z_\m(x) g(x) g^{-1}(x+\widehat{\m}) ~~~,~~~ z_\m(x) \in Z_N
\eeq
These are the thin center vortices, created by singular gauge transformations.
The field strength of such a configuration is singular on the vortex surface,
but of course this singularity can be smoothed out, in a typical
vacuum configuration, by the $\tA_\m(x)$ contribution.

   In the center vortex picture, confinement is simply a matter of
short range correlations in the center flux on minimal surfaces.  Suppose
that we can subdivide any plane on the lattice into $L\times L$ squares
of area $\S_{min}$, such that if $C_1$ and $C_2$ are $L\times L$ loops
around any two different squares, then
\beq
\lla Z(C_1) Z(C_2) \rra \approx \lla Z(C_1) \rra \lla Z(C_2) \rra
\label{vsrc}
\eeq
Then consider a planar loop $C$ of area $\S$ which encloses a number
$\S/\S_{min}$ of the $L\times L$ squares $\S_i$, each bounded by a loop $C_i$.
Using the commutative property of the $Z(C_i)$, and the assumed
property of short range correlations among areas $\S_i$
\bea
     \lla \chi_r[\U(C)] \rra &=& \lla \chi_r\left[\prod_{i=1}^{\S/\S_{min}}
         Z^{k_r}(C_i) I_N \right] \rra
\non \\
  &\approx&  \chi_r[I_N] \prod_{i=1}^{\S/\S_{min}} \lla Z^{k_r}(C_i) \rra
\non \\
            &=& d_r \exp[-\s(k_r) \S]
\eea
where
\beq
       \s(k) = -{\log\Bigl[\lla Z^k(C_i) \rra \Bigr] \over \S_{min} }
\eeq
is the string tension of the loop, depending only on the N-ality.

\subsection{Vortices as Local Minima of the Lattice Action}

\subsubsection{$N > 4$}

   We begin by showing that center vortices are local minima of
the Wilson lattice action, and that they also must exist as local
minima of long-range effective lattice actions.  Confinement is then
a matter of whether or not vortex surfaces
percolate through the lattice volume.
The proof for the Wilson action, which is due to Bachas and Dashen
\cite{BD}, is quite simple:
Any thin center vortex in SU($N$) lattice gauge theory
is gauge equivalent to
\begin{equation}\label{thin}
U_\mu(x)\ =\ Z_\mu(x) I_N
\end{equation}
with
\[
\nonumber
Z_\mu(x)=\exp\left(\frac{2\pi i n_\mu(x)}{N}\right),
\quad
n_\mu(x)=1, 2, \dots, N-1.
\]
Now write a small deformation of this configuration as
\begin{equation}
U_\mu(x)=Z_\mu(x) V_\mu(x),
    \qquad V_\mu(x)=e^{i A_\mu(x)}
\label{deform1}
\end{equation}
with
\begin{equation}
    A_\mu(x)=\sum_a A_\mu^a(x)t_a,\qquad \vert A_\mu^a(x)\vert \ll 1.
\end{equation}
Substituting (\ref{deform1}) into the Wilson action
\begin{equation}
S=\frac{\beta}{2N}\sum_P
    \left(2N-\mbox{Tr}[U_P]-\mbox{Tr}[U^\dagger_P]\right)
\end{equation}
we obtain
\begin{equation}
S=\frac{\beta}{2N}\sum_P
    \left(2N-Z_P\mbox{Tr}[V_P]-Z^\ast_P\mbox{Tr}[V^\dagger_P]\right)
\end{equation}
Writing the product of $V$-link variables around a plaquette in terms of
a field strength
\begin{equation}
V_P=e^{i F_P}=I_N+i F_P -\oh F_P^2 +{\cal{O}}(F_P^3),
\label{deform2}
\end{equation}
and also writing for $Z_P$
\begin{equation}
Z_P=\exp\left(\frac{2\pi i n_P}{N}\right)
\label{deform3}
\end{equation}
the Wilson action becomes
\beq
S=\frac{\beta}{2N}\sum_P
    \left[2N\left(1-\cos\left(\frac{2\pi  n_P}{N}\right)\right)
  +\cos\left(\frac{2\pi  n_P}{N}\right) \mbox{Tr}[F_P^2]
    +{\cal{O}}(F_P^3)\right].
\eeq
This action has a local minimum at $\mbox{Tr}[F_P^2]=0$ if the
coefficient in front of $\mbox{Tr}[F_P^2]$, at each plaquette $P$,
is positive.  That requirement is trivially satisfied at non-vortex
plaquettes ($n_P=0$).  For vortex plaquettes ($n_P\ne 0$), we have a
positive coefficient, and hence a local minimum at $\mbox{Tr}[F_P^2]=0$,
providing that
\beq
\cos\left(\frac{2\pi n_P}{N}\right)>0 ~~~,~~~
\mbox{i.e.\ }\quad\frac{n_P}{N}<\frac{1}{4}
    \quad \mbox{\ or\ } \quad
    \frac{N-n_P}{N}<\frac{1}{4}.
\label{condition}
\eeq
This  condition cannot be satisfied for $N\le4$.  However,
beginning with $n_P=1, 4$ at $N=5$, vortices appear as local minima
of the Wilson action already at the classical level.

   This simple result can of course be extended \cite{kstring}
beyond the simple Wilson action, and therein lies its physical
relevance.  It is obvious that the action of thin vortices, stable
or not, is singular in the continuum limit, and their contribution
to the functional integral can be neglected. The configurations of
physical interest are vortices which have some finite thickness
$d$ in physical units.  In order to investigate the stability of
vortices of some thickness $d$ (or of lesser thickness, but
including quantum fluctuations of the vortex surface up to that
scale), starting from a lattice action at spacing $a$, we imagine
following the renormalization group approach, successively
applying blocking transformations of the form
\begin{equation}
e^{-S'[U']}=\int DU\;\delta[U'-F(U)]e^{-S[U]}
\label{RG}
\end{equation}
where $U'_\m$ are links on the blocked lattice, and $F(U)$ is the blocking
function, until the lattice scale of the blocked action equals $d$.

    In this kind of approach, one usually assumes that only a few contours
(plaquettes, 6 and 8-link loops, etc.) are important in the effective
action on the blocked lattice. Given an effective action at length scale $d$,
it is possible to study ``thin'' (thickness$=d$) vortices at that scale.
For simplicity, consider
lattice actions which consist of plaquette ($P$) and $1\times 2$ rectangle
($R$) terms:
\beq
S_I = c_0 \sum_{P} \Bigl( N-\mbox{ReTr}[U(P)] \Bigr)
    + c_1 \sum_{R} \Bigl( N-\mbox{ReTr}[U(R)] \Bigr)
\eeq
This class of actions has been widely discussed in the lattice literature,
and includes the tadpole-improved action \cite{tadi}, the Iwasaki action
\cite{Iwasaki}, and two-parameter approximations \cite{deForc2} to the
Symanzik action \cite{Symanzik} and the DBW2 action \cite{DBW2}.

    The thin vortex configuration, eq.\ (\ref{thin}), is easily
shown to be a stable minimum of the action $S_I$ providing both $c_0$
and $c_1$ are positive; the proof is as simple as in the case of the
Wilson action. But for most improved actions of this type one has
$c_0>0$ and $c_1<0$, and the proof of stability is a little more
involved.  It can be
shown that the condition for vortex stability is,
first, that the trivial vacuum $U=I$ is a global minimum of $S_I$,
which requires
\begin{equation}
c_0+8 c_1>0
\end{equation}
and, second, that the vortex stability condition for the Wilson
action, eq.\ \rf{condition} also holds.  The proof of this
statement can be found in ref.\ \cite{kstring}.

   So it appears that the result first derived by Bachas and Dashen \cite{BD}
is very robust, and holds for a large class of improved actions in
coupling-constant space.  It seems unlikely that this result would be
altered by adding a few more contours, as in certain proposed perfect
actions \cite{perfect}.  If, in fact, center vortices are local minima
of the effective action all along the renormalization trajectory, then
their effects must become apparent at some scale.  This is because
the entropy factor increases with vortex surface area as
\beq
\exp[+\mbox{const}\cdot
(\mbox{Vortex Area})]
\eeq
while the Boltzmann suppression factor
goes like
\beq
\exp[-(\mbox{Vortex Area})/\kappa^2(d)]
\eeq
As $d$ increases, so does $\kappa^2(d)$.
Eventually entropy wins over action, and
vortices at that scale will percolate through the lattice.

\subsubsection{$N \le 4$}

   We next consider $N \le 4$.  Vortices are certainly not minima
of the Wilson or two-parameter actions in this case, but there are
good reasons to believe that they are local minima of the effective action
at large scales.  Suppose we have carried
out RG blocking transformations up to some scale $La$ which is well beyond
the adjoint string-breaking scale.  Define the probability distribution
for loop holonomies $U'(C)$ around contours $C$ on the blocked lattice
\bea
        P_C(g) &=& \lla \d[g,U'(C)] \rra
\non \\
               &=& \left\langle \sum_k \sum_{r_k}
                     \chi_{r_k}[U'(C)] \chi_{r_k}[g^\dg] \right\rangle
\non \\
               &=&  \sum_k \sum_{r_k} W_{r_k}(C) \chi_{r_k}[g^\dg]
\label{PC1}
\eea
where the sums are over N-ality $k$, and over group representations $r_k$ of
a given N-ality, and the $W_{r_k}(C)$ are Wilson loop expectation values
on the blocked lattice.  The leading terms (largest $W_{r_k}(C)$) come from
the lowest $k,N-k$ values, and, for each $k$, the representation of lowest
dimension.  Keeping only the leading terms at $k=0,1,N-1$ we have
\bea
      P_C(g) &\approx& 1 + W_A(C) \chi_A(g) + W_F(C)(\chi_F(g) + \mbox{c.c.})
\non \\
    &=& 1 + e^{-\m \mbox{\scriptsize Perim}(C)  - c_0} \chi_A(g)
          + e^{-\s_F \mbox{\scriptsize Area}(C)
               -\m_F \mbox{\scriptsize Perim}(C) - c_1}
                 (\chi_F(g) + \mbox{c.c.})
\label{PC2}
\eea
where $A,F$ refer to the adjoint and fundamental representations.

   $P_C(g)$ is the probability density that a given loop holonomy has the
SU(N) group value $g$.  Because of color screening, the coefficient of
$\chi_A(g)$ is far greater than that of $\chi_F(g)$ at large blocking scales.
Since $\chi_A(1) = \chi_A(z)$, this means that the loop probability density
has local maxima at center elements.  The only configurations which locally
maximize all loop probabilities and are consistent with the Bianchi
constraint are the center vortex configurations.  We conclude that the
effective action at this blocking scale has local minima at the vortex
configurations.

   This conclusion can be derived explicitly in strong-coupling lattice
gauge theory, where the blocking transformation
\beq
       e^{-S'[U']} = \int DU ~ \prod_{l'}
           \d[U'_{l'}-(UUU...U)_{l'}] e^{-S[U]}
\eeq
from a lattice of spacing $a$ to a lattice of spacing $La$
can be carried out analytically \cite{scup}.
If $g$ is the lattice coupling constant
at lattice spacing $a$, one finds
\bea
     -S'[U'] &=&
  \sum_{p'} \left\{ 2 N\left({1\over g^2 N}\right)^{L^2}
  \mbox{ReTr}_F [U'(p')]
 + 4(D-2) \left({1\over g^2 N}\right)^{4(4L-4)}
  \mbox{Tr}_A [U'(p'] \right\}
\non \\
  & & + \mbox{larger adjoint loops with $e^{-P}$ coefficients}
\eea
The adjoint loops are non-planar contributions, and arise from the
``tube'' diagrams shown in Fig. \ref{scup}.  The non-local effective action
can be expressed
(with some additional complications)
as a local action containing a number of adjoint Higgs fields \cite{scup},
but for simplicity we will simply consider the above action
truncated to the leading plaquette terms, i.e.
\beq
   -S'_{trunc}[U'] = \sum_{p'} \Bigl\{ c_0 \mbox{ReTr}_F[U'(p')] +
                                   c_1 \mbox{Tr}_A[U'(p')] \Bigr\}
\eeq
\ni with
\beq
       c_0 \sim \exp[-\s \mbox{~Area}(p')]  ~~~,~~~
   c_1 \sim \exp[-4 \s \mbox{~Perimeter}(p')]
\eeq
Consider small fluctuations around a vortex configuration
\rf{deform1},\rf{deform2},\rf{deform3}
\beq
  S'_{trunc} = \mbox{const} + {1\over 2} \sum_{p'} \left\{
       c_0 \cos\left({2\pi n_{p'}\over N}\right) \mbox{Tr}_F[F_{p'}^2]
     + c_1 \mbox{Tr}_A[F_{p'}^2] \right\}
\eeq
It is clear that at large blocking $L$, $c_1$ is much greater than
$c_0$. Then $S'_{trunc}$, at the scale where $c_1 \gg c_0$, is
minimized at $F_{p'}=0$.  This means that at this scale, vortex
configurations are local minima of the coarse-grained action
even for $N=2,3,4$.

\subsubsection{Center Monopoles}

   Thin center vortices on a classical SU(N) vacuum background, which can
be written in the form \rf{thin}, can obviously be identified
with configurations in $Z_N$ lattice gauge theory.  In $D$ dimensions
these vortices form $D-2$ dimensional surfaces.  The Bianchi identities
of $Z_N$ lattice gauge theory require that these surfaces are either
closed, or else three or more surfaces may meet at $D-3$ dimensional
vertices,  which are the monopoles of
$Z_N$ lattice gauge theory \cite{Yoneya,thooft1}.

   It is easiest to visualize matters in D=3 dimensions.  A $Z_N$ vortex
pierces a given plaquette $p$ if $Z(p) \ne 1$.
We introduce a {\bf dual lattice}
whose sites lie in the center of unit cubes of the original lattice.
Links of the dual lattice run through the middle of plaquettes on the
original lattice.  The vortex which pierces a set of plaquettes on the
original lattice can be visualized as a path on the dual lattice.

\begin{figure}[t!]
\begin{center}
\begin{minipage}[t]{8 cm}
\centerline{\includegraphics[scale=0.35]{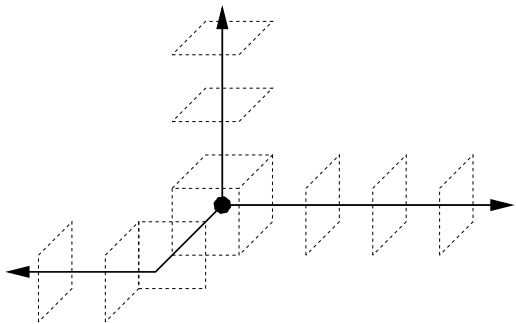}}
\end{minipage}
\begin{minipage}[t]{16.5 cm}
\caption{Three vortices (solid lines) diverging from a
monopole (solid circle) on the dual lattice, in $Z_3$ lattice
gauge theory.  The vortices pierce
plaquettes on the original lattice (dashed contours); oriented loops
around these plaquettes, with normals in the direction of the vortices,
have a value of $\exp[2\pi i/3]$.}
\label{cmon}
\end{minipage}
\end{center}
\end{figure}

   Now consider a set of plaquette loops $\{p_1,p_2,..,p_6\}$ on a unit
cube of the lattice, with the orientation of the loops chosen so that the
normal to the plaquette points out of the cube.  It is easy to verify
the Bianchi identity
\beq
         \prod_{i=1}^6 Z(p_i) = 1
\eeq
In, e.g., $Z_3$ lattice gauge theory, this identity allows three
vortices to diverge from the dual lattice site at the center of the cube,
as indicated in Fig.\ \ref{cmon}.  The dual lattice site is then the position
of a $Z_3$ monopole.  Of course this generalizes to higher $N$: several
vortices can diverge from a $Z_N$ monopole providing the product of
factors $\exp[2\pi i n/N]$, associated with each vortex line, is equal
to one.  In D=4 dimensions vortices are surfaces on the dual lattice,
joining at monopole worldlines.

    The local minima of the effective SU(N) lattice actions discussed above
are simply the vortices and monopoles of a $Z_N$ lattice configuration.
It has been argued recently that center monopoles play an important role in
determining the precise N-ality
dependence of asymptotic string tensions at large N \cite{kstring}.

\subsection{Vortices as Local Minima of the One-Loop Effective Action}

   In continuum SU(N) gauge theory, the only non-trivial classical
solutions of the theory in D=4 Euclidean space are the instantons.
The situation changes however, as shown by Diakonov and Maul
\cite{Mitya}, when quantum corrections at one-loop order are
included.  Diakonov and Maul calculate the Yang-Mills effective
action, to one loop order, of the following configuration (in
cylindrical coordinates) \beq
       A_\p^a(r,\p,x_\|) = \d^{a3} {\m(r)\over r} ~,~~~
       \m(0) = 0 ~,~~~ \m(\infty) = 1
\eeq
where $\m(r)$ is a profile of the vortex cross-section.
Any large Wilson loop in the fundamental representation, located far from
$r=0$ but winding $n$-times around the vortex, has the value
$\oh \tr[U(C)] = (-1)^n$ in this configuration.  This configuration is
therefore a thick center vortex on a classical vacuum background.
By the symmetry of the
vortex configuration, the effective vortex action is proportional the extent
of spacetime $L_\|^{D-2}$ in the directions parallel to the vortex, and
orthogonal to the $(r,\p)$-plane; i.e.\
\beq
         S_{vortex} = \E L_\|^{D-2}
\eeq
where $\E$ is the action density of the vortex in the transverse directions.
Let $r_0$ be the scale of the vortex cross-section (e.g.\
$\m(r) = \exp[-r_0/r]$).  Including quantum corrections up to one loop,
Diakonov and Maul find a transverse action
density for the vortex in D=4 directions
\beq
      \E = - {A \over r_0^2}{11 \over 12\pi^2} \log[\L_{QCD}r_0 B]
\eeq
where $A,B$ are dimensionless constants which depend on the precise
functional form of $\m(r)$.  Classically the vortex
action is always minimized by sending the vortex thickness $r_0 \ra \infty$.
At the one loop level, however, the one-loop vortex action is minimized
at a finite vortex thickness
\beq
       r_0 = {\sqrt{e}\over \L_{QCD} B}
\eeq
and action density at the minimum
\beq
       \E = - \L_{QCD}^2 {11\over 12\pi^2}{AB^2 \over e}
\eeq
(where $e=2.718...$).  A similar analysis also finds a minimum thickness
in $D=3$ dimensions, when one loop quantum corrections are included.
The issue of vortex stability at one loop level is still not
completely settled however; c.f.\ Bordag \cite{Bordag}
for a discussion of this point.

\subsection{Locating Vortices in Monte Carlo Simulations}

   It is a simple matter to create center vortices on a given background;
this is accomplished by singular gauge transformations.  The more difficult
problem is to locate vortices in a given lattice configuration, but this
can be managed by the technique of center projection
in an adjoint gauge, which we now discuss.

   An adjoint gauge is a complete gauge-fixing condition for link
variables $U_{A\m}(x)$ in the adjoint representation, leaving a
residual $Z_N$ symmetry for the links in the fundamental representation.
The gauge-fixing condition may be
that some functional of the links ${\cal R}[U_A]$ is zero or extremal,
such as maximal center gauge, or it may fix the lattice to a
unitary gauge of some kind, such as
Laplacian center gauge.  We can express the fundamental
representation link variables in adjoint
gauge as
\beq
        U^{ag}_\m(x) = Z_\m(x) V_\m(x)
\eeq
where $Z_\m(x)$ is the center element which is closest, on the SU(N)
group manifold, to $U^{ag}_\m(x)$.  The mapping $U^{ag}_\m \ra Z_\m$
is known as {\bf Center Projection} \cite{IMC}.

  A center vortex is created, in an arbitrary configuration {$U$}, by making
a singular gauge transformation.  Call the result {$U'$}.
The corresponding link variables in
the adjoint representation, {$U^{A}$} and {$U'^A$},
are gauge equivalent.  It follows that in an adjoint gauge,
$U_A$ and $U'_A$ are transformed into the same configuration
$U^{ag}_A$. The original fundamental link configurations
{$U$} and {$U'$} are therefore transformed into configurations which
can only differ by center elements, i.e.
\bea
       U'^{ag}_\m(x) &=& Z'_\m(x) U^{ag}_\m(x)
\non \\
                     &=& Z'_\m(x) Z_\m(x) V_\m(x)
\eea
where $Z'_\m(x)$ is gauge-equivalent to the thin vortex created by
the singular gauge transformation.
Therefore, each time a thin vortex is created in a given SU(N) configuration,
the effect is to add a thin vortex to the corresponding center-projected
configuration.  For this reason, it is natural to interpret center-projected
configurations $Z_\m(x)$, of an SU(N) configuration in an adjoint gauge,
as locating the position of center vortices in the full SU(N) configurations.
The excitations of any $Z_N$ configuration are center vortices; the
vortices of projected configurations are known as {\bf P-vortices}, and the
negative plaquettes in the projected configurations are known as
{\bf P-plaquettes} \cite{IMC}.

   The weakness of this reasoning, which says that center projection in any
adjoint gauge will locate the same vortices,
is that it ignores the fact that center vortices in SU(N) gauge
theories would have  finite thickness, and different gauges
may be more or less sensitive to this. In addition, many gauges suffer from
the Gribov copy problem, and the center projection of different gauge
copies may differ.  Fortunately there are ways of checking whether
P-vortices in a particular gauge actually locate thick vortices
in the unprojected theory, by calculating the correlation of P-vortices
with gauge-invariant observables.  These correlations will be discussed
in the next section.  Some adjoint gauges which have been
found useful for vortex finding in SU(2) lattice gauge theory are listed below.

\subsubsection{Direct Maximal Center Gauge \cite{DMC}}

   Direct maximal center gauge, which is simply the lattice Landau gauge in the
adjoint representation, can also be regarded \cite{ER3}
as a best fit of a lattice
configuration $U$ by a thin vortex configuration
$\U_\m(x)=g(x)Z_\m(x)g^\dg(x+\m)$.

   A typical thermalized lattice $U_\m(x)$ generated in Monte Carlo
simulations, if printed out, looks like a
set of random numbers.  But of course it is not random, and locally,
at couplings $\b \gg 1$, the configurations are small fluctuations around
classical vacua
\beq
       U_\m(x) \approx g(x) g^\dg (x+\hat{\m})
\eeq
Suppose we ask for the best fit, to a given lattice configuration
$U_\m(x)$, by a pure gauge $g(x) g^\dg (x+\hat{\m})$.
We will consider, for simplicity, the SU(2) gauge group (the approach
is readily generalized to SU(N)).  A best fit will minimize the distance
on the SU(2) group manifold
\bea
     d^2_F &=&  \sum_{x,\m} \mbox{Tr} \left[
     \left( U_\m(x) - g(x)g^\dg(x+\hat{\m}) \right)
     \times \Bigl(\mbox{h.c.}\Bigr) \right]
\non \\
           &=&  \sum_{x,\m} 2 \mbox{Tr} \left[
                I - g^\dg(x) U_\m(x) g(x+\hat{\m}) \right]
\non
\eea
Define
\beq
       {}^gU_\m(x) = g^\dg(x) U_\m(x) g(x+\hat{\m})
\eeq
Minimizing $d^2$ is the same as maximizing
\beq
         \sum_{x,\m} \mbox{Tr}[{}^gU_\m(x)]
\eeq
which is the lattice Landau gauge.  Fixing to
lattice Landau gauge is therefore
equivalent to the problem of finding the best fit of the given
lattice to a vacuum configuration.

   In the same spirit, we may ask for the best fit of the lattice by a thin
vortex configuration
\beq
       \U_\m(x) = g(x) Z_\m(x) g^\dg(x+\hat{\m})
\eeq
In maximal center gauge, this is accomplished in two steps.  First,
since the adjoint representation is blind to $Z_\m$, find the best fit to the
adjoint representation lattice by a (classical) vacuum configuration.  The
best fit minimizes
\bea
     d^2_A  &=&  \sum_{x,\m} \mbox{Tr}_A \left[
     \left( U_\m(x) - g(x)g^\dg(x+\hat{\m}) \right)
     \times \Bigl( \mbox{h.c.}\Bigr) \right]
\non \\
         &=&  \sum_{x,\m} 2 \mbox{Tr}_A \left[
                I - g^\dg(x) U_\m(x) g(x+\hat{\m}) \right]
\eea
which is equivalent to maximizing the quantity
\beq
          R = \sum_{x,\m} \mbox{Tr}_A[{}^gU_\m(x)]
\eeq
The condition that $R$ is a maximum is the adjoint representation
version of Landau gauge, and is also known as {\bf Direct Maximal Center
Gauge} (DMC).

   The second step, for fixed $g(x)$ determined by DMC gauge, is
to choose $Z_\m(x)$ so as to minimize
the distance function in the fundamental representation
\beq
     d^2 =  \sum_{x,\m} \mbox{Tr} \left[
     \left( U_\m(x) - g(x)Z_\m(x)g^\dg(x+\hat{\m}) \right)
      \times \Bigl( \mbox{h.c.}\Bigr) \right]
\eeq
and this is achieved by setting
\beq
     Z_\m(x) = \mbox{signTr}[{}^gU_\m(x)]
\eeq
which is center projection in SU(2) gauge theory.

    The original and gauge-transformed lattices can be expressed,
respectively
\bea
       U_\m(x) &=& g(x)Z_\m(x) e^{i\d A_\m(x)} g^\dg(x+\hat{\m})
\non \\ \non \\
  {}^gU_\m(x) &=& Z_\m(x) e^{i\d A_\m(x)}
\eea
We interpret $Z_\m(x)$ as the vortex background,  and $\d A_\m(x)$
is the fluctuation around the background.  DMC gauge finds the
optimal  $Z_\m(x)$ minimizing $\d A_\m(x)$.

  Although this method of vortex-finding by a best fit procedure
seems very natural, there are two shortcomings.
First, in practice there is a Gribov copy problem.  In general there is
no technique for finding the global maximum of $R$; instead there are various
methods (over-relaxation, simulated annealing) which find local maxima.
Secondly, one must be aware that the
action density of a thin vortex is singular at P-plaquettes
and therefore the fit to a thermalized lattice is very bad
at those locations.  In general
\beq
      \oh \mbox{Tr}[U(p)] \approx 1 ~~~~~~~~ \b \gg 1
\eeq
then
\beq
 \oh \mbox{Tr}[U(p)]  = Z(p)
    \mbox{Tr}[\prod_{links \in p}e^{i\d A_\m(x)}] \approx 1
\eeq
which implies
\beq
 \oh \mbox{Tr}[\prod_{links \in p}e^{i\d A_\m(x)}] \approx -1
       ~~~\mbox{when}~~~Z(p)=-1
\eeq
It follows that $\d A_\m(x)$ is not small at these P-plaquette
locations (and is in fact singular in the continuum limit).

\subsubsection{Indirect Maximal Center Gauge \cite{IMC}}

   Maximal abelian gauge (MAG) was introduced in ref.\ \cite{mag}, in order
to identify and study abelian monopoles in lattice configurations.
In SU(2) lattice gauge theory, the gauge is defined as maximizing
\beq
       R = \sum_{x,\m} \tr[U_\m(x) \s_3 U^\dg_\m(x)\s_3]
\eeq
which leaves a residual U(1) gauge symmetry.   SU(2) link variables in
this gauge can be decomposed as
\beq
        U_\m(x) = C_\m(x) D_\m(x)
\eeq
where $D_\m$ is the diagonal part of the link variable, rescaled to
restore unitarity
\bea
         D_\m &=&
 {1 \over \sqrt{\Bigl|[U_\m]_{11}\Bigr|^2 + \Bigl|[U_\m]_{22}\Bigr|^2}}
                    \left[ \begin{array}{cc}
                    [U_\m]_{11} & 0 \cr
                      0         & [U_\m]_{22}
                   \end{array} \right]
\non \\
       &=& \left[ \begin{array}{cc}
                    e^{i\th_\m} & 0 \cr
                      0         & e^{-i\th_\m}
                   \end{array} \right]
\eea
It can be seen that $D_\m(x)$ transforms like a U(1) gauge field under
the residual U(1) gauge symmetry.

   Indirect maximal center (IMC) gauge begins with maximal abelian gauge
fixing, and then uses the residual U(1) gauge symmetry
to maximize
\beq
       \widetilde{R} = \sum_{x,\m} \cos^2\Bigl(\th_\m(x)\Bigr)
\label{Rproj}
\eeq
leaving a residual $Z_2$ gauge symmetry.  Vortex configurations are
then identified by projection
\beq
      Z_\m(x) = \mbox{sign}\Bigl[\cos(\th_\m(x))\Bigl]
\eeq
and this procedure can be interpreted as a best fit of vortex configurations
to the abelian configuration $D_\m(x)$ extracted from $U_\m(x)$.

   Indirect center gauge has the advantage that both abelian monopoles
and vortices can be identified and correlated.  It has the same disadvantages
as maximal center gauge; i.e.\ there is a Gribov copy problem, and the
best fit is a bad fit at P-plaquette locations.

\subsubsection{Laplacian Center Gauge \cite{LC}}

   Consider a Yang-Mills theory with two Higgs fields $\p_1,~\p_2$
in the adjoint representation.
A unitary gauge, fixing color components
$\p_1^c(x)=\rho(x) \d^{c3}$, leaves a residual
U(1) symmetry.  This residual symmetry can be used to set color components
$\p^2_2(x)=0,\p^1_2(x)>0$ leaving a remnant $Z_2$ symmetry.  The idea of
Laplacian Center gauge, in pure Yang-Mills theory with no elementary
scalars, is to replace the scalars $\p_{1,2}$ with the two lowest-lying
eigenstates (no sum over $\a$, which labels eigenstates)
\beq
      \sum_y \D_{ij}(x,y) f^\a_{j}(y) = \l_\a f^\a_i(x)
\label{llap1}
\eeq
of the lattice Laplacian operator in adjoint representation
\beq
\D_{ij}(x,y)  = - \sum_{\mu}\left(
         [U_{A\m}(x)]_{ij}\delta_{y,x+\hat\mu}
       + [U_{A\m}(x-\widehat{\m})]_{ji}\delta_{y,x-\hat\mu}\
      -  2\delta_{xy}\delta_{ij}\right)
\label{llap2}
\eeq
There are standard numerical algorithms for obtaining these eigenstates.
This construction has the great advantage that there is no Gribov copy
problem.

   The use of eigenstates of the Laplacian operator
for gauge fixing was originally proposed by Vink and Wiese
\cite{vink}, in order to define a version of Landau gauge which
avoids the Gribov copy problem. The use of the single lowest-lying
eigenstate of the Laplacian operator in adjoint representation, to
reduce the gauge symmetry from SU(2) to U(1), was proposed by van
der Sijs \cite{Arjan}, and is known as Laplacian abelian gauge.
de Forcrand and co-workers \cite{LC} took this idea one step
further, and used a second eigenstate of the adjoint Laplacian to
also fix the U(1) symmetry, leaving only a remnant center
symmetry.  The extension of Laplacian center gauge to SU(N)
theories is straightforward, and is described in ref.\ \cite{LC}.

   Langfeld et al.\ \cite{ILC}
have employed a combination of the LC and DMC
gauges, first fixing to a unique configuration using the Laplacian center
gauge, and subsequently relaxing to a nearby local maximum of the
direct maximal center gauge.  This amounts to selecting a particular
Gribov copy of the DMC gauge, and is similar to the next center gauge
we shall discuss.

\subsubsection{Direct Laplacian Center Gauge \cite{DLC}}

   The direct laplacian center (DLC) gauge begins from a Gribov-copy
free modification of adjoint Landau gauge.  The idea is to first maximize
\beq
       R_M = \sum_{x,\m} \mbox{Tr}[M^T(x) U_{A\m}(x)
         M(x+\hat{\m})]
\eeq with $U_{A\m}$ in the adjoint representation.  However,
instead of imposing the condition (as in DMC gauge) that $M(x)$ is
an orthogonal matrix, we require only that $M(x)$ is a real
symmetric matrix with the orthogonality condition only imposed on
average: \beq
     {1\over \V} \sum_x M^T(x) M(x) = I
\eeq
where $\V$ is the lattice volume.

   This problem is solved by finding the three lowest eigenvalues,
and corresponding eigenfunctions, of the lattice Laplacian eigenvalue
equation \rf{llap1}.  Then
\beq
       M_{ab}(x) = f_a^b(x)
\eeq
is the matrix field maximizing $R_M$, where
$f^1_i,f^2_i,f^3_i$ are the three lowest-lying eigenstates.
However, $M_{ab}$ is not an SO(3) gauge transformation, so
we next map
$M_{ab}(x)$ to the nearest SO(3) matrix-valued field
\beq
 [g_A(x)]_{ij} = \tilde{f}_i^j(x)
\eeq
which also satisfies a Laplacian equation
\beq
      \sum_y \D_{ij}(x,y) \tilde{f}^a_{j}(y) = \L_{ac}(x) \tilde{f}^c_i(x)
\label{Lambda}
\eeq
This is done by polar decomposition
\beq
         M(x) = \pm \Omega(x) P(x)
\eeq
where $\Omega(x)$ is an SO(3)-valued field,
followed by over-relaxation from $\Omega$ to the
nearest local maximum $g_A$ of the DMC condition
($\widetilde{f}^b_a = [g_A]_{ab}$ then satisfies eq.\ \rf{Lambda}).
Finally transform $U_\m(x) \ra {}^gU_\m(x)$ (determined by $g_A$ up to an
irrelevant $Z_2$ gauge transformation), and center project
\beq
       Z_\m(x) = \mbox{signTr}[{}^gU_\m(x)]
\eeq
to locate the P-vortices.

   All of the center gauges described here yield qualitatively similar results,
but the DLC gauge and the method of ref.\ \cite{ILC}, which combine
versions of
Laplacian gauge fixing with relaxation to the DMC gauge, have the best
center dominance properties, to be discussed below.

\section{Numerical Investigations of Center Vortices}

   Monte Carlo investigations of the center vortex
mechanism have been mainly carried out in SU(2) lattice gauge
theory, and the results displayed below were obtained for that gauge
group.\footnote{Some SU(3) results in maximal center gauge
are found in ref.\ \cite{SU3}, and in Laplacian center gauge in ref.\
\cite{LC}.} Center projection, unless otherwise
noted, is carried out in the direct Laplacian center gauge.

\subsection{Center Dominance}

   The first question is whether center projection in DLC gauge provides
a candidate for the decomposition
\beq
       U_\m(x) = \U_\m(x) V_\m(x)
\eeq
where $\U_\m(x)=Z_\m(x)$ (determined by center projection)
carries all of the confining field fluctuations.
The requirement is that the string tension extracted from the
$\U=Z$ projected configuration agrees with the usual asymptotic string tension,
while the asymptotic string tension extracted from the $V$ configuration
vanishes. We can then identify $\tU(C)=V(C)$ as the non-confining part of loop
holonomies.

   Let $Z(I,J)$ denote a Wilson loop $Z(C)$ in the projected configuration,
along an $I\times J$ rectangular contour.  The center-projected
Creutz ratio is defined as
\beq
       \chi_{cp}[I,J] \equiv -\log\left\{ {Z[I,J] Z[I-1,J-1] \over
                                           Z[I-1,J] Z[I,J-1]} \right\}
\eeq
The limit of this quantity, for large loops, is the asymptotic string
tension of the center-projected Wilson loops $Z(C)$.

\begin{figure}[htb]
\begin{center}
\begin{minipage}[t]{8 cm}
\centerline{\scalebox{1.15}{\includegraphics{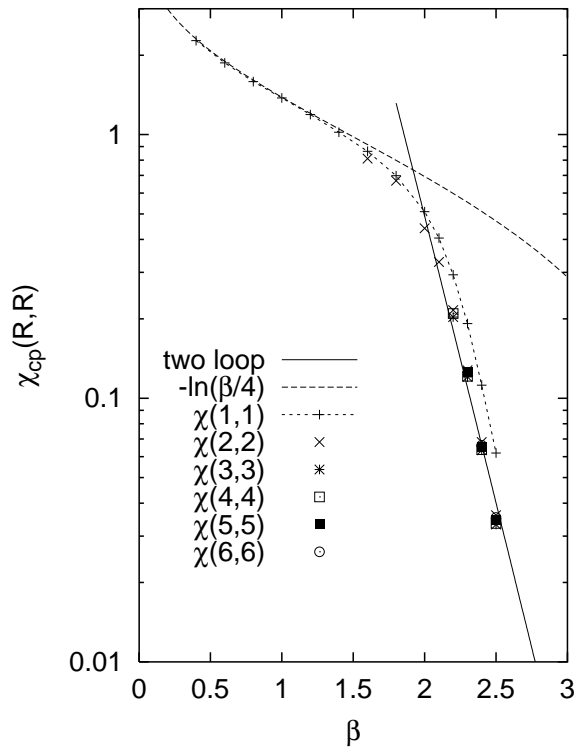}}}
\end{minipage}
\begin{minipage}[t]{16.5 cm}
\caption{Creutz ratios of center-projected Wilson loops.  From
 Faber et al., ref. \cite{DLC}.}
\label{nchi}
\end{minipage}
\end{center}
\end{figure}

\begin{figure}[htb]
\begin{center}
\begin{minipage}[t]{8 cm}
\centerline{\scalebox{0.8}{\includegraphics{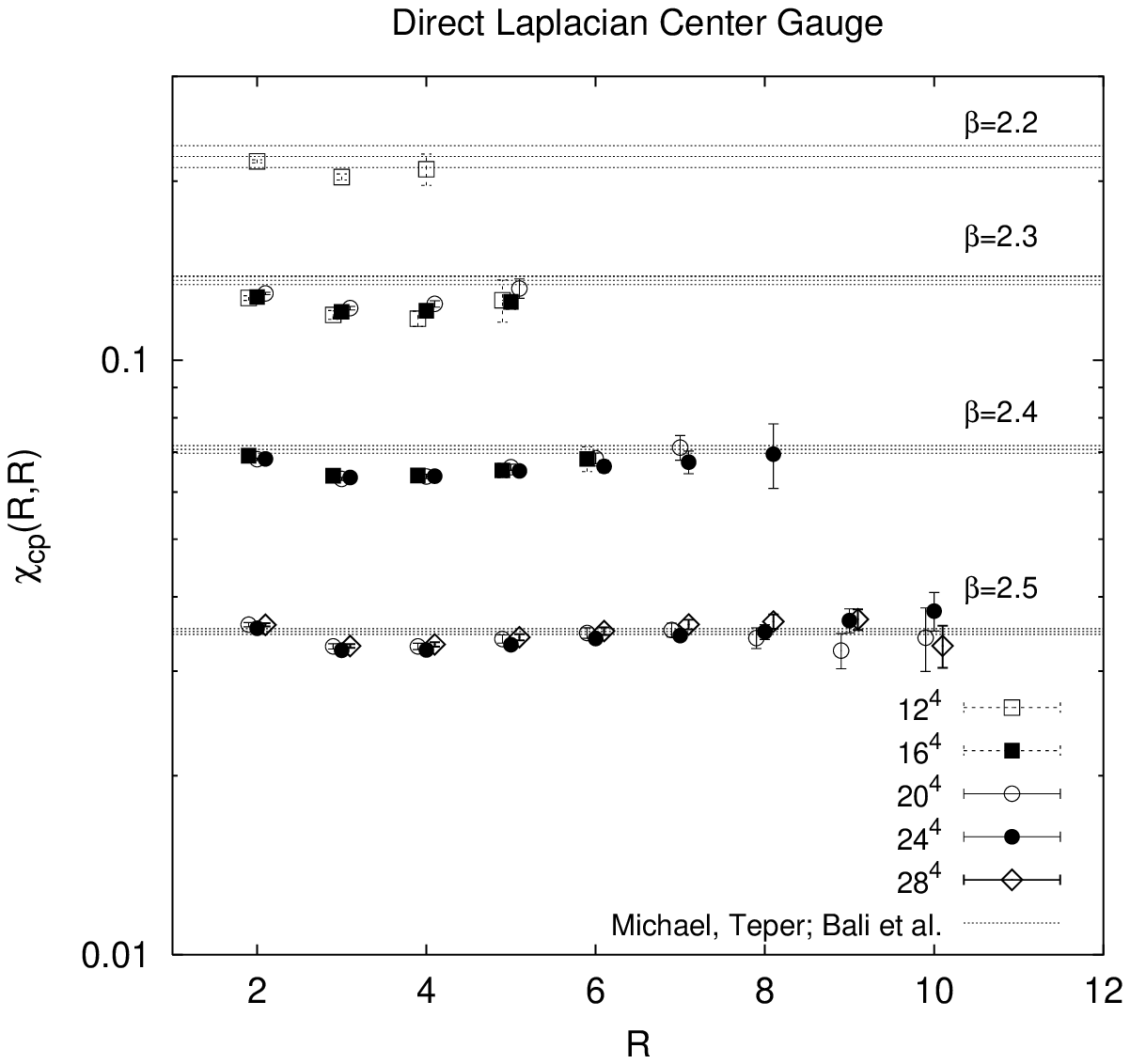}}}
\end{minipage}
\begin{minipage}[t]{16.5 cm}
\caption{Combined data, at $\b=2.2-2.5$, for center-projected
Creutz ratios obtained after direct Laplacian center gauge fixing.
Horizontal bands indicate the accepted values of asymptotic string tensions
on the unprojected lattice, with the corresponding errorbars \cite{BSS}.
From Faber et al., ref.\ \cite{DLC}.}
\label{all_beta}
\end{minipage}
\end{center}
\end{figure}

\begin{figure}[t!]
\begin{center}
\begin{minipage}[t]{8 cm}
\centerline{\scalebox{0.8}{\includegraphics{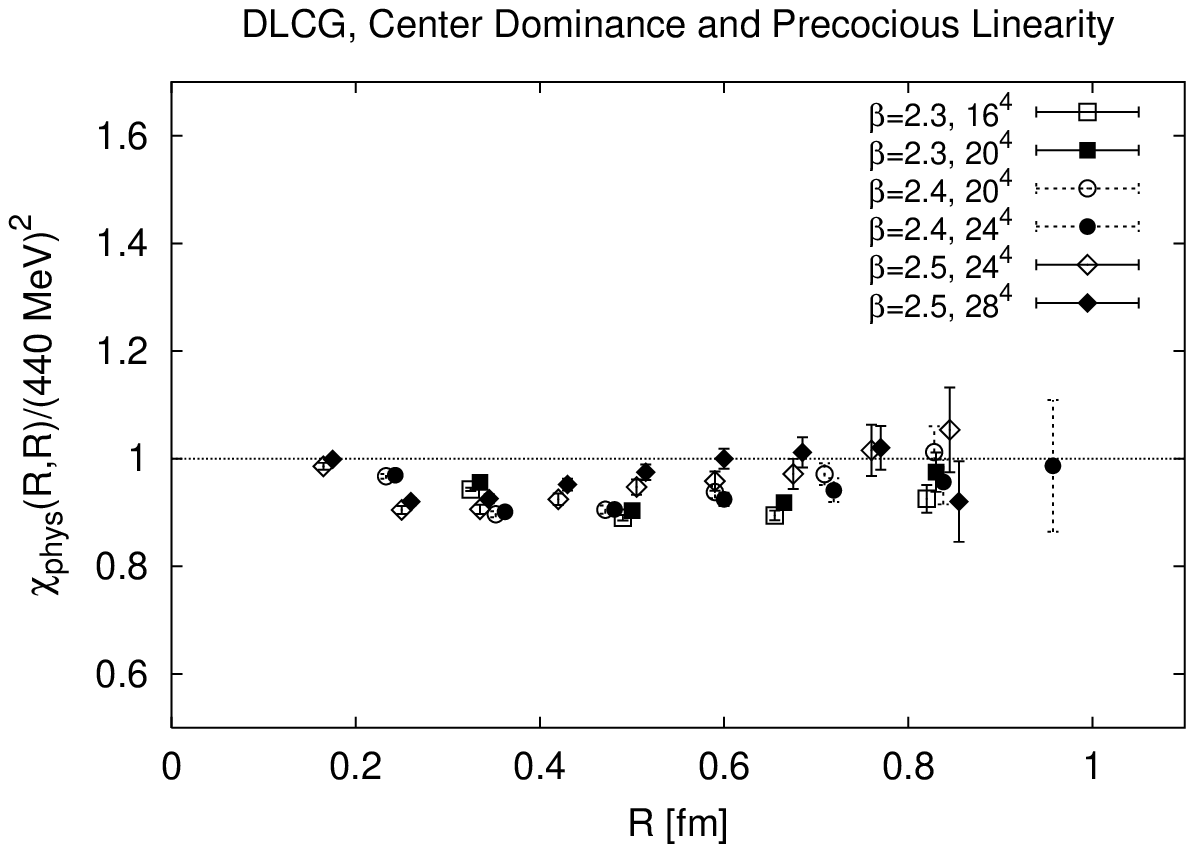}}}
\end{minipage}
\begin{minipage}[t]{16.5 cm}
\caption{Ratio of center projected and physical string tensions,
vs.\ quark separation in physical units.  From Faber et al., ref.\
\cite{DLC}.}
\label{plin}
\end{minipage}
\end{center}
\end{figure}

    The results for center-projected Creutz ratios $\chi_{cp}(R,R)$,
in SU(2) lattice gauge theory are shown in Figs.\ \ref{nchi} and
\ref{all_beta}.  What is rather striking about these results is
not only the fairly good agreement with the asymptotic string
tension derived from unprojected loops, which is known as {\bf
Center Dominance}, but also the fact that the projected Creutz
ratios are almost independent of $R$, which is known as {\bf
Precocious Linearity}. Precocious linearity implies that the
projected potential is linear already at two lattice spacings.
Another way of displaying both center dominance and precocious
linearity is to plot the ratio \beq {\chi_{cp}(R,R) \over
\s_L(\b)}  = {\chi_{phys}(R,R) \over \s_{phys}(\b)} \eeq as a
function of the distance $R_{phys} = R a(\b)$.  The result, for
several values of $\b$ and lattice size, is shown in Fig.\
\ref{plin} ($\chi_{phys}$ is $\chi_{cp}$ in physical units, $\s_L$
is in lattice units, and $\sqrt{\s_{phys}}=440$ MeV). In general,
center dominance is good to about 10\%.

   The influence of a vortex on a large Wilson loop is to contribute a factor
of a center element, but this influence is only fully realized for loops whose
dimensions exceed the thickness of a thick vortex surface.  The large loops
are sensitive to the asymptotic field of vortex, which is a singular gauge
transformation.  Since the asymptotic effect is only seen in large loops,
the linear potential, sensitive only to N-ality, is also only seen for large
loops.  Center projection, however, essentially collapses the thick vortex
surface (or ``core'') onto a thin surface only one lattice spacing thick.
This means that for the projected configuration, the full asymptotic
effect of center vortices is felt on any distance scale greater than
one lattice spacing.  If the locations of P-plaquettes on a plane
are  completely
uncorrelated, then there must be a linear potential at short distances on the
projected lattice.  This is the origin of precocious linearity.
If precocious linearity is not observed, it means that either the P-vortex
surface is very rough, or else some large fraction of the P-plaquettes
belong to P-vortices of small extent.  Either case leads to correlations
among the P-plaquettes, and corresponds to some high-frequency effects not
directly correlated with the long-range physics.

   Center dominance indicates that $\U$ is a confining background, and at
least contains the relevant confining fluctuations.  Precocious linearity
is evidence that $\U$ contains \emph{only} the relevant fluctuations.

\begin{figure}[t!]
\begin{center}
\begin{minipage}[t]{8 cm}
\centerline{\scalebox{0.9}{\includegraphics{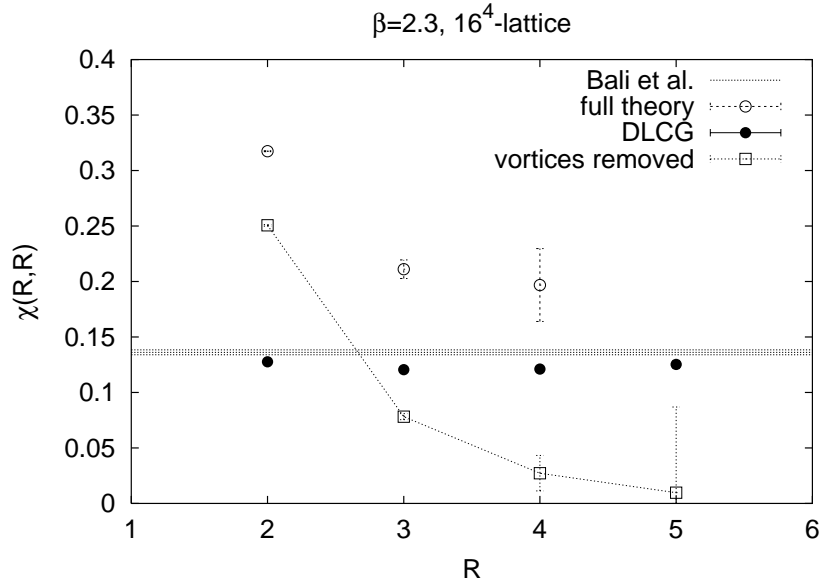}}}
\end{minipage}
\begin{minipage}[t]{16.5 cm}
\caption{Creutz ratios in the unprojected configuration (open circles),
in the center projected configurations in DLC gauge (full circles),
and in the unprojected configurations with vortices removed (open squares).
The horizontal band indicates the accepted asymptotic string tension of
the unprojected theory. From Faber et al., ref.\ \cite{DLC}.}
\label{vgone}
\end{minipage}
\end{center}
\end{figure}

    In the decomposition $U=\U V$, we have assumed that the $V_\m(x)$
fluctuations are, by themselves, non-confining.  This is simple to check,
as first done by de Forcrand and D'Elia in ref.\ \cite{dF2}.
We observe that for SU(2) lattice gauge theory, with $Z_\m$ the projected
configuration in an adjoint gauge,
\beq
      V_\m(x) = Z_\m(x) U_\m(x)
\eeq
$V$ is the ``vortex-removed'' configuration.  Figure \ref{vgone}
displays the Creutz ratios obtained from the $V$ configuration,
as compared to the center projected Creutz ratios, unprojected Creutz
ratios, and the asymptotic string tension.  Creutz ratios in the
vortex-removed configurations tend to zero, indicating that vortex
removal also removes the confinement property.

\subsection{Vortex-Limited Wilson Loops}

   We define a vortex-limited Wilson loop $W_n(C)$, in SU(2) lattice
gauge theory, to be a planar Wilson loop in the fundamental representation,
evaluated in the subensemble of configurations in which the minimal area
of the loop is pierced by precisely $n$ P-vortices (i.e.\ there are
$n$ P-plaquettes in the minimal area).  Here the center projection is
used only to select the data set; the loop itself is evaluated using
unprojected link variables.

   In practice, in a Monte Carlo simulation, the procedure for evaluating
$W_n(C)$ for a rectangular $I\times J$ loop is as follows:  For each
independent thermalized lattice configuration, fix to an adjoint gauge such
as DMC or DLC gauge, and compute the projected configuration.  For each
$I\times J$ loop on the projected lattice, count the number of P-plaquettes.
If this number equals $n$, use the unprojected link variables along the
loop to compute $\tr[U(C)]$, and include this value in the data set used
to evaluate $W_n(C)$.

   Vortex-limited loops are useful, because they allow us to check whether
P-vortices on the projected lattice really locate the middle of thick
center vortices in the unprojected lattice.  If that is the case, then
we expect in the limit of large loop area
\beq
        {W_n(C) \over W_0(C)} \ra (-1)^n
\label{Wnratio}
\eeq
The argument is that if center projection actually does locate correct
number of vortices piercing a Wilson loop, then
the vector potential in the neighborhood of loop
$C$ can be decomposed as
\beq
        A_\m^{(n)}(x) = g \d A^{(n)}_\m(x) g^{-1} - i g\pa_\m g^{-1}
\label{decomp}
\eeq
where $g$ is a singular gauge transformation associated with the $n$ vortices,
and $\d A^{(n)}$ is a perturbation around that
background.  It is understood that the delta function associated
with the discontinuity of $g\pa g^{-1}$ is dropped on the rhs of eq.\
\rf{decomp}.  Then
\beq
       W_n(C) = (-1)^n \left\langle \tr \exp \left[i\oint dx^\m \d A_\m^{(n)}
                                         \right] \right\rangle
\eeq
It is assumed that the fluctuations $\d A_\m^{(n)}$ have only
short-range correlations, and are therefore
oblivious to the presence or absence of vortices in the middle of
minimal loop area, far from the perimeter.  In that case
\beq
\left\langle \tr \exp \left[i\oint dx^\m \d A_\m^{(n)}\right] \right\rangle
    \approx
\left\langle \tr \exp \left[i\oint dx^\m \d A_\m^{(0)}\right] \right\rangle
\eeq
for large loops, and eq.\ \rf{Wnratio} follows immediately.

\begin{figure}[t!]
\begin{center}
\begin{minipage}[t]{8 cm}
\centerline{\scalebox{0.9}{\includegraphics{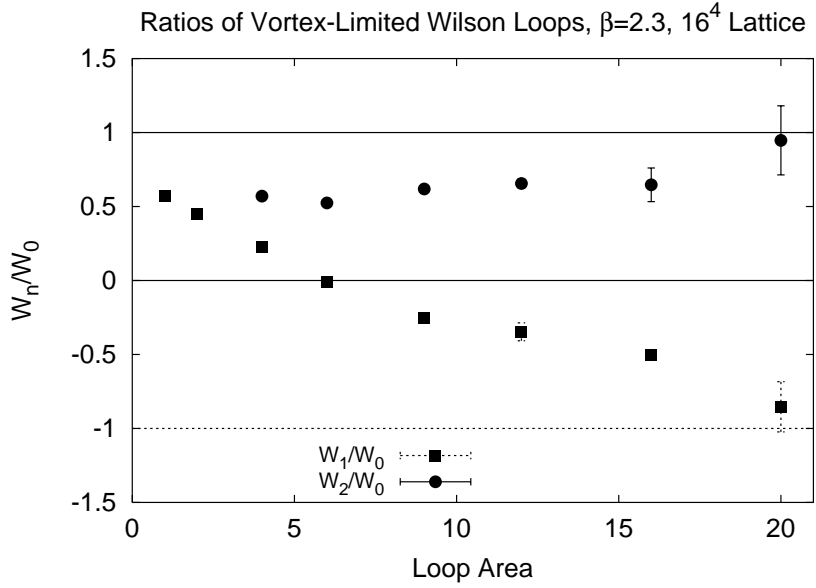}}}
\end{minipage}
\begin{minipage}[t]{16.5 cm}
\caption{$W_n/W_0$ ratios at $\b=2.3$. From Faber et al., ref.\ \cite{DLC}.}
\label{WnW0}
\end{minipage}
\end{center}
\end{figure}

    Figure \ref{WnW0} shows our results for $W_1/W_0$ and $W_2/W_0$,
for $L\times L$ and $L\times (L+1)$ loops as a function of loop area at
$\b=2.3$.  The data appears to agree quite well with eq.\ \rf{Wnratio},
supporting the assumption that
P-vortices locate thick center vortices in the unprojected lattice.
We also find that
\begin{enumerate}
\item $\chi_0(R,R) \ra 0$, where only the $W_0(C)$ loops are used
in computing Creutz ratios;
\item $W_{odd}(C)/W_{even}(C) \ra -1$, with loops limited to having an
odd or even number of P-vortices, respectively, piercing the loop;
\item $\chi_{even}(R,R) \ra 0$.
\end{enumerate}

\subsection{Scaling of the Vortex Density}

   The asymptotic scaling property of vortex surfaces was first observed
by Langfeld et al.\ in ref.\ \cite{tubby1}.
Let $N_{vor}$ denote the total number of P-vortex plaquettes
in a given lattice configuration, and $N_T$ the total number of plaquettes.
Then the fraction of P-vortex plaquettes $p$ is related to the vortex density
$\rho$ in physical units via
\bea
      p &=& {N_{vor} \over N_T} = {N_{vor} a^2 \over N_T a^4} a^2
\non \\
        &=& {\mbox{Total Vortex Area} \over 6 \times
             \mbox{Total Lattice Volume}} a^2
\non \\
        &=& {1\over 6} \rho a^2
\label{pfrac}
\eea
If $\rho$ is a physical quantity (i.e.\ constant at large $\b$) then
the prediction of asymptotic freedom (inserting $a(\b)$ into \rf{pfrac})
is that
\beq
     p = {\rho \over 6 \Lambda^2} \left( {6\pi^2 \over 11}\b \right)^{102/121}
            \exp\left[-{6 \pi^2 \over 11} \b \right]
\label{af}
\eeq
The average value of $p$ can be extracted from the expectation value of
center projected plaquettes, since it is easy to see that
\beq
      \langle Z(1,1) \rangle = (1-p) + p \times (-1) = 1-2p
\eeq

\begin{figure}[t!]
\begin{center}
\begin{minipage}[t]{8 cm}
\centerline{\scalebox{1.0}{\includegraphics{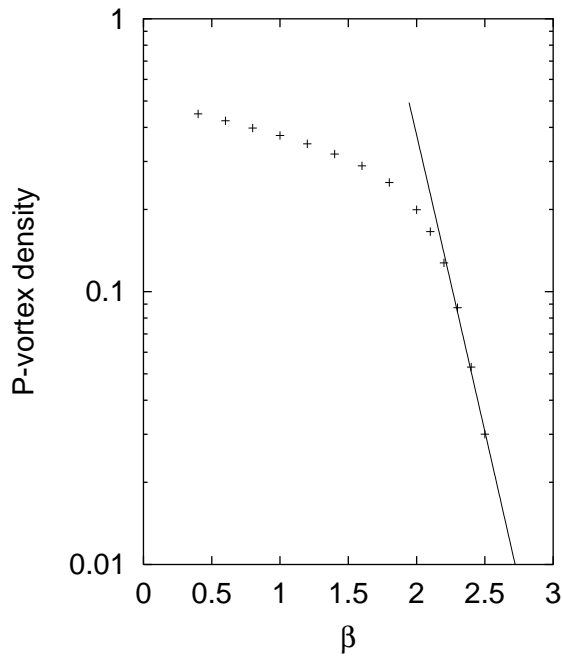}}}
\end{minipage}
\begin{minipage}[t]{16.5 cm}
\caption{Evidence of asymptotic scaling of the P-vortex surface density.
The data points are the number of P-plaquettes per unit volume,
and the solid line is the asymptotic freedom prediction with
$\sqrt{\rho/6\L^2}=50$.  From Faber et al., ref.\ \cite{DLC}.}
\label{pvor}
\end{minipage}
\end{center}
\end{figure}

   Figure \ref{pvor} is a plot of $p$ vs.\ $\b$; the solid line is
the asymptotic
freedom form \rf{af} with $\sqrt{\rho/6\Lambda^2}=50$.  The apparent scaling
of $p$ is consistent with the density of P-vortices corresponding to the
density of physical, surface-like objects (see also the recent results of
Gubarev et al.\
\cite{Gubarev} in the IMC gauge). Note that the scaling
line has the slope appropriate to a density of surfaces.  The scaling
lines for pointlike objects (e.g.\ instantons) or line-like objects
(monopoles) would have very different slopes.

\subsection{Finite Temperature}

   The argument that center vortices lead to an area law falloff for
planar loops assumes that the vortex piercings of the minimal area of
a loop are statistically independent.  It is clear that this assumption
cannot be true for vortex surfaces whose extension is smaller than the
size of the given loop.  Let us define the extension $L_{max}$
of the vortex surface to be the maximum distance between any two points
on the vortex sheet.  In that case, a piercing of a plane at point
$x$ by a vortex would be accompanied by another piercing at some distance
$L\le L_{max}$ from $x$.  This is due to the fact that vortex surfaces
are closed.   If there were an upper limit to vortex extension (apart from
that imposed by the finite size of the lattice), then it is not hard to
show that vortex effects could at most lead to a perimeter law falloff
for Wilson loops \cite{tubby2}.  This means that the vortices which lead
to a finite string tension must have an extension which is comparable to
the maximum possible separation on a finite size lattice; we therefore
expect that the vortices responsible for confinement percolate through
the lattice.

\begin{figure}[h]
\begin{center}
\begin{minipage}[t]{8 cm}
\centerline{
\epsfysize=5cm
\epsffile{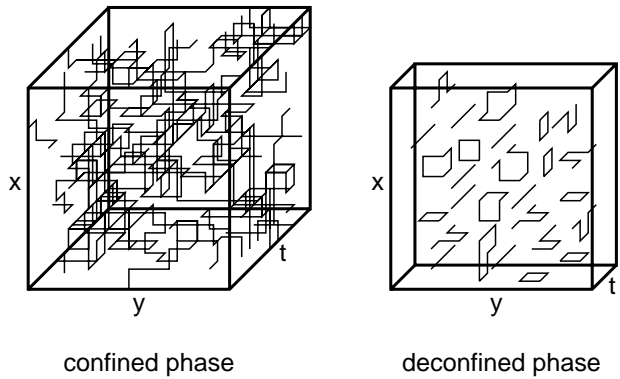}
}
\end{minipage}
\begin{minipage}[t]{16.5 cm}
\caption{Schematic picture of P-vortices in a ``space-slice'' of the lattice,
below and above the deconfinement phase transition.  From
Engelhardt et al., ref.\ \cite{tubby2}.}
\label{finT2}
\end{minipage}
\end{center}
\end{figure}

   This brings us to the question of deconfinement at finite temperature.
On time asymmetric lattices, the length of the lattice in the time direction
plays the role of inverse temperature, and it is known that a transition
from the confining to the deconfining phase in SU(2) lattice gauge theory
occurs at a temperature around $T = 220$ MeV.  Beyond this temperature,
Polyakov lines have a non-vanishing expectation value, and Polyakov line
correlators therefore tend to a finite limit with spatial separation.
If the vortex surfaces identified by center projection are responsible for
confinement, it would make sense that these surfaces cease to percolate
beyond the deconfinement transition.  Figure \ref{finT2} is a sketch of
these expectations on
the projected lattice at fixed $z$, below (left hand figure) and above
(right hand figure) the transition. In this space-slice the vortices appear
as closed loops, or as loops closed by lattice periodicity.

   On the other hand, even beyond the deconfinement transition, spacelike
Wilson loops are known to retain an area-law falloff.  Somehow the
vortex surfaces which would disorder these loops must percolate through
the lattice at any temperature.

\begin{figure}[h!]
\begin{center}
\begin{minipage}[t]{8 cm}
\centerline{
\hspace{0.5cm}
\epsfysize=7cm
\epsffile{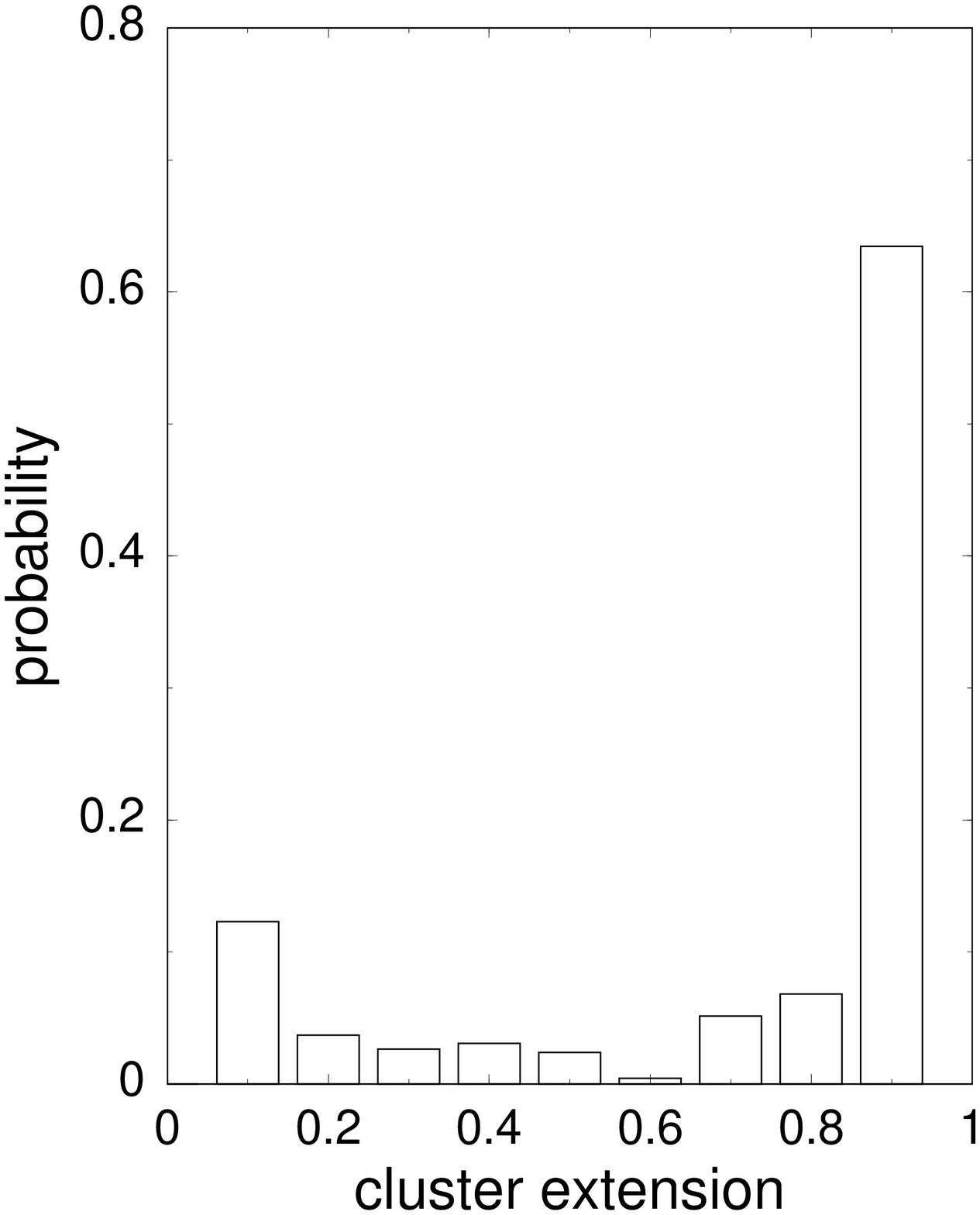}
\hspace{0.5cm}
\epsfysize=7cm
\epsffile{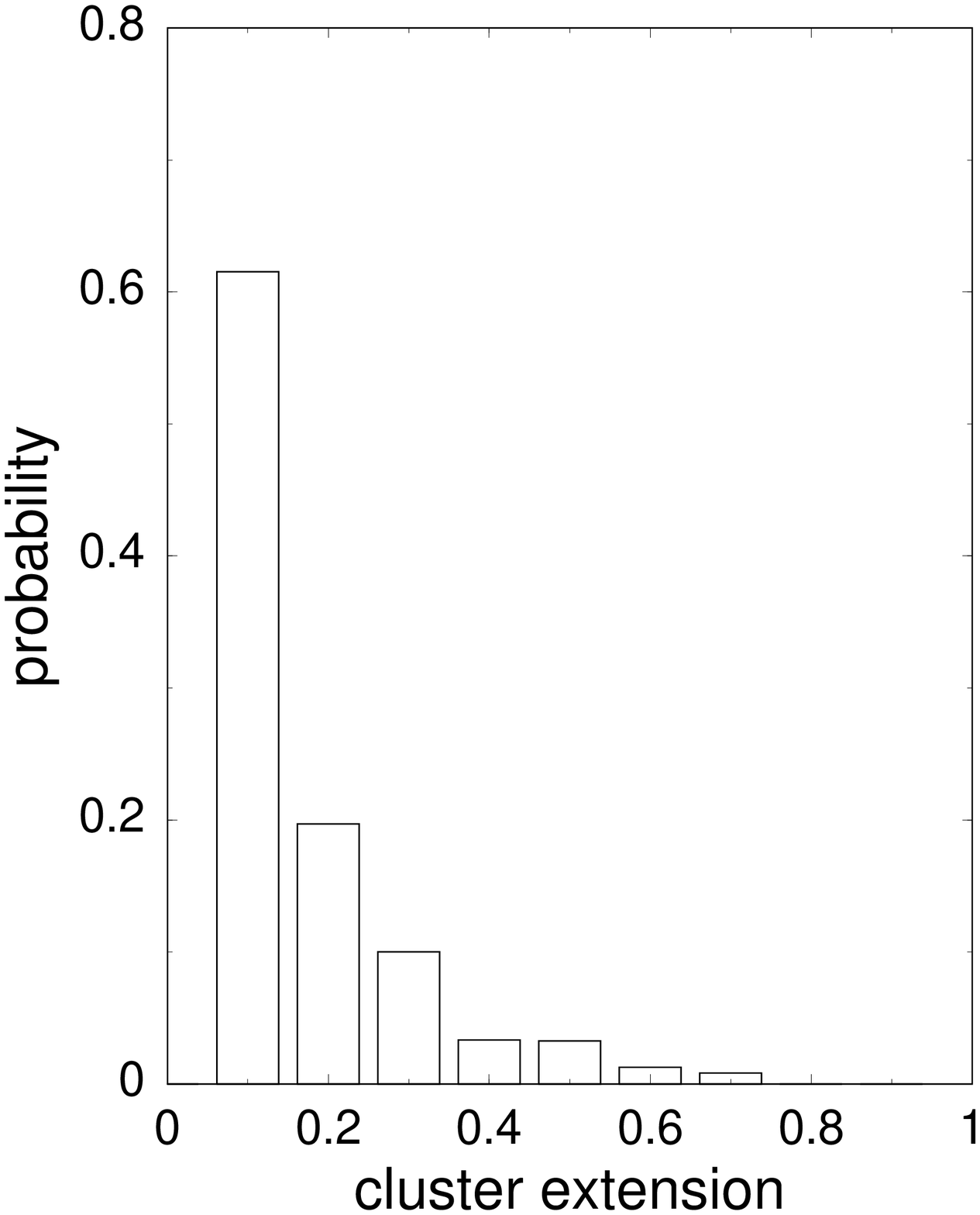}}
\end{minipage}
\begin{minipage}[t]{16.5 cm}
\caption{Histograms of vortex extension in a space-slice at
finite temperature, both below (left figure, $T=0.7 T_c$) and above
(right figure, $T=1.85 T_c$) the deconfinement phase transition.
The data is for $\b=2.4$ on a $12^3 \times 7$ (confined)
and $12^3\times 3$ (deconfined) lattices; center projection
in DMC gauge. From Engelhardt et al., ref.\ \cite{tubby2}}
\label{finT1}
\end{minipage}
\end{center}
\end{figure}

   A study of P-vortex extension across the deconfinement
phase transition was carried out by Engelhardt et al.\ \cite{tubby2}
(with earlier related work carried out by Langfeld et al.\ \cite{tubby0}
and Chernodub et al. \cite{Misha1}).  Figure \ref{finT1} displays
the main result.  Engelhardt et al.\ consider the extension of vortex
loops in a slice of the lattice with one spatial coordinate held fixed.
In this ``space-slice'', the vortices are closed loops, and the maximum
possible distance between links on a P-vortex loop is
$\sqrt{2(L/2)^2 + (L_t/2)^2}$, where $L,L_t$ is the length of the lattice in
the space and time directions respectively.  The axis labeled ``cluster
extension'' is in units of this largest distance.  Figure \ref{finT1}
is a histogram showing the fraction of P-plaquettes in the
space-slice belonging to vortices of a given extension.  The left-hand
figure is in the confined phase at $T=0.7~T_c$, and we see that almost
all vortex plaquettes belong to surfaces of maximal extension.  The right-hand
figure is at $T=1.85~T_c$ in the deconfined phase, and in this case the
vortex plaquettes belong mainly to small loops.

\begin{figure}[h!]
\begin{center}
\begin{minipage}[t]{8 cm}
\centerline{
\epsfysize=7cm
\epsffile{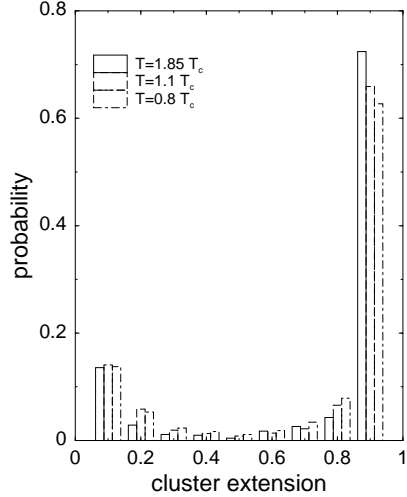}
}
\end{minipage}
\begin{minipage}[t]{16.5 cm}
\caption{Histograms of vortex extension in a time slice at
finite temperature.  Vortices are identified by center projection at
$\b=2.4$ in DMC gauge. Data at three temperatures are shown on
the same figure.
From Engelhardt et al.\, ref.\ \cite{tubby2}.}
\label{tcl}
\end{minipage}
\end{center}
\end{figure}

   Spacelike Wilson loops, however, have an area law at all $T$.  Thus,
if we take a slice of the lattice at fixed time, we expect to see that
vortices still percolate, so as to disorder these loops.  In fact, this
is just what happens, as seen in Fig.\ \ref{tcl}.  Further support of
this general picture, also obtained via center projection in DMC
gauge, is found in ref.\ \cite{Bertle}.

   The deconfinement phase transition has also been investigated
from the vortex free energy point of view
by de Forcrand and von Smekal \cite{vS}, Del Debbio, Di Giacomo, and
Lucini \cite{Adriano1}, and by Hart et al.\ \cite{Hart}.
For vortex flux in a spacelike plane (timelike vortices),
the vortex free energy goes to zero both below and above the deconfinement
transition.  However, for vortex flux in a timelike plane (spacelike
vortices), the free energy goes to zero below and infinity above the
transition temperature.   The interpretation of these results \cite{vS} is that
raising the temperature by reducing the time extension of the lattice
has the effect of ``squeezing'' spacelike vortices, whose flux cannot
spread out in all directions of a timelike plane.  This is what drives
the deconfinement phase transition.  On the other hand, timelike vortices
are not squeezed by a reduced time extension, since their flux can
spread out in a spacelike plane.  These vortices remain condensed
across the transition, and account for the spacelike string tension
above the deconfinement temperature.

\subsection{Topological Charge}

\begin{figure}[t!]
\begin{center}
\begin{minipage}[t]{8 cm}
\centerline{\scalebox{0.6}{\includegraphics{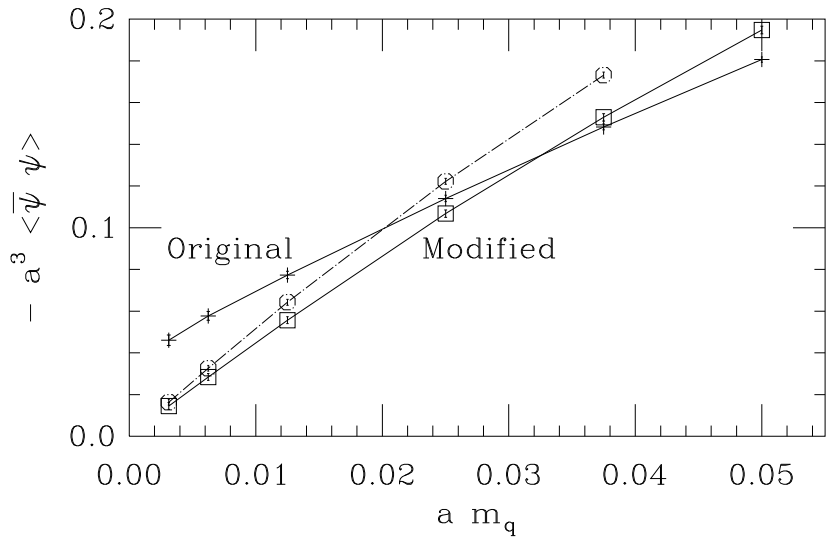}}}
\end{minipage}
\begin{minipage}[t]{16.5 cm}
\caption{The effect of vortex removal on the chiral condensate
in quenched lattice gauge theory.  Data points denoted ``Modified''
are from configurations with vortices removed in DMC gauge (open circles
come from a looser gauge-fixing criterion).  From de Forcrand
and D'Elia, ref.\ \cite{dF2}.}
\label{chiral}
\end{minipage}
\end{center}
\end{figure}

   A remarkable result, found numerically by de Forcrand and D'Elia
\cite{dF2}, is that removal of vortices from lattice
configurations removes not only the confinement property, but
chiral symmetry breaking disappears as well, and the no-vortex
field configurations $V_\m(x)$ are always in the trivial
topological sector.  In Fig.\ \ref{chiral} we show the results of
these authors for the chiral condensate $\lla \overline{\psi} \psi
\rra$ as a function of quark mass, at $\b=2.4$ in SU(2) lattice
gauge theory.  The solid line labeled ``original'' corresponds to
the condensate extracted from the $U_\m(x)$ configurations, the
solid line labeled ``modified'' is extracted from the
corresponding no-vortex configurations $V_\m(x)=Z^*_\m(x) U_\m(x)$
(the dashed line corresponds to a weaker gauge-fixing criterion).

\begin{figure}[t!]
\begin{center}
\begin{minipage}[t]{8 cm}
\centerline{\scalebox{0.4}{\includegraphics{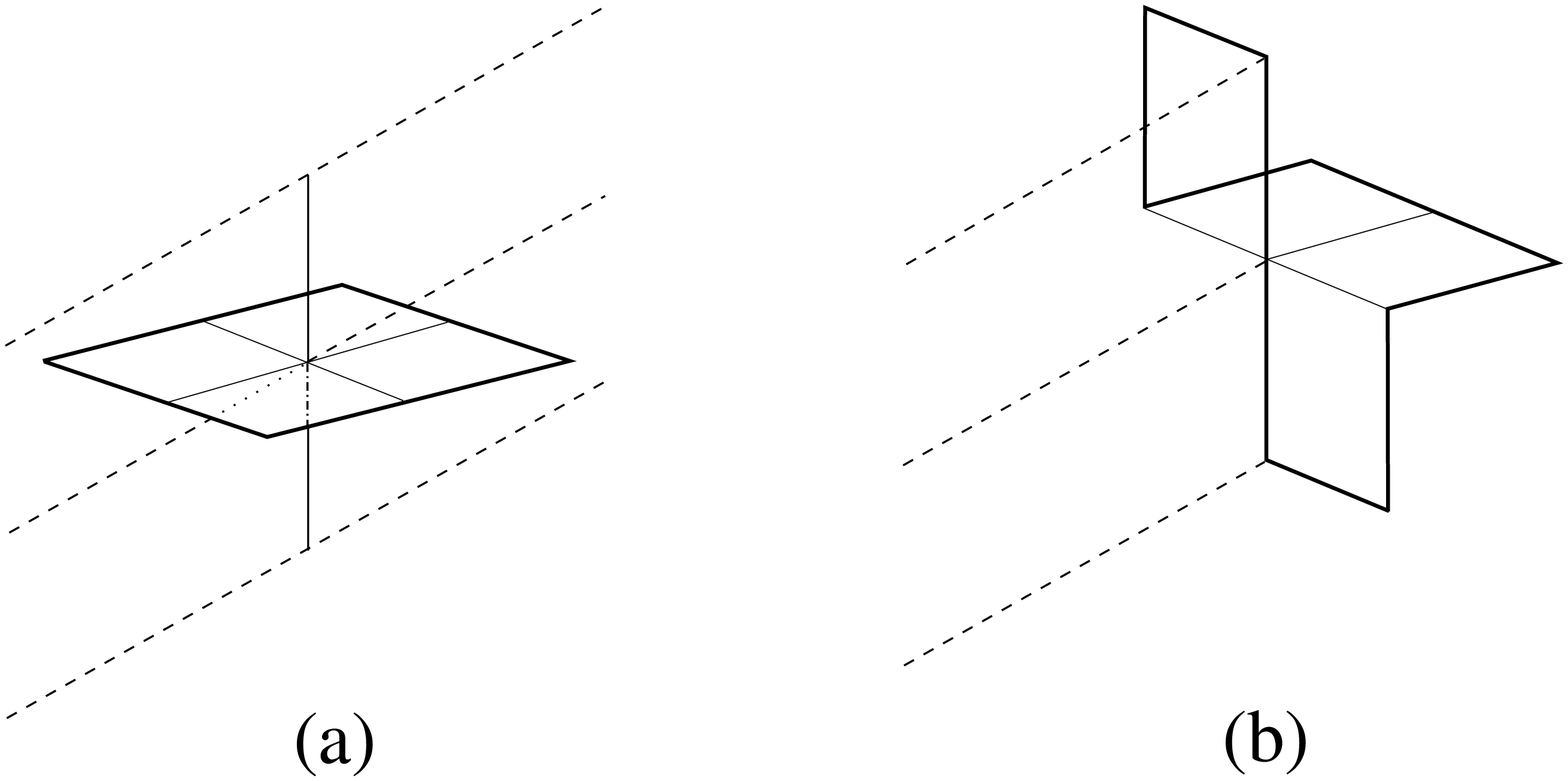}}}
\end{minipage}
\begin{minipage}[t]{16.5 cm}
\caption{Intersection points (a) and writhing points (b) which contribute
to the topological charge of a P-vortex surface. From
Reinhardt, ref.\ \cite{hugo2}.}
\label{picreinh5}
\end{minipage}
\end{center}
\end{figure}

    The fact that the topological charge of the no-vortex
configurations $V_\m(x)$ is zero suggests that this charge is entirely
carried by the (thin) vortex background $Z(C)$. The way in which
topological charge can arise at isolated points on a thin vortex surface
has been analyzed
by Engelhardt and Reinhardt in refs.\ \cite{engel1,hugo2,ER3}.
In general, topological charge arises at lattice sites at which the
set of tangent vectors to the vortex surface span all four space-time
directions.  These singular points are of two varieties:
\begin{itemize}
\item Self-intersections, where two areas of the vortex surface intersect
at a point (Fig.\ \ref{picreinh5}a).
These points carry topological charge $\pm \oh$.
\item Writhing points, in which the vortex surface twists about the point
in such a way as to produce four linearly independent tangent vectors
(Fig.\ \ref{picreinh5}b). These points carry topological charge of modulus
less than $\oh$.
\end{itemize}
Supposing the center flux on the vortex surface to lie in a Cartan
subalgebra of the gauge group, the orientation of that flux
is still relevant. Vortex
surfaces with net topological charge different from zero must be
non-orientable, consisting of surface
patches of differing orientation \cite{engel1,hugo1}.  The boundaries
of these patches can be regarded as monopole worldlines on the vortex
sheet (to be discussed further in section 7, below).  The net topological
charge of a closed surface is always an integer.
A very similar picture for the vortex origin of topological charge
was put forward independently by Cornwall in
ref.\ \cite{Corn2}.\footnote{In this work the boundary of a patch is identified
as the worldline of a ``nexus'', which appears (along with center vortices)
as a solution of an effective gauge theory containing a gluon mass term
\cite{Corn3}.}

\begin{figure}[t!]
\begin{center}
\begin{minipage}[t]{8 cm}
\centerline{\scalebox{0.9}{\includegraphics{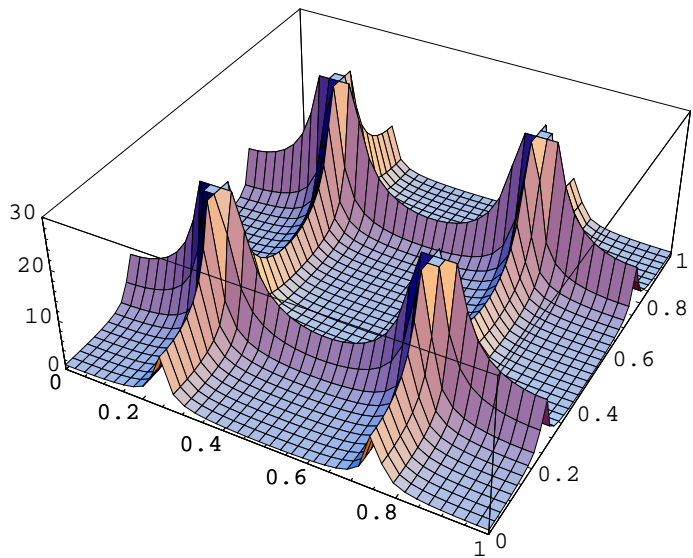}}}
\end{minipage}
\begin{minipage}[t]{16.5 cm}
\caption{Quark zero modes in a background of intersecting P-vortex
 sheets. From Reinhardt et al., ref.\ \cite{hugo1}.}
\label{tok}
\end{minipage}
\end{center}
\end{figure}

   The chiral condensate is related to the density of zero modes of
the Dirac operator by the famous Banks-Casher formula \cite{bc},
and it is well known
that quark zero modes are localized around instantons in an instanton
background.  Reinhardt et al.\ have shown that something analogous works
for topological
charge generated by vortices \cite{hugo1,hugo2}.
The concentration of quark zero modes around
the intersection points of four intersecting vortex sheets is shown in
Fig.\ \ref{tok}.

   Having associated topological charge with singular points on thin vortices,
a natural question is whether P-vortices generated by center projection
provide a topological charge density consistent with measurements of this
quantity in the unprojected theory.  Bertle, Engelhardt, and Faber have
investigated this question in ref.\ \cite{BFE}, and find that the
P-vortex result for topological susceptibility is consistent with conventional
measurements of this quantity.

   Engelhardt, Reinhardt, and Faber \cite{ER1}
have also investigated a simple random surface
model of vortex behavior, with a curvature-dominated action on a
hypercubic lattice, which was put forward in ref.\ \cite{ER}.
They report that this simple model is capable of reproducing,
both qualitatively \emph{and} quantitatively, many of the non-perturbative
features of the Yang-Mills vacuum.  Their model includes a dimensionless
constant and a lattice scale, which are fixed by fitting to the physical string
tension and deconfinement temperature.  Then the spatial string tension
at high temperature, the topological susceptibility, and the chiral condensate
are predictions of the random surface model, and these are found to be in
agreement with the values obtained in Monte Carlo simulations of SU(2)
lattice gauge theory.  Given the simplicity of the model, these results
are of course very encouraging.  A detailed discussion of the chiral
condensate in the random surface model is presented by Engelhardt in
ref.\ \cite{mengel}.

   How all of this relates to the more standard scenario \cite{mitya1}
of chiral symmetry breaking in the instanton
liquid picture remains to be seen.  Some recent data on the distribution
of topological charge on the lattice, which may be
relevant to this issue, is found in ref.\ \cite{horvath1} (see also
\cite{horvath2}).

\subsection{Casimir Scaling and Center Vortices}

   The vortex mechanism is motivated by the fact that the asymptotic
string tension depends only on the N-ality of the color charge representation.
On the other hand,
as emphasized in section 4.2, there is an intermediate distance range,
the Casimir scaling regime, in which string tension is instead proportional
to the quadratic Casimir of the color charge representation.

   The numerical data, however, also indicates that center vortices in
SU(2) lattice gauge theory are comparatively thick objects, with a surface
thickness on the order of one fermi.  This estimate of thickness is based
on three pieces of evidence: (i) the loop size at which
ratios of vortex-limited Wilson loops $W_1/W_0$ are close to $-1$;
(ii) the spatial lattice extension
at which the vortex free energy drops nearly to zero \cite{KT};
(iii) the adjoint string-breaking distance \cite{deF-P}.
In analyzing the effect of center vortices on Wilson loops of extensions
less than one fermi, the finite thickness of vortices must surely be taken into
account.

\begin{figure}[h]
\begin{center}
\begin{minipage}[t]{8 cm}
\centerline{\scalebox{.5}{\includegraphics{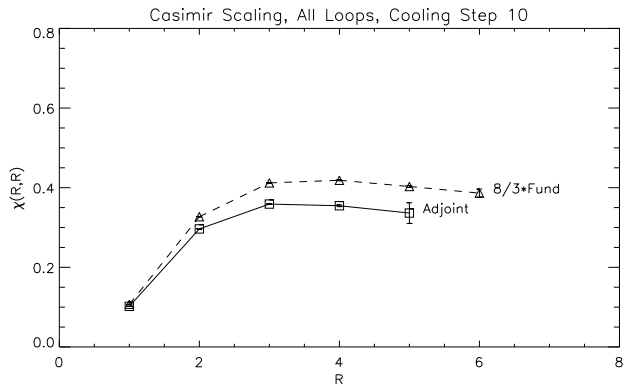}}}
\end{minipage}
\begin{minipage}[t]{16.5 cm}
\caption{Casimir scaling for the cooled lattice.
All-loop adjoint Creutz ratios $\chi^{adj}(R,R)$ (squares) compared to
the corresponding Casimir-rescaled ($\times 8/3$) data for
the all-loop fundamental Creutz ratios (triangles).
From Faber et al., ref.\ \cite{Picut}.}
\label{casall}
\end{minipage}
\end{center}
\end{figure}

\begin{figure}[h]
\begin{center}
\begin{minipage}[t]{8 cm}
\centerline{\scalebox{.5}{\includegraphics{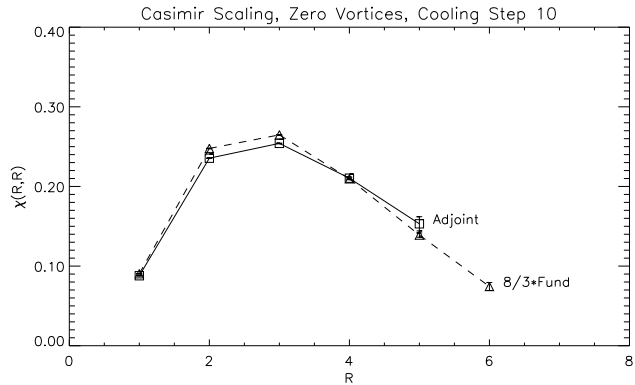}}}
\end{minipage}
\begin{minipage}[t]{16.5 cm}
\caption{Zero-vortex adjoint Creutz ratios $\chi^{adj}_0(R,R)$ (squares)
compared to
the corresponding Casimir-rescaled ($\times 8/3$) data for
the zero-vortex fundamental Creutz ratios (triangles).
From Faber et al., ref.\ \cite{Picut}.}
\label{caszero}
\end{minipage}
\end{center}
\end{figure}

   A very simple model of the effect of finite vortex
thickness was put forward in ref.\ \cite{Cas}.
If the vortex core overlaps a Wilson loop perimeter, the suggestion is to
represent its effect on the loop
by insertion of a group element $G$ somewhere in the product of link
variables around the loop,
interpolating smoothly between $-I$ and $I$.
Explicitly,
\beq
      G = S \exp[i\a_C(x) \s_3/2] S^\dg ~~~~~~~~~~~~~~~0\le \a_R(x) \le 2\pi
\eeq
\ni where $S$ is some SU(2) group element, $x$ is the position of
middle of the vortex in the plane of a (planar) loop C, and
$\a_C(x) \ra 0 ~ (2\pi)$ for $x$ in the exterior (interior) region
of a large loop, far from the perimeter.  In this simple
model, $S$ is assumed to be random on the SU(2) group manifold.
We define $f$ to be the probability for the middle of a vortex to
pierce a given plaquette.  If we then consider rectangular $R\times T$
loops ($T \gg R$) in SU(2) color representation $j$, the model predicts
\bea
      V_j(R) &=&
          - \sum_{n=-\infty}^{\infty}\ln\{(1-f)
          + f{\cal{G}}_j[\a_R(\mbox{x}_n)] \}
\non \\
     {\cal{G}}_j[\a] &=&
         {1\over 2j+1} \sum_{m=-j}^{j} \cos(\a m)
\eea
where $x_n=n+\oh$ is the coordinate in the $R$ direction.
For large loops, the string tension has
the correct asymptotic form, i.e
\beq
     \s_j = \left\{ \begin{array}{cl}
               -\ln(1-2f) & j=\mbox{half-integer} \cr
                     0    & j=\mbox{integer} \end{array} \right.
\eeq
For small loops, where $\a_R(x) << 2\pi$,
we find instead
\beq
      V_j(R) = \left\{ {f\over 6}\sum_{n=-\infty}^{\infty}
                 \a_R^2(\mbox{x}_n) \right\} j(j+1)
\eeq
which is proportional to the SU(2) quadratic Casimir (a result which
generalizes to SU(N)). However, the potential is not necessarily
linear.

   It has been found in refs.\ \cite{Cas,Deldar1} that one can choose
some reasonable form of $\a_R(x)$, interpolating between $0$ and $2\pi$,
such that there is an interval of charge separation $R$
in which the potential is roughly linear, and roughly Casimir scaling.
Without fine-tuning of $\a_R(x)$,
one can really expect no more than rough agreement from
such an obviously oversimplified model.  Nevertheless, the point which
is made is that the vortex theory is by no means incompatible with
Casimir scaling, which is an effect that may be attributed to finite
vortex thickness.\footnote{We note that the idea that the finite thickness of
vortices could produce a linear potential for adjoint-representation
particles, over some finite distance range, was actually put forward
long ago by Cornwall in ref.\ \cite{Corn1}.}

   As a test of the relation between Casimir scaling and center vortices,
one may examine the  string tension of vortex-limited
Wilson loops $W_0^{adj}(C)$ in the adjoint representation.  Excluding
vortices from the middle of the loop should greatly reduce
the adjoint string tension in the Casimir-scaling regime, if indeed
vortex cores are correlated with the Casimir scaling effect.  This
test was carried out in ref.\ \cite{Picut}, which compared
$\chi_0^{adj}(R,R)$ to $\chi^{adj}(R,R)$,
using the constrained cooling procedure \cite{cooling} for noise
reduction.  Figure \ref{casall} shows the agreement with Casimir
scaling for $\chi^{adj}(R,R)$ in the cooled configuration.
Figure \ref{caszero} displays corresponding results for the
zero vortex loops.  Again there is Casimir scaling, but this time
there is no apparent string tension, for either the fundamental or
the adjoint representation loops.  We conclude from this data
that thick center vortices seem to be associated with string tension
also in the Casimir scaling regime.

\subsection{Critique}

   One criticism that can be levelled against the center projection data
presented here
is that the agreement of the center projected and full string tensions
varies widely among different adjoint gauges, and among Gribov copies
in the same adjoint gauge \cite{KT1,vf,borny1,aborny1,borny2,Remarks,DLC}.
DMC gauge, for example, when fixed by
the over-relaxation procedure, has very good center dominance properties,
with disagreement between full and projected string tensions on the order
of a few percent \cite{aborny1}.
However, using an improved procedure of gauge fixing based on
simulated annealing, the disagreement increases to approximately 30\%
\cite{borny2}.
An alternative to maximal center gauges is the
Laplacian center gauge, which is free of Gribov copies.  In that case
the asymptotic string tensions of projected
and full configurations seem to agree reasonably well. However, in
Laplacian gauge there is no
sign of precocious linearity or asymptotic scaling of the vortex density,
and this means that the projected configurations, in this gauge, contain traces
of irrelevant short-distance fluctuations.  The two best methods to date,
the DLC gauge and the method of ref.\ \cite{ILC},
begin with two different versions of the Laplacian center gauge, and
subsequently ``cool'' away the irrelevant high-frequency fluctuations
by relaxation to the nearest local
maximum of $\tr_A[U]$. The projected configurations derived from each of these
two methods, from a given lattice configuration, are found to be
highly correlated \cite{DLC}.

   Although there are some arguments about why one adjoint gauge (or
set of gauge copies), might be better than another
\cite{LC,Remarks,DLC}, the real argument that center projection locates
vortex configurations is at present empirical; namely, the correlation of
P-vortices with gauge-invariant observables, as seen in the results
for vortex-limited Wilson loops $W_n(C)$, and the observables of
vortex-removed configurations $V_\m=Z^*_\m U_\m$.

   A second area which is not entirely satisfactory is Casimir scaling.
Vortex confinement is by no means
incompatible with Casimir scaling, as discussed in the previous subsection.
That is different, however, from
saying that the center vortex mechanism actually predicts Casimir scaling, to
the level of accuracy seen in numerical simulations \cite{Bali,Deldar}.
At the moment only crude approximations in the vortex picture,
leading to equally crude
approximations to Casimir scaling, are available.

   A third problem area concerns the L\"{u}scher $1/r$ term, which has not
been seen in the center-projected data. In view of the accuracy
needed to see the L\"{u}scher term in Monte Carlo simulations, this fact
is unsurprising.  Still, there is as
yet no good theoretical understanding of the L\"{u}scher term in the context
of vortex fluctuations.
The effect is presumably buried in small correlations
among vortex piercings in neighboring areas, leading to a small correction
to the dominant area-law falloff.  There \emph{does} seem to
be numerical evidence \cite{vrough} in center-projected configurations, of the
logarithmic growth in the width of the QCD flux tube (i.e.\ roughening),
which is a definite indication of the presence of string-like fluctuations.

\section{Monopole Confinement and the Abelian Projection}

   Confinement as an effect due to abelian monopoles is one of the
oldest proposals for quark confinement, and it is also the proposal
which has received the most attention over the years.  The idea is
motivated by the squeezing of magnetic fields into flux tubes in
type II superconductors, and by the demonstrable confinement of heavy
electric charge in a monopole plasma, which arises in compact QED
in D=3 dimensions.

   I will discuss ``dual'' superconductivity and monopole plasmas in a
little more detail below.  In order to apply the physics of these
essentially abelian examples directly to a non-abelian gauge theory,
it is crucial to single out an abelian subgroup, $U(1)^{N-1}$, of the
full $SU(N)$ gauge group.  This is usually accomplished by fixing the
gauge so as to diagonalize a certain operator, leaving a $U(1)^{N-1}$
subgroup.  The operator may be an adjoint representation scalar field
in the Lagrangian, as in the
Georgi-Glashow and Seiberg-Witten models, or else it may be a
composite field, as suggested by 't Hooft \cite{tH} for SU(N) pure
gauge theories. In either case, it is the electric quantum numbers of
the quark with respect to the abelian $U(1)^{N-1}$ subgroup which are
relevant for confinement in the abelian monopole models.

   Despite its popularity (and relative antiquity), I believe that
confinement by abelian monopoles is problematic for QCD because it
leads to the wrong representation-dependence of the string tension.
In monopole gas and dual-superconductor models, quark string tension
is proportional to $U(1)^{N-1}$ electric charge.  In contrast:
\begin{enumerate}
\item At asymptotic distances, string tension depends only on N-ality.
This means that for quarks in Yang-Mills theory with, e.g., two units
of U(1) abelian electric charge , the string tension is zero,
rather than twice the single-charge tension.
\item At intermediate distances,
the monopole prediction of
$U(1)^{N-1}$ electric charge dependence contradicts Casimir scaling.
It also leads to an incorrect multiplicity of Regge trajectories.
\end{enumerate}
These remarks will be presented in more detail below, after a brief
discussion of dual superconductivity and compact $QED_3$.

\subsection{The Abelian and Dual Abelian Higgs Models}

    When a type II superconductor is placed in an external magnetic
field, the magnetic field can only penetrate the superconductor in
cylindrical regions known as Abrikosov vortices, which carry only a
certain fixed amount of magnetic flux.  Abrikosov vortices are
magnetic flux tubes, and in principle such a flux tube could begin at
a monopole, of appropriate magnetic charge, and end on an
antimonopole.  The constant energy density along an Abrikosov vortex
then implies a linear potential between the monopole and antimonopole.
In other words, a type II superconductor, which is a system in which
bosons (the Cooper pairs) of electric charge $2e$ are condensed, is a
system which squeezes magnetic fields into tubes of quantized flux,
and confines magnetic charge via a linear potential.

    The example of type II superconductors becomes even more suggestive
in view of the electric/magnetic {\bf duality} symmetry of Maxwell's
equations.  If we are willing to allow for the existence of elementary
magnetic charges $g$,
then the corresponding Maxwell's equations
\beq
       \pa^\m F_{\m\n} = j^e_\n ~~~,~~~ \pa^\m {}^*F_{\m\n} = j^m_\n
\eeq
are symmetric with respect to the interchange of fields and currents
\beq
       \vec{E} \ra \vec{B} ~~~,~~~ \vec{B} \ra - \vec{E}
                           ~~~,~~~ j_\m^e \leftrightarrow j_\m^m
\eeq
Conservation of both the electric and magnetic currents implies separate
electric and magnetic U(1) symmetries.
It occurred independently to 't Hooft \cite{tH1} and Mandelstam
\cite{Mandelstam}, in 1976, that quark confinement in QCD might be due to
some type of {\bf dual superconductivity}, in which the QCD vacuum state
can be regarded as a condensate of a magnetically charged boson field
(magnetic monopoles), confining electrically charged particles (the quarks).

   This idea is expressed more concretely in a relativistic
generalization of the Landau-Ginzburg theory, known as the
{\bf abelian Higgs model}, the action of which is
\beq
      S = \int d^Dx ~ \Bigl( {1\over 4} F_{\m\n}F^{\m\n} +
 \Bigl|(\pa_\m + ieA_\m)\p\Bigl|^2 +{\l\over 4}(\p\p^* - v^2)^2 \Bigr)
\eeq
The abelian Higgs model has magnetic
flux-tube solutions analogous to Abrikosov vortices, which are known as
{\bf Nielsen-Olesen vortices} \cite{Holger-Poul}.  For a vortex along
the $z$-axis, the Nielsen-Olesen vortex in cylindrical coordinates has
the following asymptotic form:
\beq
     \begin{array}{ccc}
     \p(r) = \chi(r) e^{in\th} & , & A_\th(r) = A(r) \cr
          &   &  \cr
     \chi(r) \stackrel{r \ra \infty}{\longrightarrow} v
        + \mbox{(const.)} \times e^{-\sqrt{\l}vr}  & , &
    A(r) \stackrel{r \ra \infty}{\longrightarrow} n/(e r)
       +  \mbox{(const.)} \times e^{-evr}/\sqrt{r}
     \end{array}
\eeq
The topological stability of these vortices is related to the fact
that the Higgs potential has its minimum away from $\phi=0$, and that
the complex phase of the Higgs field along a closed loop around the vortex
has a non-zero winding number.  The magnetic flux carried by a
Nielsen-Olesen vortex is $2\pi n/e$.
In the dual version of the abelian Higgs
model, the Higgs field is magnetically charged, and it couples, not to
the usual vector potential $A_\m$, but rather to a ``dual'' photon field
$C_\m$.  Then the Nielsen-Olesen vortex is an electric, rather than magnetic
flux tube, and it is electric charge which is confined.

   In both the abelian and dual abelian Higgs models, the Higgs potential
has a classical minimum away from $\p=0$, and the symmetry is said to
be spontaneously broken.  But it is not so obvious what that phrase
really means, and which symmetry is actually broken.
Certainly if $\lla \p \rra \ne 0$, then it is the local
gauge symmetry which is broken, because $\p$ is not invariant under local U(1)
gauge transformations. But Elitzur's theorem assures us that in the
absence of gauge-fixing, $\lla \p \rra = 0$ in any phase.  Of course, it
is possible to fix to a unitary gauge so that $\lla \p \rra \ne 0$,
but then $\lla \p \rra \ne 0$ in both the
broken \emph{and} the Coulomb phases.  Consider instead gauge fixing to
Coulomb gauge.  In that case there are some remnant global symmetries:
a constant transformation
$g(x) = \exp[i\th]$, and the ``displacement''
symmetry
\beq
       g(x) = \exp[i \vec{d} \cdot \vec{x}]
\eeq
It is breaking of these remnant global symmetries which
separates the Higgs from
the Coulomb phase in abelian theories; c.f.\ ref.\ \cite{Lenz}
for a discussion of this point.

   There is an important limitation of the dual abelian Higgs model, which
is relevant to its applicability in a non-abelian gauge context:
The dual-abelian Higgs model does not easily accommodate
fundamental fields which are electrically charged, since the
kinetic term of an electrically-charged field involves the vector
potential of the photon, not the dual-photon field.  Despite the
beautiful electric-magnetic symmetry of Maxwell's equations with
electric and magnetic 4-currents, it is not possible to write a
local Lagrangian involving both electrically and magnetically
charged fields using only the photon, or only the dual photon,
vector potential. And as it happens, there \emph{are}
electrically-charged fields in QCD which are highly relevant to
the long-distance dynamics; these are the components of the gluon
field which are electrically charged with respect to the
$U(1)^{N-1}$ subgroup.

\subsection{Confinement in Compact $QED_3$}

   Compact electrodynamics on the lattice contains lattice-scale magnetic
monopoles, with masses on the order of $1/e^2$ times
the inverse lattice spacing.  These massive objects
interact with each other via a Coulombic potential.  In D=4 dimensions
the monopole worldlines form small closed loops at weak couplings, with
little or no effect on the long distance physics, which is described by
free massless photons.  In D=3 dimensions the situation is quite different.
Monopoles are instantons in that case, and the result, as shown by Polyakov
\cite{Polyakov}, is an area law falloff for large Wilson loops.

   In the Villain version of compact QED, a series of field transformations
results in the monopole contribution to the partition function \cite{BMK}
\beq\label{j0}
Z_{mon} = \sum_{m(r)=-\infty}^{\infty}
\exp\Bigl[ -\frac{2\pi^2}{g^2a} \sum_{r,r'}
m(r') G(r-r') m(r) \Bigr],
\eeq
where $m(r)$ is an integer valued monopole field at the (dual) lattice site
$r$, and $G(r-r')$ is the lattice Coulomb  propagator
($\nabla^2 G(r-r')=-\d_{rr'}$) in three
dimensions.  This is the partition function of a monopole Coulomb gas.
The propagator can be replaced by a Gaussian
functional integral:
\bea
Z_{mon} &=& \int \prod_r d\chi(r) \; \exp\Bigl[-\frac{g^2a}{4\pi^2}
\sum_{r} \oh (\Delta_\m \chi(r))^2\Bigr]
\non \\
        & & \times \sum_{m(r)=-\infty}^\infty
        \exp\Bigl[-{2\pi^2 \over g^2 a}G(0)\sum_r m^2(r)
        + i\sum_r m(r) \chi(r) \Bigr].
\label{plasma}
\eea
and in the weak coupling limit we need only
to maintain the first terms $|m(r)| \leq 1$ in the sum, so that
\beq
Z_{mon} \approx \int \prod  d\chi(r) \; \exp \Bigl[-\frac{g^2a}{4\pi^2}
\sum_r\Bigl( \oh (\Delta_\m \chi)^2 - M_0^2 \cos \chi(r) \Bigr)\Bigr],
\eeq
where
\beq
M^2_0 = \frac{8\pi^2}{g^2a}\, \exp\Bigl[-\frac{2\pi^2}{g^2a} \,G(0)\Bigr].
\eeq
Going to continuum notation, the action is
\beq
Z_{mon} \approx \int \cD \chi(r) \;
\exp \Big[ -\frac{g^2}{4\pi^2}
\int d^3r \;\Bigl( \oh (\pa_\m \chi)^2 - M^2 \cos \chi(r) \Bigr)\Bigr].
\label{Zm}
\eeq
where $M=M_0/a$.

   The electromagnetic field due to the monopole density
is given by
\beq
\pa_\m H_\m(r) = 2\pi \, m (r),~~~{\rm i.e.}~~~
H_\m(r) = \oh\int d^3r' \;\frac{(r-r')_\m}{|r-r'|^3}\; m(r'),
\eeq
Then if $C$ denotes a closed curve and $S(C)$
a surface with $C$ as boundary, we have that
\beq\label{j6}
\oint_C dr_\m A_\m(r) = \int_{S(C)} dS_\m(r)\; H_\m(r) =
\int d^3r \;\eta_{S(C)}(r)\; m(r),
\eeq
where
\beq\label{j7}
\eta_{S(C)}(r) =
-\oh \frac{\pa}{\pa r_\m} \int_{S(C)} dS_\m(r') \;\frac{1}{|r-r'|}.
\eeq
The monopole contribution to a Wilson loop carrying $n$ units of
the elementary electric charge is
\beq
 U_n(C)  \equiv  e^{in \oint dr_\m\; A_\m(r)}  =
    e^{ i n\int d^3r \;\eta_{S(C)}(r)\;m(r)}
\eeq
Inserting this expression into the partition function for the monopole
Coulomb gas \rf{plasma}
\beq
\langle U_n(C)\rangle =
\frac{1}{Z_{mon}}
\int \cD  \chi(r) \;
\exp \Bigl[-\frac{g^2}{4\pi}
\int d^3r \;\Bigl( \oh (\pa_\m (\chi-n
\eta_{S(C)})^2 - M^2 \cos \chi(r) \Bigr)\Bigr],
\eeq

  The functional integral is then evaluated by a saddlepoint method,
solving the equations of motion for $\chi$ in the presence of a
source $n\eta_{S(C)}$.  The calculation for $n=1$ was first carried
out by Polyakov \cite{Polyakov,BMK} and the result is
\beq
\langle U_1 (C) \rangle \approx \exp\Bigl[ - \s \; {\rm area}(C)\Bigr]
   ~~~,~~~ \s = {2g^2 M \over \pi^2}
\eeq
where area($C$) is the minimal area of the loop.
The Polyakov result has a simple generalization to arbitrary integer
charges, found in ref.\ \cite{j2},
\beq
 \langle U_n (C) \rangle \approx \exp\Bigl[ - n \s \; {\rm area}(C)\Bigr]
\eeq
which will be of some importance in our discussion of the monopole mechanism
in non-abelian gauge theories.  We note that compact QED can also
be regarded as a particular limit of the dual
abelian Higgs model \cite{Misha3,Smit1}.

   For compact QED in $D=4$ dimensions,
the integer-valued monopole field $m(r)$ in the monopole action
\rf{j0} is replaced by an integer-valued, conserved monopole
current $k_\m(r)$, defined on links of the dual lattice.  Smit and
van der Sijs \cite{Smit1} proposed that an abelian monopole action
of this type, which in addition includes a local monopole mass
term, could be regarded as the effective long-range action of a
pure non-abelian gauge theory in D=4 dimensions. This effective
action is deduced from the action of multi-monopole configurations
in an underlying Yang-Mills theory. In the absence of a Higgs
field, BPS-like monopole configurations \cite{BPS} in the pure
Yang-Mills theory were considered. In this approach the concept of
a monopole "container," i.e.\ a finite volume with certain monopole
boundary conditions, is introduced to specify the relevant
configurations in the quantized theory, and to compute some
of their properties \cite{Smit1,Smit2}.

\subsection{The D=3 Georgi-Glashow Model}

   The Georgi-Glashow model, aka the SU(2) adjoint Higgs model, is an
SU(2) gauge theory with a Higgs field in the adjoint representation.
In $D=3$ dimensions the action is
\beq
    S = \int d^3x \left[ \oh \tr[F_{\m\n}^2] + \oh (D_\m \p^a)^2
           + {1\over 4} \l (\p^a \p^a - v^2)^2 \right]
\eeq
The model has charged W-bosons of mass $M_W=gv$, a massive Higgs field
of mass $M_H = \sqrt{2\l v^2}$, a (perturbatively) massless ``photon,''
associated with the unbroken U(1) symmetry, and 't Hooft-Polyakov
monopoles (which are instantons in D=3 dimensions), with action
\beq
      S_{mon} = {2\pi M_W \over g^2} \e\left({M_H\over M_W}\right)
\eeq
where $\e(r)$ is a slowly varying function of order one.
In a unitary gauge, where the Higgs field is rotated to point in the
positive color 3 direction, the $A_\m^3$ field is the photon, and
the charged W-bosons are the $W^\pm_\m = A_\m^1 \pm i A_\m^2$ fields.
In unitary gauge, the 't Hooft-Polyakov monopole has
the form of a Dirac monopole in the $A^3_\m$ field component
\beq
   F_\m^3 = \e_{\m\n\l} \pa_\n A_\l^3 = {1\over g} {x_\m \over r^3}
\eeq at distances far from the monopole center (where $|\p|=v$).
The 't Hooft-Polyakov monopoles again interact via a Coulombic
potential, and the corresponding partition function has the form
of eq.\ \rf{Zm}, except that now the parameter $M$, which diverges
in compact $QED_3$ in the lattice spacing $a\ra 0$ limit, is
replaced by a finite constant, dependent on the coupling $g$ and
W-mass $M_W$.  Wilson loops can be computed as in compact $QED_3$,
and a finite string tension $\s \sim \exp[-S_{mon}]$ is obtained.
It would seem that in the Georgi-Glashow model, the goal of
separating the gauge field into a piece $\A$, which carries only
the confining fluctuations, and the remaining non-confining
fluctuations $\tA = A-\A$ has been accomplished unambiguously: The
instruction is to go to unitary gauge, identify the monopole
positions $\{x^\m_k\}$ from the zeros of the Higgs fields, and let
$\A_\m^a(x)=\d^{a3}\A_\m $ be the vector potential of Dirac
monopoles at those positions, so that
\bea
       F^3_\m(x) &=& \d_{a3} \e_{\m\n\l} \pa_\n \A_\l
\non \\
        &=& \sum_k q_k {(x-x_k)_\m \over |\vec{x}-\vec{x}_k|^3}
\eea
where $q_k=\pm 1/g$.
This identification of the confining fluctuations is certainly correct
for compact $QED_3$.  It is not correct, however, in the Georgi-Glashow
model.

   The problem is connected with loops of multiple electric charge.
Consider, in unitary gauge, loops of the form
\beq
      U_n(C) = \exp\left[in \oh \oint_C A^3_\m(x) dx^\m\right]
\eeq If the confining fields are those generated by a monopole
Coulomb gas, then the charge-dependence of the string tension is
the same as for compact $QED_3$, namely
\beq
        \lla U_n(C) \rra \sim e^{-n \s A(C)}
\eeq
with $A(C)$ the minimal area.
But this answer cannot be correct in the Georgi-Glashow model, as pointed
out in ref.\ \cite{j2}, because
there are charged $W^\pm$ bosons carrying two units of the minimum electric
charge.  This means that static charges carrying $n$ units of electric
charge are screened to $n$ mod 2 units of charge, and therefore
\beq
        \lla U_n(C) \rra \sim \left\{ \begin{array}{cl}
            e^{- \s A(C)} & \mbox{~~$n$ odd} \cr
            e^{- \m P(C)} & \mbox{~~$n$ even} \end{array} \right.
\eeq
The vector potential obtained from the monopole Coulomb gas action leads to the
wrong answer for these observables, and from this it follows that confining
magnetic flux $\A^a_\m(x)$ is not distributed in accordance with a monopole
plasma.

   What we learn from this is that the infrared effects
of W-bosons cannot be neglected in constructing an effective action
for the Georgi-Glashow theory.  The usual justification for ignoring
the W-bosons is that these are massive objects, and therefore cannot
affect the dynamics at large scales.  For a confining theory, that
argument is simply wrong.  Moreover, confining fluctuations must organize
themselves at large scales in such a way that $n=$ even charged loops
have a vanishing string tension.  The most obvious way to satisfy this
condition is for the confining flux to be collimated into $Z_2$ vortices,
rather than being distributed as in a monopole Coulomb gas \cite{j2}.

   On the other hand, the monopole Coulomb gas
effective action \rf{Zm} ought to have \emph{some} range of validity
in the D=3 Georgi-Glashow model.
On the basis of simple energetics arguments, the electric flux tube between
static sources carrying two units of U(1) electric charge will break by
W pair creation only when the energy of the flux tube is greater than the
combined mass of the two W bosons; i.e.\ when $L \s_2 > 2 m_W$.
The double-charge string tension $\s_2$ is zero beyond the string-breaking
distance; however, up to that point, the monopole
Coulomb gas result
that $\s_2=2\s_1$ should hold.
Then the string-breaking distance is $L\approx m_W/\s_1$, and
up to that length scale a monopole Coulomb gas analysis ought to give a good
account of the dynamics.

    For the D=3 Georgi-Glashow model
there exists a proposal for the infrared effective theory,
 in terms of elementary
vortex field operators, which reduces to the monopole action \rf{Zm} in a
certain approximation. W-bosons, in this dual formulation, appear as solitons
of the effective action.  This effective theory was originally proposed by
't Hooft in ref.\ \cite{thooft1}, and has been investigated extensively
by Kovner and co-workers \cite{Alex1}.  The action
is motivated by
't Hooft's observation that the analog of the SU(N) vortex creation operator
$B(C)$, in 2+1 dimensions, is an operator $V(x)$ defined such that
\beq
        V(x) |A_k,\p \rangle = |A'_k, \p' \rangle
\eeq
where $A',\p'$ are related to $A,\p$ via a singular gauge transformation,
defined such that
\beq
   W(C) \ra e^{\pm 2\pi i/N} W(C)
\eeq
if $C$ is topologically linked to point $x$ (with the $\pm$ sign depending
on the clockwise or counterclockwise orientation of the loop).  In the
Georgi-Glashow model, suitably generalized to an SU(N) gauge group,
$V(x)$ creates a topologically stable soliton, and the number of solitons
minus antisolitons is conserved mod $N$ (the worldline of a such a soliton
is a center vortex in D=3 dimensions).  This means that SU(N) adjoint Higgs
theory in $D=3$ dimensions
has a global $Z_N$ topological symmetry associated
with this conservation law, and
't Hooft argued that confinement is associated with the spontaneous breaking
of this dual topological symmetry.\footnote{Of course the usual $Z_N$ global
center symmetry, discussed in section 4.2, is
unbroken in this phase}.  On these grounds, 't Hooft proposed an effective
action for the soliton field with a Lagrangian
\beq
       L = \pa_\m V \pa^m V^* - \l\Bigl(V^* V -\m^2\Bigr)^2
                + m\Bigl(V^N + (V^*)^N\Bigr)
\label{tHK}
\eeq
Kovner et al.\ \cite{Alex1} point out that $L$ can be related to Polyakov's
effective action in eq.\ \rf{Zm}
by taking $N=2$ and $\l$ very large, so that the modulus of $V$ is almost
frozen.  Writing
\beq
        V(x) = \m \exp\left[{i\over 2} \chi(x)\right]
\eeq
and inserting this expression into eq.\ \rf{tHK} yields the Polyakov action
in \rf{Zm}, provided one ignores the fact that $\chi(x)$ is actually a
phase with a certain periodicity, rather than a continuous field.
In the 't Hooft action \rf{tHK}, as opposed to the Polyakov action,
the field $\chi(x)$ is allowed to have quantized discontinuities,
which are essentially vortex configurations of the vortex field itself.
It turns out that these vortices in the vortex field are associated with
charged W-bosons, whose importance to infrared dynamics has already been
noted.   Further details concerning this vortex operator approach to
confinement in the $D=3$ Georgi-Glashow model can be found in the review
articles by Kovner et al.\ \cite{Alex1,Alex2}.

\subsection{The Abelian Projection}

   In the Georgi-Glashow model, the abelian U(1) subgroup is singled
out by the adjoint Higgs field.  In the unitary gauge $\phi^a = \r \d^{a3}$,
the ``photon'' of the theory is simply the $A^3_\m$ component of the
gauge field, and U(1) magnetic
monopoles are associated with singularities in the
gauge-fixing condition; i.e. points (D=3) or lines (D=4) where $\phi=0$.
In QCD, on the other hand, there are no scalar fields in the Lagrangian,
and no obvious way of distinguishing an abelian subgroup.

   In ref.\ \cite{tH} 't Hooft suggested that the adjoint scalar field
used to identify the abelian subgroup could be a composite operator $X$, formed
from the gauge field.  Operators such as $X=F_{\m\n}^2$ or $F_{12}$, which
transform under a gauge transformation $g$ via
\beq
      X(x) \ra X'(x) = g(x)X(x)g^{-1}(x)
\eeq
are candidates for this role.  The corresponding unitary gauge, in an
SU(N) gauge theory, is the
gauge in which $X$ is diagonalized, i.e.
\beq
       X = \left[ \begin{array}{ccccc}
                  \l_1 &      &      &      &    \cr
                       & \l_2 &      &      &    \cr
                       &      &  .   &      &    \cr
                       &      &      &  .   &    \cr
                       &      &      &      & \l_N \cr \end{array} \right]
\eeq
and the eigenvalues ordered $\l_1<\l_2<...<\l_N$.
This gauge choice leaves a remnant $U(1)^{N-1}$ gauge symmetry,
generated by the Cartan subalgebra of the SU(N) gauge group.  The ``photons''
corresponding to the remnant abelian gauge symmetry are the diagonal
elements $A_\m^{aa}(x)$ of the matrix-valued SU(N) vector potential,
while the ``W-bosons'' correspond to the off-diagonal elements.  Singularities
of the gauge-fixing condition are identified with points (D=3) or lines
(D=4) where two of the eigenvalues $\l_i$ of $X$ coincide; these singularities
are identified as magnetic monopoles of the abelian subgroup.  All this
is in complete analogy to the Georgi-Glashow model, except that a
composite operator $X$ replaces the adjoint Higgs field.  As in the
dual superconductor picture and in compact QED, confinement would be
associated with the condensation of monopoles, and the confining fluctuations
that we have denoted by $\A_\m$ would be entirely
contained in the diagonal (``photon'') components of the vector potential.

    In translating this idea to the lattice formulation, and testing
it by Monte Carlo simulations, the first issue is the choice of gauge.
In 't Hooft's original article, the emphasis was on
``non-propagating'' gauge choices, which are free of massless ghosts.
Within this class, any gauge choice would suffice.  This gauge
independence does not appear to survive lattice tests, where strong
correlations have been found between, e.g., the gauge choice and the
string tension arising from the diagonal (``photon'') piece of the
vector potential.  Most lattice investigations have been carried out for the
SU(2) gauge group, and we will now focus on this particular case.  The
most widely used and successful gauge choice in SU(2) lattice gauge
theory is the {\bf Maximal Abelian gauge} \cite{mag}, defined
as the gauge which maximizes the quantity
\beq
      R = \sum_{x} \sum_{\m=1}^D \tr[U_\m(x) \s_3 U^\dg_\m(x) \s_3]
\eeq The trace can be recognized as the $33$ component of the
$U_\m(x)$ link variable in the adjoint representation, and the
condition that $R$ is maximized is simply the condition that the
link variables are, on average, as diagonal as possible.  As with
maximal center gauge, there is no known method for finding the
global maximum of $R$. Instead one obtains local maxima (Gribov
copies) from a combination of simulated annealing and
over-relaxation techniques.  Locally, in any Gribov copy of
maximal abelian gauge, the operator \beq
      Y(x) = \sum_{\m=1}^D \Bigl[ U_\m(x) \s^3 U^\dg_\m(x) +
      U_\m^\dg(x-\widehat{\m}) \s^3 U_\m(x-\widehat{\m})  \Bigr]
\eeq
is diagonal.  In the continuum, the diagonality condition is
\beq
       (\pa_\m \pm ig A^3_\m) A^{\pm}_\m = 0
\eeq
On the lattice, maximal abelian gauge allows the residual U(1) gauge
transformations
\beq
       U_\m(x) \ra e^{i\p(x) \s_3} U_\m(x) e^{-i\p(x+\widehat{\m}) \s_3}
\eeq
Define $u_\m(x)$ as the diagonal part of the link variable in maximal
center gauge, rescaled so as to restore unitarity.  Writing SU(2) link
variables in terms of Pauli matrices
in the usual way
\beq
      U_\m(x) = a_0 I + i \vec{a} \cdot \vec{\s}
\eeq
this means
\bea
      u_\m(x) &=& {1 \over \sqrt{a_0^2 + a_3^2}}
            \Bigl[ a_0 I + i a_3 \s_3 \Bigr]
\non \\
         &=& \left[ \begin{array}{cc}
              e^{i\th_\m(x)} & \cr
                 & e^{-i\th_\m(x)} \end{array} \right]
\label{alink}
\eea
The full link variable $U_\m(x)$ can be factored into the diagonal
matrix $u_\m$ and another matrix $C_\m$ which contains off-diagonal
elements
\bea
      U_\m(x) &=&  C_\m u_\m(x)
\non \\
              &=& \left[ \begin{array}{cc}
              \Bigl(1-|c_\m(x)|^2\Bigr)^{1/2} &  c_\m(x) \cr
        - c^*_\m(x)   & \Bigl(1-|c_\m(x)|^2\Bigr)^{1/2} \end{array} \right]
              \left[ \begin{array}{cc}
              e^{i\th_\m(x)} & \cr
                 & e^{-i\th_\m(x)} \end{array} \right]
\eea
It is not hard to see that under the remnant gauge transformation,
the diagonal link variables $u_\m$ transform like an abelian gauge
field
\beq
   u_\m(x) \ra e^{i\p(x) \s_3} u_\m(x) e^{-i\p(x+\widehat{\m}) \s_3}
\eeq
while $c_\m(x)$ transforms like a matter field with two units of
electric charge
\beq
        c_\m(x) \ra e^{2i\p(x)} c_\m(x)
\eeq
After maximal abelian gauge fixing, the mapping from the original
SU(2) link variables to the U(1) link variables
\beq
       U_\m(x) \ra u_\m(x)
\eeq
is known as the {\bf ``Abelian Projection''}.  Wilson loops
constructed from the abelian projected link variables $u_\m(x)$
are referred to as ``abelian projected'' (or simply abelian) Wilson loops.

   {\bf Abelian Dominance} is the property that the asymptotic string
tension extracted from abelian projected Wilson loops agrees with
usual asymptotic string tension.  This property was first demonstrated
by Suzuki and Yotsuyanagi \cite{Yots}; the most accurate measurements
to date find an abelian projection string tension at $(92\pm 4)$\%
of the full SU(2) string tension \cite{Bali_Borny}.

   Abelian dominance is not yet proof of a monopole confinement mechanism;
the confining fluctuations could still, e.g.,
take the form of vortex configurations.\footnote{Actually,
even a ``naive'' abelian projection,
carried out with no prior gauge-fixing whatever, has the property of
abelian dominance.  This is for rather trivial group-theoretic reasons,
however, which seem to have nothing to do with the confinement mechanism
\cite{mog}.} Nor does the abelian projection, by itself, separate the
(infrared) confining from the (ultraviolet) perturbative field
fluctuations, since the abelian projected potential at short distances,
which is due presumably to one-photon exchange, is not linear.

   An attempt to go beyond abelian projection, and to really isolate
the confining background in the abelian-projected lattice
configuration, was initiated by Shiba and Suzuki in ref.\ \cite{SS},
and by Stack, Nieman, and Wensley in ref.\ \cite{Stack}.
These authors begin by isolating monopoles in the abelian-projected
lattice configuration according to the De Grand-Toussaint criterion.
Let
\bea
         f_{\m\n}(x) &=& \pa_\m \th_\n(x) - \pa_\n \th_\m(x)
\non \\
                     &=& \overline{f}_{\m\n}(x) + 2\pi n_{\m\n}(x)
\eea
where $\th_\m(x)$ is related to the abelian link variables according
to eq.\ \rf{alink}, $\pa_\m$ is the forward lattice derivative,
$\overline{f}_{\m\n}(x)$ is in the range $[-\pi,\pi]$, and $n_{\m\n}$
is an integer valued ``monopole string'' variable.  The monopole current on the
lattice, according to De Grand and Toussaint \cite{deGrand}, is given by
\beq
       k_\m(x) = {1\over 4\pi} \e_{\m\a\b\gamma} \pa_\a
                 \overline{f}_{\b\gamma}(x)
\label{moncurrent}
\eeq

   The next step is to construct the lattice configuration corresponding
to a monopole Coulomb gas, with the monopole sources specified by
the current $k_\m(x)$ above.  This configuration is determined by the
monopole string variables $n_{\m\n}$ as follows
\bea
       u^{mon}_\m(x) &=& \exp[i\th^{mon}_\m(x)]
\non \\
       \th^{mon}_\m(x) &=& - \sum_y D(x-y) \pa'_\n n_{\m\n}(y)
\label{mdom}
\eea
where $\pa'_\m$ is the backward lattice derivative and $D(x)$ is the
lattice Coulomb propagator. The ``photon'' contribution is then
defined as the difference between the abelian projected and monopole
configurations
\bea
       u^{ph}_\m(x) &=& \exp[i\th^{ph}_\m(x)]
\non \\
           \th^{ph}_\m(x) &=& \th_\m(x) - \th^{mon}_\m(x)
\eea

\begin{figure}[h]
\begin{center}
\begin{minipage}[t]{8 cm}
\centerline{\scalebox{.7}{\includegraphics{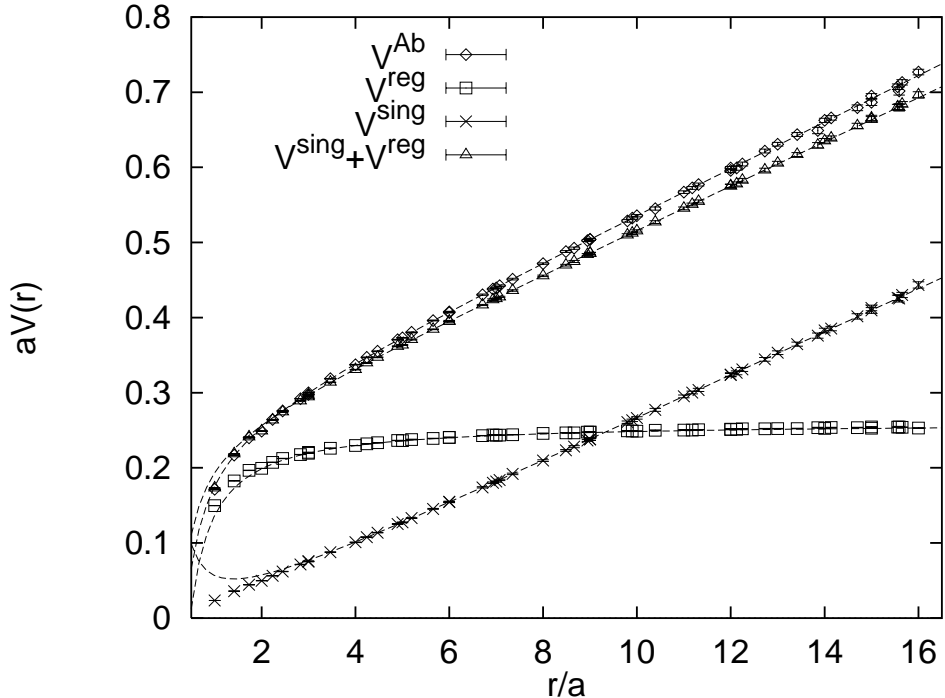}}}
\end{minipage}
\begin{minipage}[t]{16.5 cm}
\caption{Static potentials computed from monopole fields only ($V^{sing}$),
and photon fields ($V^{reg}$) only, in units of lattice spacing $a=0.081$ fm.
The potential computed from the abelian projected lattice is denoted
$V^{Ab}$.  From Bali, ref.\ \cite{Bali1}.}
\label{potfin}
\end{minipage}
\end{center}
\end{figure}

   Figure \ref{potfin} shows the static potential derived from abelian
projected configurations ($V^{Ab}$), the monopole Coulomb gas configuration
($V^{sing}$), and the photon ``remainder'' field ($V^{reg}$)
\cite{Bali_Borny}.  This figure
certainly suggests that the monopole Coulomb gas picture of Shiba and
Suzuki is on the right track.  The monopole potential is linear starting
from one lattice spacing, which means there are no extraneous high-frequency
fluctuations, and the associated string tension
is about 95\% of the abelian projection string tension (itself about
92\% of the full SU(2) string tension).  This close agreement is known
as {\bf Monopole Dominance}.  The Wilson loops constructed
from $u^{ph}_\m$ alone seem to have no string tension
at all, which would appear to confirm their role as non-confining,
mainly high-frequency, fluctuations.  Although total monopole density doesn't
scale, the density of monopoles belonging to the largest cluster of monopole
worldlines does scale according to asymptotic freedom \cite{HT}.
Moreover, the position of monopoles
identified by the abelian projection is highly correlated with the
gauge-invariant action density of the full SU(2) gauge fields at those
locations \cite{Misha2}.

   Further investigations have gone in several directions.  Suzuki and
co-workers have concentrated on constructing effective actions in terms of
monopole currents and their Coulombic interactions, which reproduce
the monopole dominance results obtained from abelian-projected configurations
\cite{Suzuki}. There have also been detailed
comparisons of the electric flux tube, found
in abelian-projected lattice configurations derived from lattice Monte Carlo
simulations, with the form which is expected in the dual abelian Higgs
models \cite{cc1,Dick,Bali_mag}.
These comparisons work out rather well.  For details,
the reader is referred to the reviews in refs.\ \cite{Bali1,Dick1,Misha3}

\subsection{Monopole Operators}

   There has been some effort, largely by the Pisa group \cite{Adriano},
to verify dual superconductivity by demonstrating that a suitably
defined monopole creation operator, denoted $\m$, acquires an
expectation value in the confinement phase of gauge theories at
finite temperatures. The idea is that if $\m$ is magnetically
charged, and acquires a VEV, then a magnetic U(1) symmetry is
spontaneously broken.

  The $\m(x,t)$ operator is defined so as to shift any given gauge field
configuration $A_\m(y,t)$ at a given time $t$ by the field $b_k(y;x)$
of an abelian monopole  located at point $x$.  The $U(1)^{N-1}$ subgroup
of SU(N), required to define the monopole, is
chosen according to some abelian projection.  In practice, what is
calculated is
\beq
       \rho = {d \over d\b} \ln\Bigl[\lla \m \rra \Bigr]
            = \lla S \rra_S - \lla S + \D S \rra_{S+\D S}
\eeq
where $S + \D S$ is the action obtained by shifting the link variables
by a monopole field at time $t$.  $\lla S \rra_S$ is the usual plaquette
action, and $\lla S + \D S \rra_{S+\D S}$ is the expectation value
of the shifted action, where the shifted action is also
used for the Boltzmann distribution.
The calculation has been carried out for lattice gauge theories at
finite temperatures, and it is found that the transition from
$\lla \m \rra \ne 0$ to $\lla \m \rra = 0$ indeed
coincides with the deconfinement
transition, for gauge theories with no matter fields.\footnote{Actually,
the same result is found when $\m$ is defined to add a vortex configuration,
rather than a monopole field, c.f.\ ref.\ \cite{Adriano1}.}  Related results
were obtained by Cea and Cosmai \cite{cc2}.

  Since the $\m$ operator makes no obvious reference to the center of
the gauge group, a natural question is what happens when matter fields
are added to break the global center symmetry explicitly.  Let us
again consider the Fradkin-Shenker model. If $\lla \m \rra = 0$
corresponds to deconfinement, then, since there is no
Higgs/confinement transition in this model, one would expect that
$\lla \m \rra = 0$ throughout the phase diagram.  Unfortunately no
results have yet been reported for the Fradkin-Shenker theory.  There
are new results, however, obtained recently by the Pisa group, for
$\lla \m \rra$ in QCD with dynamical quarks \cite{Adriano2}.  It is
found that $\lla \m \rra \ne 0$ at low temperatures, which is contrary
to what ought be the case in the Fradkin-Shenker model.  A transition
is also found to $\lla \m \rra = 0$ at finite temperature in full QCD,
in a region of the phase diagram where it is generally believed that
only crossover behavior, rather than a true transition occurs.  It is
probably fair to say that the situation is not completely clear at the
present time.\footnote{Fr\"{o}hlich and Marchetti have pointed out
that the Pisa monopole operator actually depends on the position of
the associated Dirac string \cite{Marchetti}.  They have suggested a
related construction for the monopole operator which avoids this
problem, and which makes use essential use of center vortices.  Some
numerical work, based on this construction, has been reported in ref.\
\cite{belavin}.}

\subsection{Objections}

   As discussed at some length in section 4, string tension varies with
the group representation of the color charges.  At asymptotic distances,
the string tension depends only on the N-ality of the representation,
while in an intermediate regime of charge separation,
up to the onset of color screening,
the string tension is proportional to the quadratic Casimir of the
representation.  These dependencies are different from what is expected
(and found) in monopole Coulomb gas and dual abelian Higgs models, where
the string tension is simply proportional to the abelian
electric charge relative
to the  $U(1)^{N-1}$ gauge group.  Moreover, the monopole
and dual-superconductor models predict a multiplicity of different types
of electric flux tubes, in the intermediate distance regime,
which should not exist in QCD.  We will consider the intermediate and
asymptotic distance scales separately.

\subsubsection{Intermediate Distances}

   As shown in ref.\ \cite{j2}, the string tension of a charge-anticharge
pair in compact $QED_3$, where the charge of each source is $n$ units
of the minimum charge, is simply
\beq
        \s_n = n \s_1
\label{sn}
\eeq
The result can be related to a picture of $n$ independent
electric flux tubes running between the two charges.  A similar result
is expected in U(1) monopole Coulomb gas models in D=4 dimensions, and in
dual (and real) superconductors.  Numerical simulations of double-charged
abelian-projected Wilson loops, i.e.
\beq
        W_2^{ab} = \lla \exp\left[ 2i \th(C) \right] \rra
\eeq
do in fact display string tensions which are roughly double the single-charge
string tension \cite{Poulis,Bali1}, where $\th(C)$ is the lattice ``loop
integral'' of link angles $\th_\m(x)$ around loop $C$.

   The problem is that if the confining force is sensitive only to abelian
electric charge, then it is difficult to understand Casimir scaling, and
in particular how there could ever be a linear static potential between
color charges in the adjoint representation \cite{Us1}.
Quarks in the adjoint
representation of the SU(2) color group come in three different varieties,
with $+2,~0,~-2$ units of the minimal U(1) electric charge, respectively.
The zero electric charge component should completely dominate the behavior
of the adjoint Wilson loop, and these, according to dual
superconductor and monopole gas ideas, are impervious to the confining
force. In the abelian projection
\bea
      \lla \tr[U_A(C)] \rra &\ra& \sum_{m=-1}^1 \lla \exp[2i m \th(C)] \rra
\non \\
        &=&  1 + \lla e^{2i \th(C)} \rra + \lla e^{-2i\th(C)} \rra
\non \\
        &=&  1 + 2 e^{-2\s_1 \mbox{\scriptsize Area}(C)}
\eea
The zero-charge component contributes a constant (one) to the trace
in the adjoint representation.  The neglected off-diagonal gluons would
presumably convert the constant to a perimeter-law falloff, but in any case
there is no reason to expect an area-law falloff for zero N-ality Wilson
loops, much less Casimir scaling.

    A second problem at intermediate distance scales,
pointed out by Douglas and Shenker in ref.\ \cite{DS}, is
that in the SU(N) case the corresponding abelian projection is to
a $U(1)^{N-1}$ theory, which in the dual superconductor picture would
involve $N-1$ copies of the dual abelian Higgs model, each with its own
solitonic flux tube.  The number of flux tubes of each type would be
a conserved quantity, which leads in principle
to physically distinct Regge trajectories
which are not expected in SU(N) gauge theory, and not observed in QCD.

\subsubsection{Asymptotic Distances}

   The problem at asymptotic distances is that the string tension $\s_2$
of double-charged abelian Wilson loops should be zero, not
$2\s_1$. This is because, in addition to monopoles and
``photons,''  the theory contains off-diagonal gluons carrying two
units of the minimal U(1) electric charge.  Eventually it becomes
energetically favorable for the off-diagonal gluons to screen the
static double charges, and the string tension beyond that
screening length vanishes.  In general, for the U(1) projection of
SU(2) gauge theory, the asymptotic string tension between sources
carrying $n$ units of U(1) charge is \beq
        \s_n = \left\{ \begin{array}{cc}
                       \s_1 & ~~n~\mbox{odd} \cr
                         0  & ~~n~\mbox{even}
                       \end{array} \right.
\label{assn}
\eeq
It follows that the abelian field distributions characteristic of
a dual superconductor, or a monopole Coulomb gas,
cannot be the distribution of confining fields in the abelian
projection.

\begin{figure}[h]
\begin{center}
\begin{minipage}[t]{8 cm}
\centerline{\scalebox{0.8}{\includegraphics{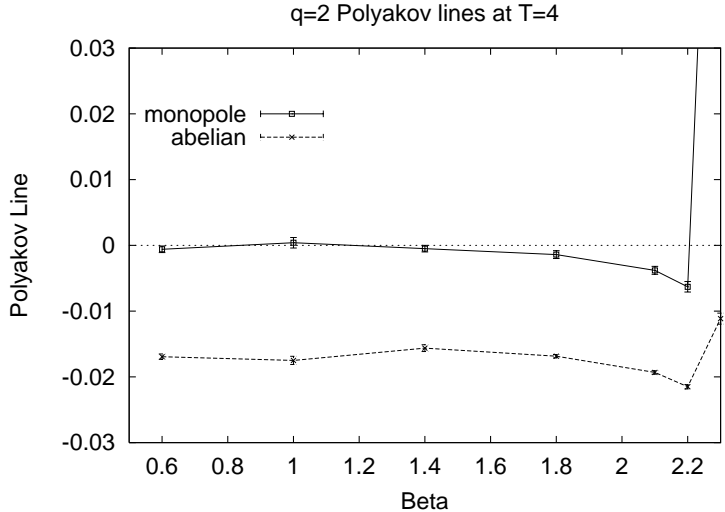}}}
\end{minipage}
\begin{minipage}[t]{16.5 cm}
\caption{Double-charged Polyakov lines, computed in the abelian projection
and in the monopole dominance approximation, on a $12^3 \times 4$
lattice.  From Ambj{\o}rn et al., ref. \cite{j3}.}
\label{P2}
\end{minipage}
\end{center}
\end{figure}

   It is difficult to check double charge screening numerically; this has
only recently been done for SU(2) adjoint Wilson loops \cite{deF-K}, as
already noted.  Instead, the expectation values of double-charged Polyakov
lines have been computed for abelian projected configurations, and
in the corresponding monopole dominance configurations (i.e.\ fields
obtained from eq.\ \rf{mdom}, from monopole string variables identified in the
abelian projection).  The abelian projection and monopole dominance results
for the double-charged Polyakov line are drastically different, as
shown in Fig.\ \ref{P2}.  The data shown in this figure were
obtained on a $12^3 \times 4$ lattice, at various values of $\beta$.
The deconfinement phase transition, on a lattice of four lattice spacings
in the time direction, occurs around $\b=2.2$.
The monopole dominance data is consistent (or nearly so) with a vanishing
double-charged Polyakov line, implying confinement of quarks with two
units of electric charge.  In the abelian-projected configurations, in
which the abelian field distribution is determined by all the
degrees of freedom, including the off-diagonal gluons, it is very clear
that the double-charged Polyakov line has a non-zero expectation value
in the confined phase; i.e.\ double electric charges are screened rather
than confined.

   What is going wrong, in the monopole Coulomb gas and dual abelian
Higgs models, is that they neglect the long-range effects of the
off-diagonal gluons (the ``W-bosons'').  It is assumed in these
models that the off-diagonal bosons acquire a mass, and therefore
their effects on the abelian gauge field can be neglected at large
scales.  This is clearly incorrect.  So what would happen if the
off-diagonal gluons are properly included?  A confining theory
which contains abelian magnetic monopoles \emph{and} dynamical
electric charges is neither a monopole Coulomb gas, nor a simple
dual superconductor at large scales.\footnote{In particular, there
are no \emph{local} Lagrangians which are able to couple an
abelian gauge field to both electrically charged and magnetically
charged matter fields.}  In the case of Yang-Mills theory in the
abelian projection formulation, the vacuum distribution of the
abelian projected field $A^3_\m$ is determined by an effective
action \beq
       \exp\Bigl[-S_{eff}[A^3]\Bigr] = \int DA_\m^\pm D(\mbox{ghosts}) ~
                  e^{-S_{YM}-S_{gf}}
\eeq
so that the vacuum distribution of the ``photon'' field $A^3_\m$ is determined,
in part, by virtual ``W-bosons'' $A_\m^\pm$.  This is in contrast to
the monopole dominance approximation \rf{mdom}, where only the distribution
of monopoles, but not the distribution of the fields emanating from those
monopoles, is affected by the W-bosons.  In general, it is clear that
the W-bosons must have a back reaction on
the vacuum distribution of the ``photon'' field $A^3_\m$, in such a way
that the vacuum fluctuations of the $A^3_\m$ field are consistent
with the $Z_2$ dependence shown in eq.\ \rf{assn}.

   The argument that the confining abelian configurations generated
by $S_{eff}[A^3]$ must be vortex configurations is the same as in section 5.
We assume that vacuum configurations can be regarded as the sum of a
high-frequency fluctuation $\tA^3_\m$, responsible for most of the
perimeter-law behavior, on a confining background $\A^3_\m$,
which is mainly responsible for the area-law
behavior.  Then, since $\s_{2n}=0$,
it must be that the confining fluctuations $\A^3_\m$ have almost no
effect on the expectation values of even powers of the abelian
loop holonomy $U(C)=\exp[i\oint \oh A^3]$, which
is accomplished if
\beq
       U^{2n}(C) \approx \tU^{2n}(C) ~~\Longrightarrow~~ U(C) = Z(C)\tU(C)
\eeq
where in this case $Z(C)=\pm 1$.  So the confining fluctuations $\A^3_\m$
are those for which $\U(C)=Z(C)$ for every large loop.  These, of course,
are $Z_2$ center vortices, and on the lattice correspond to link variables
$\U_\m(x) = \pm \exp[i(\phi(x)-\phi(x+\m)]$.

   An instructive example is the case of compact $QED_4$ at strong coupling.
The theory can be rewritten in terms of monopole currents interacting via
a Coulombic potential (as in the D=3 case discussed in section 7.2), and
confinement can again be understood in terms of a monopole Coulomb gas.
However, the situation is changed qualitatively by the addition of a
charge-2 matter field.  For simplicity we consider a charged scalar
matter field $\rho$ of fixed modulus $|\rho|=1$, and
\beq
  Z =  \int D\rho  D\th_\m ~
           \exp\left[ \b \sum_p \cos(\th(p))
       + \oh \l \sum_{x,\m} \left\{\rho^*(x)
              e^{2i\th_\m(x)} \rho(x+\widehat{\m}) + \mbox{c.c}
          \right\} \right]
\eeq
with $\b \ll 1$ (confinement) and $\l \gg 1$. Integrating over the
matter field generates a set of charge-2 Wilson loops in the induced
action, which screen any external even-charged Wilson loop.

   This gauge+matter theory can be rewritten, following Chernodub
and Suzuki \cite{Maxim}, in terms of monopole currents interacting with
charge-2 electric currents induced by the matter field.  Again, it is the
electric currents which screen an external charge-2 Wilson loop.
In this case, however, rewriting the theory in monopole variables actually
obscures the underlying physics.  The confining field
configurations are no longer Coulombic fields emanating from monopole
charges.  Rather, the confining configurations are thin $Z_2$ vortices
$-$ a fact which is invisible in the monopole formulation.
To see this, go to the unitary gauge $\rho=1$, which preserves a
residual $Z_2$ gauge invariance, and make the field
decomposition
\beq
       \exp[i\th_\m(x)] = z_\m(x) \exp[i\tth_\m(x)]
\label{decompose}
\eeq
where
\beq
        z_\m(x) \equiv \mbox{sign}[\cos(\th_\m(x))]
\eeq
and
\beq
     Z =   \prod_{x,\m} \sum_{z_\m(x)=\pm 1}
           \int_{-\pi/2}^{\pi/2} {d\tth_\m(x) \over 2\pi}
          \exp\left[ \b \sum_p Z(p) \cos(\tth(p)) +
                     \l \sum_{x,\m} \cos(2\tth_\m(x)) \right]
\eeq
This decomposition separates lattice configurations into $Z_2$ vortex
degrees of freedom (the $z_\m(x)$), and small non-confining
fluctuations around these vortex configurations, strongly
peaked at $\tth=0$.  One can easily show, for $\b \ll 1,~\l \gg 1$, that
\beq
      \Bigl\langle \exp[in\th(C)] \Bigr\rangle \approx \lla Z^n(C) \rra
                        \Bigl\langle \exp[in\tth(C)] \Bigr\rangle
\eeq
with
\bea
     \lla Z^n(C) \rra &=& \left\{ \begin{array}{cl}
         \exp[-\s A(C)] & n \mbox{~odd} \cr
               1        & n \mbox{~even} \end{array} \right.
\non \\
      \Bigl\langle \exp[in\tth(C)] \Bigr\rangle
             &=& \exp[-\m n^2 P(C)]
\eea
which establishes that the confining fluctuations, in this coupling
range, are entirely due to thin vortices identified by the
decomposition \rf{decompose} in unitary gauge.  It is clear that the
addition of a charge-2 matter field has resulted in a qualitative
change in the physics of confinement.  Yet the transition
from a monopole Coulomb gas mechanism to a vortex dominance
mechanism is essentially invisible if the gauge+matter theory
is rewritten in terms of monopole + electric current variables
\cite{Maxim}, which in this case tend to obscure, rather than reveal,
the nature of the confining fluctuations.

\begin{figure}[h]
\begin{center}
\begin{minipage}[t]{8 cm}
\centerline{\scalebox{0.5}{\includegraphics{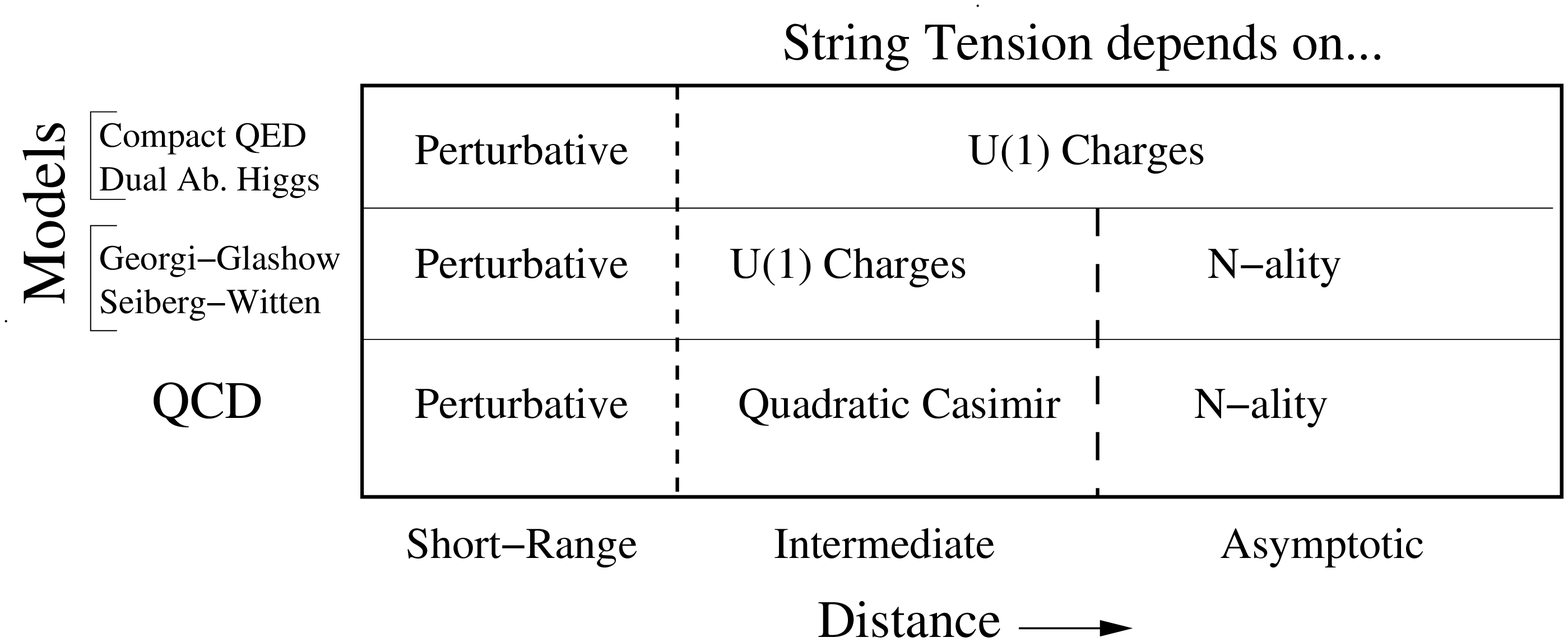}}}
\end{minipage}
\begin{minipage}[t]{16.5 cm}
\caption{Monopole confinement models in SU(N) gauge theories
predict that confinement forces are proportional to $U(1)^{N-1}$ electric
charges.  Center vortex theories predict a dependence only on N-ality.}
\label{reptab}
\end{minipage}
\end{center}
\end{figure}

   For a variety of theories, the dependence of string tension on
group representation is summarized in Fig.\ \ref{reptab}.  In compact $QED_3$
with \emph{no} dynamical matter fields, the
static potential is perturbative at short distances, followed by a
confining region where the string tension is roughly
proportional to the electric charge.  This dependence of string tension
on electric charge, which is characteristic of monopole Coulomb gas
mechanisms, also applies to the dual abelian Higgs model.  In the D=3
Georgi-Glashow
model and Seiberg-Witten models (see below), which both have matter fields
in the adjoint representation,
there is an intermediate distance range
in which string tension is proportional to $U(1)^{N-1}$ electric charge,
followed by an asymptotic regime where string
tension depends on N-ality.  Finally, in QCD, there is no $U(1)^{N-1}$
region at all.  If there were, then adjoint representation Wilson loops would
be dominated by their $U(1)^{N-1}$ charge neutral component, and would have
zero string tension at all distance scales.
The transition is from Casimir scaling at intermediate
distances to N-ality dependence at large distances.  In QCD, there is
no distance interval where the representation-dependence of
string tensions can be directly
attributed to a dual abelian Higgs model, or a monopole Coulomb gas.

\subsection{The Seiberg-Witten Model}

   The important work of Seiberg and Witten on duality and confinement
in supersymmetric gauge theories is beyond the scope of this review,
and the reader is referred to the original article \cite{SW} and also to
ref.\ \cite{SWrevs} for details.  A few comments about their work,
however, are relevant at this point \cite{j3}.

   ${\cal N}=2$ super Yang-Mills theory, like the Georgi-Glashow model,
has a scalar field in the adjoint representation, which can be used to
single out a compact abelian subgroup: U(1) for super Yang-Mills theory,
$U(1)^{N-1}$ for the SU(N) gauge group. Seiberg and Witten
were able to show that there is a point in moduli space at which the
magnetic monopoles of this theory become massless. On adding a soft
supersymmetry breaking term, which reduces the ${\cal N}=2$ supersymmetry
to ${\cal N}=1$, the theory goes into a confining phase, and this transition
is associated with a condensation of the monopole field.  Seiberg and Witten
derive in this case an effective low energy action which is a supersymmetric
generalization of the dual abelian Higgs model.

   However, just as in Polyakov's monopole gas treatment of the D=3
dimensional Georgi-Glashow model, the Seiberg-Witten low-energy effective
action neglects the W-bosons of the theory, on the grounds that they
are massive and should not contribute to long-range physics.  But we have
already seen that the W-bosons are very relevant to long-distance physics;
the N-ality dependence of string tensions $\s_r$ cannot be obtained
without them.  This means that the Seiberg-Witten effective action is
not the whole story at large scales.  This effective action is
obtained keeping only local terms with no more than two derivatives
of the bosonic fields.  It is not obtained by actually integrating out
the massive W-bosons, for otherwise the effective action would have the correct
N-ality dependence built in.  At distance scales on the order of the color
screening length,
$L \approx m_W/\s$, the dual abelian Higgs action arrives at the
wrong N-ality dependence for Wilson loops,
and is simply inadequate to describe the large-scale vacuum fluctuations.
This situation is essentially as depicted in Fig.\ \ref{reptab}: there is
an intermediate range of distances, up to the screening scale, for
which the Seiberg-Witten effective action gives a good account of the
physics.  Beyond that scale, the contributions of virtual W-bosons to
the effective action can no longer be ignored.

\subsection{Monopoles and Vortices}

\begin{figure}[h!]
\begin{center}
\begin{minipage}[t]{8 cm}
\centerline{\scalebox{.5}{\includegraphics{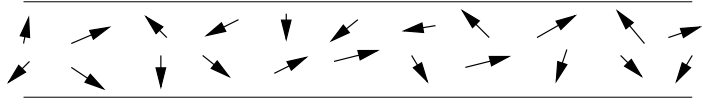}}}
\end{minipage}
\begin{minipage}[t]{16.5 cm}
\caption{Vortex field strength before gauge fixing.  The arrows indicate
direction in color space.}
\label{avort1}
\end{minipage}
\end{center}
\end{figure}

\begin{figure}[h!]
\begin{center}
\begin{minipage}[t]{8 cm}
\centerline{\scalebox{.5}{\includegraphics{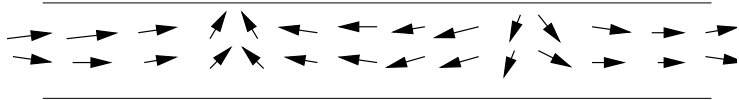}}}
\end{minipage}
\begin{minipage}[t]{16.5 cm}
\caption{Vortex field strength after maximal abelian gauge fixing.  Vortex
strength is mainly in the $\pm \s_3$ direction.}
\label{avort2}
\end{minipage}
\end{center}
\end{figure}

\begin{figure}[h!]
\begin{center}
\begin{minipage}[t]{8 cm}
\centerline{\scalebox{.5}{\includegraphics{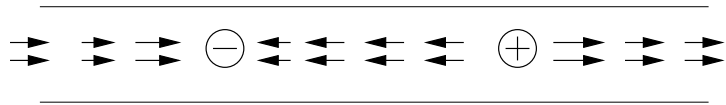}}}
\end{minipage}
\begin{minipage}[t]{16.5 cm}
\caption{Vortex field after abelian projection.}
\label{avort3}
\end{minipage}
\end{center}
\end{figure}

\begin{figure}[h!]
\begin{center}
\begin{minipage}[t]{8 cm}
\centerline{\scalebox{.8}{\includegraphics{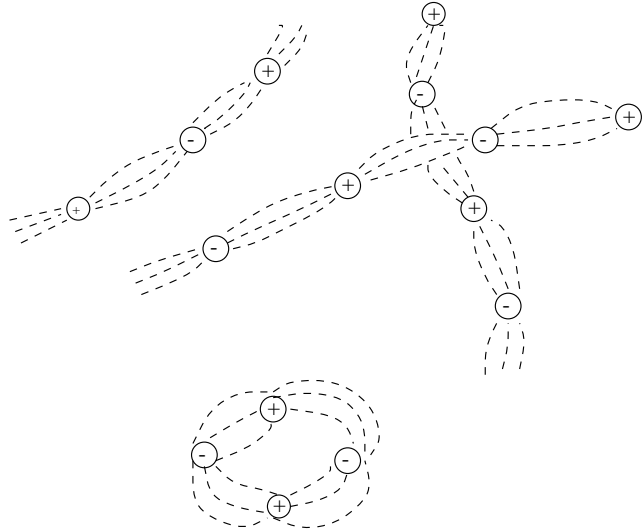}}}
\end{minipage}
\begin{minipage}[t]{16.5 cm}
\caption{Hypothetical collimation of monopole/antimonopole flux into
center vortex tubes on the abelian-projected lattice.}
\label{avort5}
\end{minipage}
\end{center}
\end{figure}

   Let us now consider how center vortices would appear in the abelian
projection (cf.\ refs. \cite{zako,j3}, and also ref.\ \cite{LC}).
We begin with a center vortex at a fixed time, as indicated
schematically in Fig.\ \ref{avort1}.  In the absence of gauge fixing,
the vortex field strength points in arbitrary directions in color space.
Upon fixing to maximal abelian gauge, the vortex field strength tends
to line up, in color space, in the $\pm \s^3$ direction.  But there are
still going to be regions along the vortex where the field rotates
from the $+\s^3$ to the $-\s^3$ direction, as shown in Fig.\ \ref{avort2}.
Upon abelian projection, these transition regions show up as monopoles
or antimonopole (Fig.\ \ref{avort3}).  The end result is that a center
vortex appears, after abelian projection, as a monopole-antimonopole chain,
with a typical vacuum configuration indicated very schematically in
Fig.\ \ref{avort5}.  If this picture is correct, then the $\pm 2\pi$
monopole flux is not distributed symmetrically on the abelian-projected
lattice, as one might expect in a Coulomb gas.  Rather, it is collimated
in units of $\pm \pi$ along the vortex line.  Because of this collimation,
monopole magnetic fields will not affect double-charged abelian Wilson
loop (or any abelian loop with an even number of units of electric charge).

   It is important to see whether this collimation of field strength
along the vortex line, and the tendency of abelian monopoles to line
up along vortex lines in a monopole-antimonopole chain, can actually
be observed in Monte Carlo simulations.  This question was studied
in refs.\ \cite{zako,j3} in SU(2) lattice gauge theory fixed to the
indirect maximal center gauge,
which has the advantage that abelian monopoles can be identified
after the first step of maximal abelian gauge fixing, and center vortices
can be identified after the final center projection.

   We concentrate on static monopoles (monopole current in the time
direction) on a given time slice.  It is found (at $\b=2.4$) that
almost all monopole and antimonopoles in the time slice (97\%, on
average) lie on P-vortex lines.  Along the vortex lines, the
predicted alternation of monopoles with anti-monopoles is observed
in the Monte Carlo data. Moreover, the (gauge-invariant) action
around a monopole cube in a fixed time slice is highly asymmetric:
almost all of the excess action is in the plaquettes pierced by a
P-vortex!

\begin{figure}[h!]
\begin{center}
\begin{minipage}[t]{8 cm}
\centerline{\scalebox{0.5}{\includegraphics{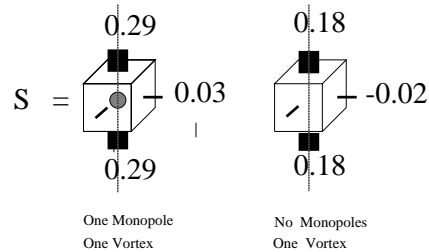}} }
\end{minipage}
\begin{minipage}[t]{16.5 cm}
\caption{Excess plaquette action distribution on a
monopole cube (left figure) pierced by a single P-vortex.  For comparison, the
excess action distribution is also shown for a no-monopole cube (right figure)
pierced by a P-vortex. From Ambj{\o}rn et al., ref. \cite{j3}.}
\label{acube3}
\end{minipage}
\end{center}
\end{figure}

\begin{figure}[h!]
\begin{center}
\begin{minipage}[t]{8 cm}
\centerline{\scalebox{0.5}{\includegraphics{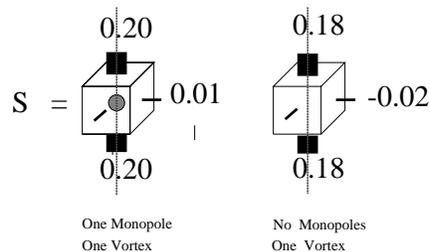}} }
\end{minipage}
\begin{minipage}[t]{16.5 cm}
\caption{Same as Fig.\ \ref{acube3}, but for ``isolated'' static monopoles,
with no nearest-neighbor monopole currents.
From Ambj{\o}rn et al., ref. \cite{j3}.}
\label{acube4}
\end{minipage}
\end{center}
\end{figure}

   The situation found in Monte Carlo simulations (again for the SU(2)
group at $\b=2.4$) is shown in Figs.\ \ref{acube3} and \ref{acube4}.
We consider cubes
on the lattice, at fixed time, containing static monopoles (shaded
circles), and which are also pierced by P-vortex lines (thick strip).
To avoid inhomogeneities due to neighboring monopoles, we also
consider ``isolated'' monopoles with no
nearest-neighbor monopole cubes on the time slice.  Define the excess
action for plaquettes on a monopole cube as
\beq
   S = S_0 -
\Bigl\langle \oh \tr[UUU^\dg U^\dg]_{\mbox{cube plaq}} \Bigr\rangle
\eeq
where $S_0$ is the usual (gauge-invariant) plaquette expectation
value, and the $U$-link variables are the links on the unprojected
lattice.  We can calculate $S$ separately for the cases that the cube
plaquettes are pierced (dark strips), or not pierced (dark lines) by
P-vortices on the corresponding center projected lattice.  What is
found is that the excess action is overwhelmingly concentrated along
the line of the P-vortex (Fig. \ref{acube3}).  Moreover, we can also
calculate the excess action distribution for plaquettes on cubes which
are pierced by a P-vortex, but which do not contain a monopole.  These
are also far above the average plaquette value.  When we look in particular
at isolated monopoles (Fig.\ \ref{acube4}), we find that the
excess action distribution
does not much depend on whether a monopole is, or is not, inside a
cube.  The correlation of excess action is with vortex piercings, not
monopoles.

   It is clear that the gauge-invariant field strength around a
monopole, far from being Coulombic, is instead highly collimated in
the vortex direction.  Similar measurements have been carried out for
cubes which are three or four lattice spacings on a side, and which
contain either a single static monopole, or no monopole current
\cite{j3}.  Instead of excess action, the observable is the fractional
deviation of Wilson loop expectation values, for loops running around
a face of the 3 or 4-cube.  The results are qualitatively similar to
what is seen in Fig.\ \ref{acube4} for 1-cubes: There is a large
effect for faces of the cube pierced by center vortices, and very
little correlation with whether or not the cube also contains a static
monopole.

   Recently the P-plaquette excess-action calculation has been
repeated by Gubarev et al.\ \cite{Gubarev} at a variety of couplings.
These authors confirm that monopole worldlines lie on vortex sheets,
that the excess action is directed along the vortex, and that vortex
density scales.  Further, they report that the excess action $\b S$ is
a lattice-independent constant. This finding can be interpreted as
implying the existence of a sheet of singular action density
in the middle of center vortices in the continuum limit.

   The strong directionality of field strength along vortex lines whether
or not monopoles are present, which is found not just on 1-cubes but
on larger cubes as well \cite{j3},
supports the general picture of vortex confinement, in which confining
flux is collimated (at fixed time) into tube-like structures.  For isolated
monopoles at $\b=2.4$, the
monopole locations do not appear to play a crucial role as far
as the action distribution is concerned.  We take note of an
interesting recent argument by Kovner, Lavelle and McMullan \cite{KLM}, who
point out that in SU(2) lattice gauge theory there are residual
permutation gauge transformations in abelian projection gauges, which
can cause monopole-antimonopole pairs to appear or disappear along
vortex lines.  Those pairs, at least, are certainly gauge artifacts.
Kovner, Lavelle and McMullan also argue that monopole charge is an
ambiguous observable in pure SU(2) lattice gauge theory, due to the absence
of a charge conjugation operator in the theory.

\section{Coulomb Energy, and the Large N Limit}

   In this section I will discuss some
aspects of confinement in Coulomb gauge, and in the large N
limit of SU(N) gauge theory.

\subsection{Confinement in Coulomb Gauge}

    The Yang-Mills Hamiltonian in Coulomb gauge has the following
form:
\beq
H = \oh \int d^3x ~ (\vec{E}^{a,tr} \cdot \vec{E}^{a,tr}
 + \vec{B}^a \cdot \vec{B}^a) + \oh \int d^3x d^3y ~
 j_0^a(x) \V^{ab}(x,y) j_0^b(y)
\eeq
where $j_0^a$ is the zero-component of the conserved color current
in eq.\ \rf{Noether}, $\vec{E}^{a,tr}$ is the transverse color electric field
operator, and
\beq
       \V^{ab}(x,y) = \left[ M^{-1}(-\pa^2)M^{-1}
                      \right]_{x,y}^{ab}
\eeq
where
\beq
         M^{ac} = - \pa_i D_i^{ac}(A) = -\pa^2 \d^{ac} - \e^{abc} A_i^b \pa_i
\eeq
is the Faddeev-Popov operator, and $A_\m^a$ is in
Coulomb gauge.  Sandwiching the $\V$ operator between static color
sources in representation $r$, located at points separated by a distance
$R$, defines the static Coulomb potential
\beq
       V_{coul}^{(r)}(R) = \lla \tr[t_r^a \V^{ab}(0,R) t_r^b] \rra
\label{coul}
\eeq
which is readily seen to be proportional to the quadratic Casimir,
since $\lla \V^{ab} \rra \propto \d_{ab}$.  We see that
the static Coulomb potential automatically satisfies Casimir scaling.
This is not surprising, since the static Coulomb potential can also
be understood as arising from instantaneous one-gluon exchange
in Coulomb gauge.  From here on we will consider the Coulomb potential
for fundamental representation sources in particular.

   There is a scenario for confinement, originally due to Gribov
\cite{Gribov}, and advocated in recent years by Zwanziger
\cite{Dan1}, which makes essential use of Coulomb gauge and
focuses on the static Coulomb potential. The idea is roughly as
follows:  In ``minimal'' Coulomb gauge, the path integral is
restricted to configurations such that the Faddeev-Popov operator
$M$ has only positive eigenvalues.  The boundary of this region in
configuration space, where the Faddeev-Popov operator acquires
negative eigenvalues, is known as the {\bf Gribov Horizon}.  Since
the dimension of configuration space is very large (on the order
of the lattice volume, in a lattice formulation), the bulk of
configurations will be located close to the horizon (just as the
volume measure $r^{d-1} dr$ of a ball in $d$-dimensions is sharply
peaked near the radius of the ball).  Because it is the inverse of
the $M$ operator which appears in the Coulomb potential \rf{coul},
we may expect that the near-zero eigenvalues of this operator,
typical of configurations near the horizon, will enhance the
magnitude of the Coulomb potential, possibly resulting in a
confining potential at long distances.

   As Zwanziger has recently pointed out in ref.\ \cite{Dan2},
the Coulomb potential is, at the very least, an upper bound on the static
quark potential.  The argument is rather simple.  In Coulomb gauge,
we can construct a physical state consisting of very massive
charged sources (``quarks'') separated by a distance $R$
\beq
       |\Psi_{qq} \rangle = q^{\dg \a}(0) q^\a(R)  |\Psi_0 \rangle
\eeq
where $|\Psi_0 \rangle$ is the vacuum state and index $\a$ denotes the
quark color indices.  This is, of course, not the most general physical
state containing two static sources; other states would contain also gluon
operators acting on the ground state.  It
does serve, however, as a trial state from which one can extract the Coulomb
energy. Defining
\beq
       Q(R,t) = q^{\dg \a}(0,t) q^\a(R,t)
\eeq
the energy expectation value of the state $\Psi_{qq}$, above the
vacuum energy, is
\bea
        \E_{qq} &=& { \lla \Psi_0| Q^\dg(R,0) H Q(R,0) |\Psi_0 \rra
               \over  \lla \Psi_0| Q^\dg(R,0)   Q(R,0) |\Psi_0 \rra }
                  - \lla \Psi_0| H |\Psi_0 \rra
\non \\
               &=& E_{se} + V_{coul}(R)
\eea
where $E_{se}$ is the quark self-energy,
and the $R$-dependent part of the energy expectation value, for very massive
sources, can be identified with the instantaneous Coulomb contribution.
Although the quark self-energy is divergent in the
continuum formulation, it is of course regulated on the lattice.

   Next, we note the identity \cite{GH}
\beq
       \E_{qq} = - \lim_{T \ra 0} {d\over dT} \log[G(R,T)]
\eeq
where
\beq
       G(R,T) = \lla Q^\dg(R,T) Q(R,0) \rra
\eeq
is a Euclidean vacuum expectation value.  In Coulomb gauge, as in any
physical gauge, the existence of a transfer matrix implies that
\beq
       G(R,T) = \sum_n c_n \exp[-E_{qq}^{(n)} T]
\eeq
where $E_{qq}^{(n)}$ is the energy (above the vacuum energy) of the n-th
energy eigenstate having a finite overlap with the trial state $\Psi_{qq}$.
From this we have the obvious inequality
\beq
       E_{qq}^{(0)} = - \lim_{T \ra \infty}
                       {d\over dT} \log[G(R,T)] \le \E_{qq}
\eeq
which is just the statement that the
energy expectation value of the trial state $\Psi_{qq}$
is an upper bound on the ground state energy of the static quark-antiquark
system.  This ground state energy defines the usual static potential
\beq
      E_{qq}^{(0)} =  E'_{se} + V(R)
\eeq
If the static potential $V(R)$ is linearly rising, then for sufficiently
large $R$ the static
quark self-energies $E_{se},E'_{se}$ are negligible compared
to $V(R)$, at least on the lattice.  From this it follows that,
asymptotically,
\beq
       V(R) \le V_{coul}(R)
\label{ineq}
\eeq
Therefore, if the static potential in fundamental representation
is confining, the corresponding Coulomb potential
is also confining.  Zwanziger and Cucchieri \cite{Dan3}
have further suggested that the bound is saturated, and that the Coulomb
potential \emph{is} the static confining potential.  If true, this would
be a little puzzling, for how would string-like properties emerge?
If we accept that the confining potential is simply due
to instantaneous one-gluon exchange (equivalent to the Coulomb energy
term in the Hamiltonian), it is a little hard to understand the origin of
roughening, or the L\"{u}scher term.

\begin{figure}[t!]
\begin{center}
\begin{minipage}[t]{8 cm}
\centerline{\scalebox{0.5}{\includegraphics{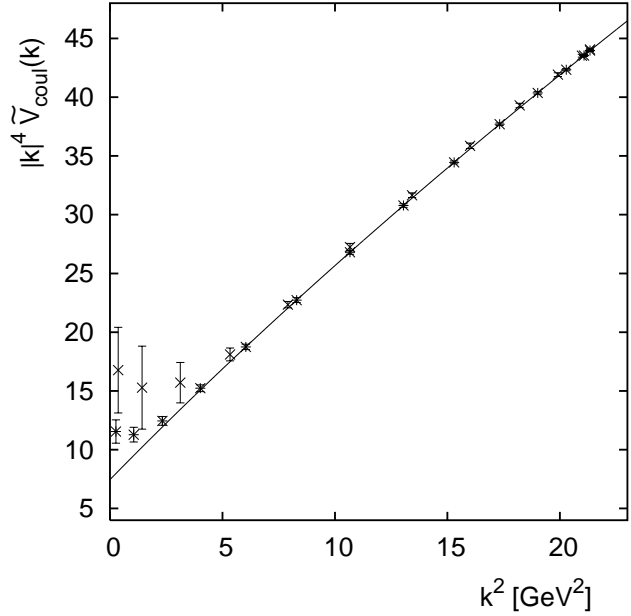}}}
\end{minipage}
\begin{minipage}[t]{16.5 cm}
\caption{Coulomb energy $\times |\bf k|^4$ in momentum space, at
$\b=2.5$ in SU(2) lattice gauge theory.  The solid line is a fit to
eq.\ \rf{danfit}.  From Zwanziger and Cucchieri, ref.\
\cite{Dan3}.}
\label{dan_fig}
\end{minipage}
\end{center}
\end{figure}

    There are a number of different approaches to calculating $V_{coul}(R)$
via lattice Monte Carlo simulations.  One method is to compute the
$D_{00}=\lla A_0 A_0 \rra$ component of the gluon propagator in
Coulomb gauge, using the standard lattice definition
\beq
     A_\m(x) = {1\over 2i}[U_\m(x) - U^\dg_\m(x)]
\eeq
In the continuum formulation, the instantaneous part of $D_{00}(x,t)$
is proportional to the Coulomb energy.
A drawback of this approach is that the modulus $|A_\m|$ of the vector
potential is bounded
in the lattice formulation,
and therefore $D_{00}(R)$ cannot possibly have an asymptotically
linear growth on a large lattice.\footnote{This objection does not apply
in Landau gauge, since the gluon propagator in position space is
non-singular in the infrared \cite{Williams}.  The effect of vortex
removal on Landau gauge propagators, and its possible implications for
confinement, has been investigated recently by Langfeld et al.\
in ref. \cite{Kurt}.}
A better method, advocated recently by Zwanziger and Cucchieri
in \cite{Dan3}, is to invert the lattice-regulated Faddeev-Popov
operator $M$ numerically, in each thermalized lattice configuration,
and from that inverse operator compute the expectation value on the rhs of
eq.\ \rf{coul}.  The result in momentum space $(k=|\vec{k}|)$,
for $k^4 V_{coul}(k)$ at $\b=2.5$ in SU(2) gauge theory, is plotted in
Fig.\ \ref{dan_fig}, together with a best fit to
\beq
       k^4 V_{coul}(k) = A + {B k^2 \over W^2 + \log(1 + k^2/\L^2)}
\label{danfit}
\eeq
For $A \ne 0$, the Coulomb potential has linear confining behavior.
The intercept of the data with the $y$-axis, which gives the value for
$A$, does indeed appear to be non-zero from the data.  According
to ref.\ \cite{Dan3}, the resulting string tension scales, and appears
to nearly saturate the inequality of eq.\ \rf{ineq}.

   A third approach gives a somewhat different result, and also indicates
a strong relationship of the Coulomb energy to the center vortex theory.
On the lattice,
the expectation value $G(R,T)$ is simply a correlator of two timelike
Wilson lines in Coulomb gauge.  Define
\beq
       L(x,T) = U_0(x,0) U_0(x,1) ... U_0(x,T-1)
\eeq
so that
\beq
      G(R,T) =  \langle \tr[ L^\dg(0,T) L(R,T)] \rangle
\eeq
We also define the lattice logarithmic derivative
\beq
      V(R,T) = \log \left[ {G(R,T) \over G(R,T+1)} \right]
\eeq
(with $G(R,0)\equiv 1$).  It follows that, as the continuum limit
is approached at large $\b$,
\beq
       V_{coul}(R) =  V(R,0) - E_{se}
\eeq
is the Coulomb energy, while
\beq
       V(R) = \lim_{T \ra \infty} V(R,T) - E'_{se}
\eeq
is the static quark potential.

\begin{figure}[h!]
\begin{center}
\begin{minipage}[t]{8 cm}
\centerline{\scalebox{1.0}{\includegraphics{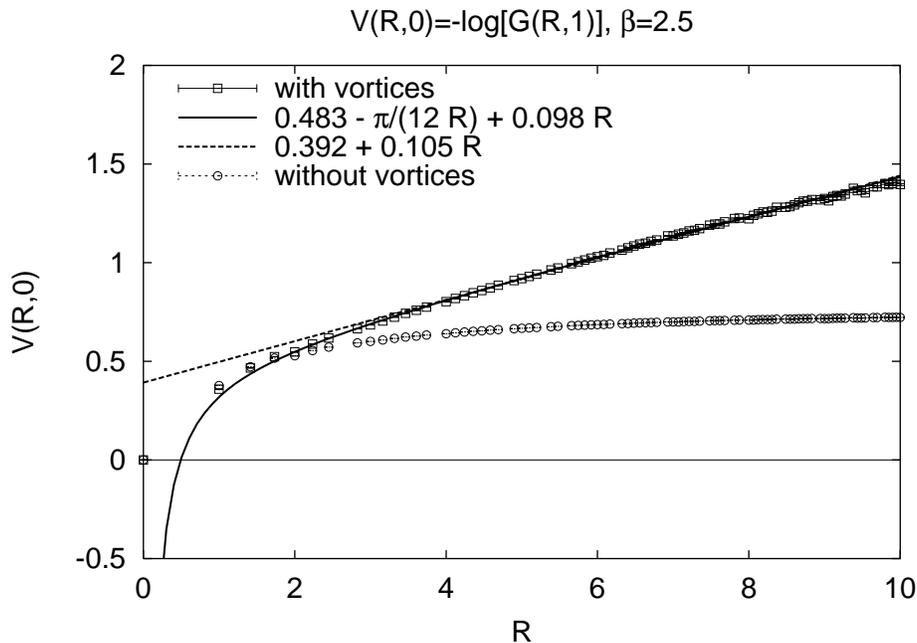}}}
\end{minipage}
\begin{minipage}[t]{16.5 cm}
\caption{$V(R,0)$ at $\b=2.5$. Open squares are derived from
timelike link correlators $G(R,1)$ in Coulomb gauge; open circles are
the same data extracted from lattices with center vortices removed.
From Greensite and Olejn\'{\i}k, ref.\ \cite{GO1}.}
\label{v02p5}
\end{minipage}
\end{center}
\end{figure}

\begin{figure}[h!]
\begin{center}
\begin{minipage}[t]{8 cm}
\centerline{\scalebox{0.75}{\includegraphics{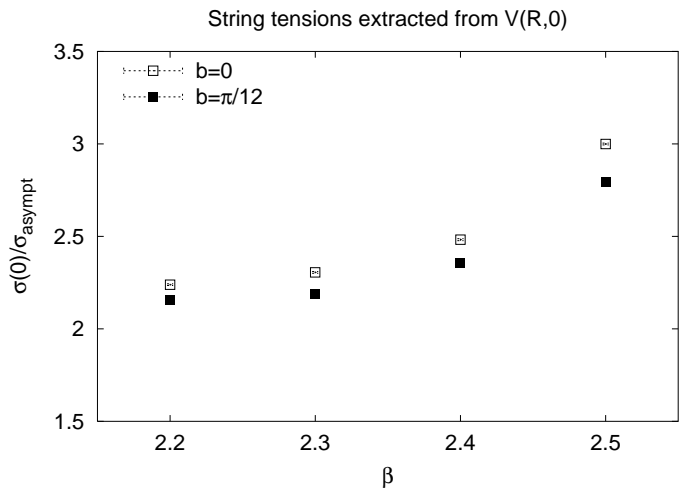}}}
\end{minipage}
\begin{minipage}[t]{16.5 cm}
\caption{Ratio of $\s(0)/\s$, where $\s(0) \ra \s_{Coulomb}$ in the
continuum limit, and $\s$ is the usual asymptotic string tension.
The string tensions are extracted from fits to $a - b/R + \s R$,
with either $b=0$, or the string-motivated value of $b=\pi/12$.
From Greensite and Olejn\'{\i}k, ref.\ \cite{GO1}.}
\label{sigmavsbeta}
\end{minipage}
\end{center}
\end{figure}

   Assuming $V(R,T)$ is asymptotically linear, with string
tension $\s(T)$, we must have
\beq
          \begin{array}{ll}
          \s(0) \ra \s_{coul} & \mbox{large ~} \b \cr
          \s(T) \ra \s        & \mbox{large ~} T\end{array}
\eeq
Then saturation of the confining potential by the Coulomb potential,
$\s \approx \s_{coul}$, implies that $\s(T)$ is approximately $T$-independent
at large $\b$.  This idea has been tested very recently in ref.\
\cite{GO1}.  In Fig.\ \ref{v02p5} we display the result for $V(R,0)$
at $\b=2.5$, together with two different fits to a linear potential
(with and without a L\"{u}scher term, whose presence in this case
is not well motivated).   From the figure, the Coulomb potential is
clearly linear.  However, the associated string tension is almost three times
that of the asymptotic string tension.  In Fig.\ \ref{sigmavsbeta} we show
all of our results for $\s(0)/\s$, where $\s$ is
the accepted value for the asymptotic string tension.
Instead of converging to one, the ratio is actually
increasing in the range of $\b=2.2-2.5$
as $\b$ increases.  This evidence argues that the Coulomb string tension is
substantially greater than the asymptotic string
tension.

   An interesting question, in view of the previous discussion of confinement
mechanisms, is whether center vortices are also responsible for the
Coulomb string tension.  This question can be addressed
by applying the vortex-removal
procedure described in section 6.1 above.  With vortices removed, the
modified configuration is fixed to Coulomb gauge by a standard over-relaxation
method, and $V(R,T)$ is computed in the no-vortex ensemble.  This result,
for $V(R,0)$ at $\b=2.5$ in the no-vortex ensemble, is also
shown in Fig.\ \ref{v02p5} (open circles).  From this figure it is clear that
removing center vortices from the lattice
also removes the Coulomb string tension completely.

\subsection{Confinement at Large N}

   It was pointed out long ago by 't Hooft \cite{tH2} that the Feynman
diagram expansion in SU(N) gauge theories could be organized as a double
expansion in powers $1/g^2N$ and $1/N$.  The leading diagrams
in $1/N$ for, e.g., a scattering process or a fundamental Wilson loop,
are planar diagrams; i.e.\ Feynman diagrams which can be drawn
on a plane surface without having any two propagators cross one another
(except at a vertex).   A similar $1/N$ organization exists in the lattice
strong-coupling expansion.  It is hoped that the case of $N=2$ or $N=3$
can be regarded as a relatively small perturbation of the $N=\infty$
limit.  Perhaps this limit is more tractable analytically; perhaps it
leads us to new insights.

   There is one striking property of the $N=\infty$ limit which seems quite
unlike the situation at $N=2,3$; this is the property of factorization.
Let $O_1$ and $O_2$ be any two gauge-invariant operators.  Then
to leading order in $1/N^2$
\beq
      \lla O_1 O_2 \rra = \lla O_1 \rra \lla O_2 \rra
\eeq
Factorization, applied to Wilson loops, implies that Casimir scaling
is exact in the $N\ra \infty$ limit, as already noted in section 4.2
above.\footnote{One might say that in this limit,
the Casimir scaling region in Fig.\ \ref{reptab}
expands, and the N-ality dependent
region recedes, out to infinite distances.}
But this property also implies, since $\D O^2 = \lla O^2 \rra - \lla O \rra^2$,
that the rms deviation of any gauge-invariant operator from its mean value
vanishes as $N \ra \infty$.  This has an astonishing consequence.  If we
would imagine performing a numerical simulation of SU(N) lattice gauge theory
at some enormous value of $N$, such that all non-leading powers of $N$
could be neglected, then every thermalized configuration would give
almost the same value for any given observable.  This led Witten \cite{Witten}
to propose the idea of a large-N master field; i.e.\ a single configuration
$A_\m^{master}(x)$ in the $N=\infty$ limit with the property that for
any gauge-invariant functional $O[A]$ of the gauge field,
\beq
       \lla O \rra = O[A_\m^{master}]
\label{master}
\eeq
The master field is certainly not unique, since any thermalized configuration
will do.  There exists, in fact, a fairly simple equation whose solution
is known to yield
QCD master field configurations \cite{Marty1}.
Unfortunately, although the equation can be solved perturbatively and
reproduces the planar Feynman diagram results, it seems
to be no easier to solve at the non-perturbative level than any other
formulation of QCD.

\begin{figure}[t!]
\begin{center}
\begin{minipage}[t]{8 cm}
\centerline{\scalebox{0.4}{\includegraphics{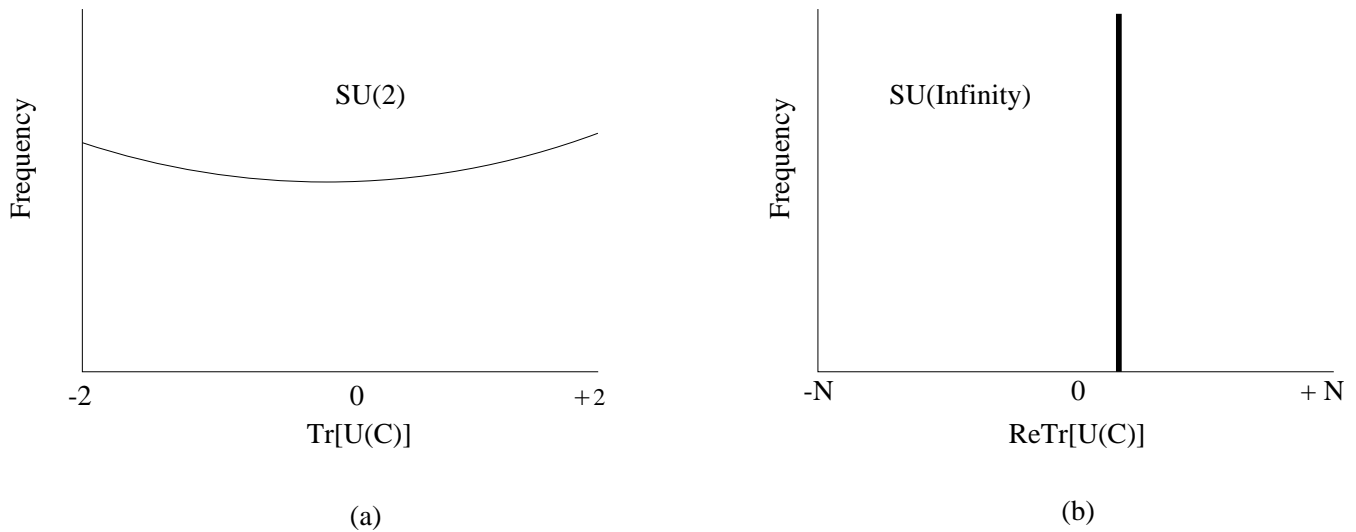}}}
\end{minipage}
\begin{minipage}[t]{16.5 cm}
\caption{Schematic Wilson loop frequency distributions for
a) SU(2), and b) SU($\infty$).  The SU(2) distribution is peaked
at the center elements, and the finite value of Wilson loops is
due to a tiny asymmetry between $\tr(g)$ and $\tr(-g)$. At $N=\infty$,
every evaluation of ReTr$[U(C)]$ yields the same number.}
\label{prob}
\end{minipage}
\end{center}
\end{figure}

   SU(2) and SU(3) lattice gauge theory certainly do not come close
to having the $\D O = 0$ property for Wilson loops.  Consider the
frequency distribution of $\tr[U(C)]$, which
can be obtained in Monte Carlo simulations by plotting a histogram
for the values obtained for $\tr[U(C)]$ around loop $C$ in
thermalized configurations.  In SU(2) lattice gauge theory, for loops
whose extension is comparable to the confinement scale, the probability
distribution is very flat, as sketched in Fig.\ \ref{prob}.  The exponentially
small value of the Wilson loop is obtained in the average not because
the loop holonomy typically has a small trace, but rather because positive and
negative values of the trace
cancel almost entirely in the average over configurations.
In contrast, the frequency distribution at $N=\infty$ would have to be
a delta function.  We must clearly be cautious in extrapolating
$N=\infty$ physics to $N=2$ or $N=3$.

\begin{figure}[h!]
\begin{center}
\begin{minipage}[t]{8 cm}
\centerline{\scalebox{0.40}{\rotatebox{270}
 {\includegraphics{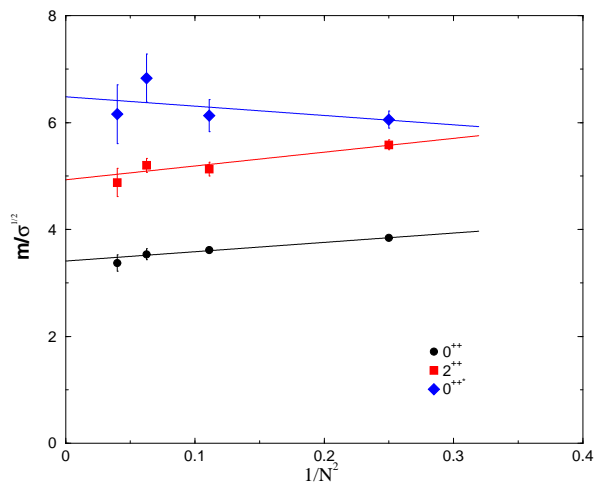}}}}
\end{minipage}
\begin{minipage}[t]{16.5 cm}
\caption{Masses of the lowest lying scalar, tensor, and first excited
scalar glueballs, from lattice simulations with $N=2-5$ colors,
in units of $\sqrt{\s}$.  The $m/\sqrt{\s}$ ratio is plotted
vs $1/N^2$.  From Lucini and Teper, ref.\ \cite{Teper2}.}
\label{sun_glueball}
\end{minipage}
\end{center}
\end{figure}

\begin{figure}[h!]
\begin{center}
\begin{minipage}[t]{8 cm}
\centerline{\scalebox{0.40}{\rotatebox{270}{\includegraphics{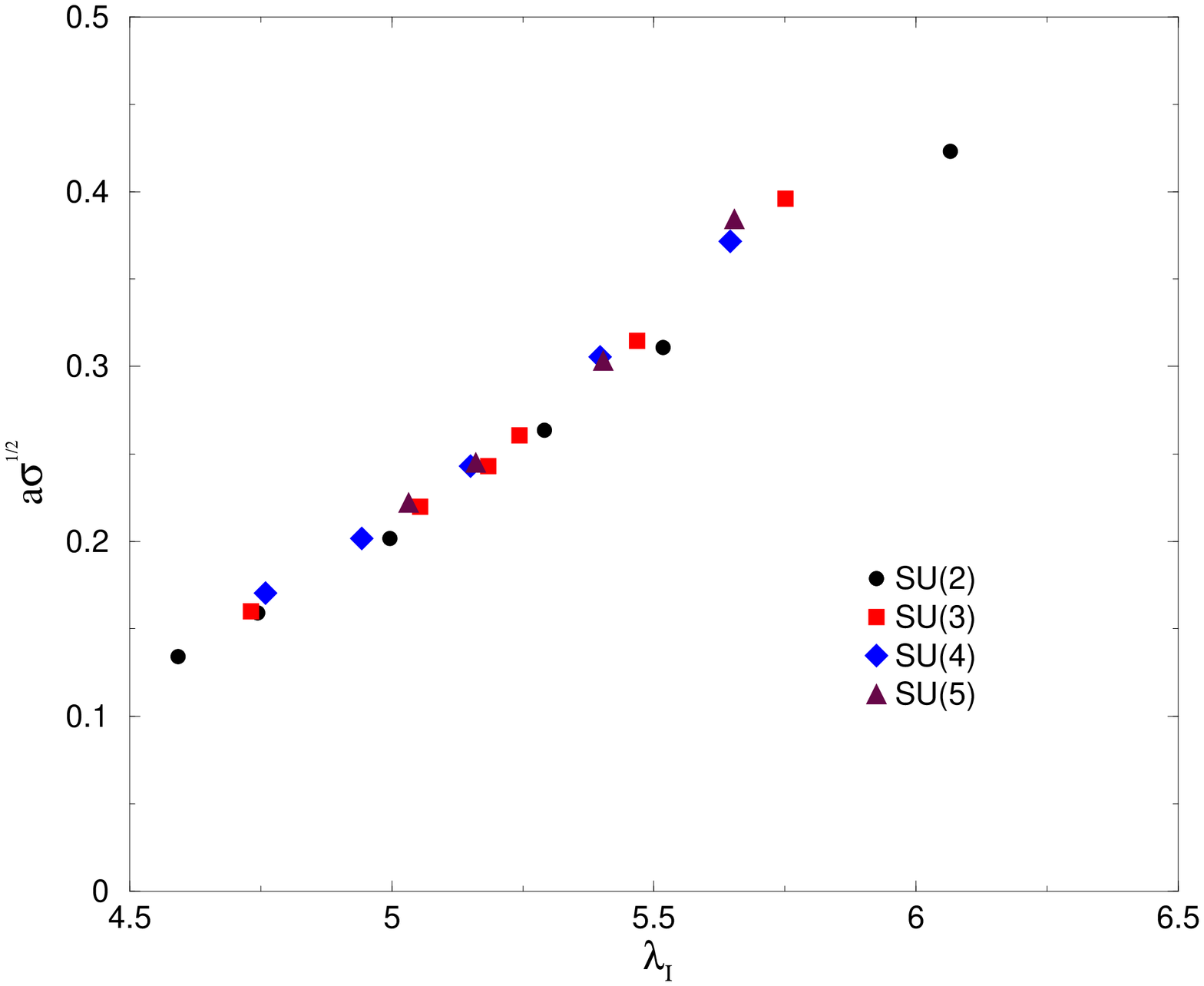}}}}
\end{minipage}
\begin{minipage}[t]{16.5 cm}
\caption{Square root of the lattice string tension vs.\ the (tadpole-improved)
lattice 't Hooft coupling $\l_I=g_I^2 N$.  From
Lucini and Teper, ref.\ \cite{Teper2}.}
\label{lambdaim}
\end{minipage}
\end{center}
\end{figure}

    Nevertheless, in some ways SU(N) gauge theory at small $N$ does
seem to be close to the $N=\infty$ limit, in the sense that ratios of
physical quantities converge rather rapidly as $N$ increases. This fact
can be seen in Fig.\ \ref{sun_glueball}, which
shows the results of Lucini and Teper
\cite{Teper2} for the masses of the lowest-lying $0^{++},~2^{++}$,
and first excited $0^{++}$ glueballs, divided
the square root of the string tension, for $N=2-5$.  Figure \ref{lambdaim}
displays the string tension as a function of the tadpole-improved lattice
version of the 't Hooft coupling $g^2 N$
\beq
        \l_I = g^2_I N = {2 N^2 \over \b_I} =
            {2 N^2 \over \b {1\over N}\lla \mbox{ReTr}[U(p)]\rra }
\eeq
where $\b$ is the usual Wilson action lattice coupling.  At fixed
lattice 't Hooft coupling, there is obviously not much variation
in the string tension for different $N$, indicating that there are only
small deviations from the planar limit, at least for this quantity.

\subsubsection{k-string Tensions at Medium N}

   {\bf k-string tensions} are the asymptotic string tensions
of the lowest dimensional representation with N-ality $k$, represented
by a single column of $k$ boxes in the Young tableau.  Because of
color screening, a $k$-string tension is also the asymptotic string tension
of any color representation of N-ality $k$.

   Douglas and Shenker \cite{DS}, generalizing the work of Seiberg
and Witten \cite{SW} to the SU(N) gauge symmetry, derived
from softly broken ${\cal N} = 2$ SUSY a supersymmetric
version of the dual abelian Higgs model, in which the (dual) $U(1)^{N-1}$
symmetry group is spontaneously broken. In the dual theory there are
distinct chiral superfields $M_n,\widetilde{M}_n,n=1,..,N-1$ with
magnetic monopole components, and vacua at
\beq
      \langle M \widetilde{M} \rangle_n \propto \sin[\pi n/N]
\eeq
This in turn leads to $N-1$ distinct string tensions $T_n$ associated with
each of the $N-1$ magnetic $U(1)$ symmetries, with
\beq
       T_n \propto \sin[\pi n/N]
\eeq
Let $q^a$ be the color components of a quark field in the fundamental
representation of SU(N).  The product of components $q^1 q^2 q^3 ... q^k$,
which has N-ality $k$, is charged only under the k-th
magnetic U(1) factor, and
is neutral with respect to the other factors \cite{Strassler1}.  The
upshot is that the representation of lowest dimension of N-ality $k$,
which is an antisymmetric product of $k$ $q^a$ factors, would have
an asymptotic string tension $\s(k) = T_k$.  This is the Sine Law prediction
for the N-ality dependence of asymptotic string tensions
\beq
 R(k,N) \equiv {\s(k) \over \s(1)} = {\sin(\pi k /N) \over \sin(\pi /N)}
\eeq The same dependence is predicted in a related M-theory
version of supersymmetric gauge theory known as MQCD
\cite{Strassler1}.

   For comparison, suppose that Casimir scaling were exact in
the intermediate distance regime.  The $k$-representations cannot
be screened by gluons, so it may be that for these representations
the string tension in the Casimir scaling regime is also the string
tension in the asymptotic regime.  Under those two assumptions, the
Casimir scaling prediction is that
\beq
        R(k,N) = {k (N-k) \over N-1}
\eeq

   With this motivation, the authors of refs.\ \cite{Teper1} and
\cite{DD} have calculated $R(k,N)$ for the
SU(4) and SU(6) gauge groups, for ordinary non-supersymmetric lattice
gauge theory, with the results shown in Table \ref{Rkn}.  For $D=2+1$
dimensions, Casimir scaling seems preferred, while in $D=3+1$ dimensions
there appears to be remarkable agreement with Sine Law scaling.\footnote{The
results for $D=3+1$ dimensions for the SU(5) gauge group,
reported in ref.\ \cite{Teper1}, are not
in such good agreement with the Sine Law, but the results of ref.\ \cite{DD}
are claimed to be more accurate, so that is what we quote here.}

\begin{table}[h!]
\begin{center}
\begin{minipage}[t]{16.5 cm}
\caption{Casimir Scaling and Sine Law Ratios vs.\ data in 2+1 and 3+1
dimensions.}
\label{Rkn}
\end{minipage}
\end{center}
\begin{center}
\begin{tabular}{|l|cc|c|c|} \hline
R(k,N) &  Cas      & Sin  &  Data, $D=2+1$ \cite{Teper1} &
                             Data, $D=3+1$ \cite{DD} \\ \hline
R(2,4) &  1.33     & 1.41 &  1.355(6)     & 1.403(15)    \\
R(2,6) &  1.60     & 1.73 &  1.616(9)     & 1.72(3)      \\
R(3,6) &  1.80     & 2.00 &  1.808(25)    & 1.99(7)      \\ \hline
\end{tabular}
\end{center}
\end{table}

   The interpretation of these results is a little obscure, in part
because the Sine scaling law is non-universal, and is subject to
$1/N^2$ corrections even in the case of softly broken ${\cal N}=2$ SUSY
\cite{Konishi}.  There is no obvious reason that the Sine Law should hold
very accurately at, say, N=4, especially in a non-supersymmetric theory.
Nor does the supersymmetric case cast much light
on the approximate Casimir scaling of the non-supersymmetric theory,
where the string tension at intermediate distances depends on the
quadratic Casimir of the color representation, rather than
the abelian electric quark charges.

   In terms of the vortex picture, if we assume short-range correlations of 
center vortex fields in a plane (eq.\ \rf{vsrc}), and assume 
either Casimir or Sine-Law scaling for the k-string tensions, then
it is not hard to work out the vortex densities
at a given $N$ \cite{kstring}.  These have no obvious pathologies
in the large-N limit, and in particular do not grow with $N$.

   As $N$ increases, the Casimir and Sine scaling laws converge
to a common limit for the k-string tensions, for $N \gg k$
\beq
       R(k,N) \approx k
\eeq
It is interesting that this limit has a simple explanation in the
context of the center vortex mechanism \cite{kstring}.
Stated briefly, the confining dynamics at asymptotic distances
are described at large $N$ by a center monopole Coulomb gas, in which
the analogue of electric charge is N-ality. We recall
that for any $N>2$ there exist center monopoles as well as
center vortices; these monopoles are the joining point of
N $k=1$ vortices, and go over to U(1) monopoles in the $N\ra \infty$ limit.
The $k$-string tension in a monopole Coulomb gas
is proportional to the electric charge, which in this case is $k$,
because there are $k$ independent flux tubes
stretching between the quark and antiquark.

\subsubsection{The QCD String at Large N}

\begin{figure}[t!]
\begin{center}
\begin{minipage}[t]{8 cm}
\centerline{\scalebox{0.45}{\includegraphics{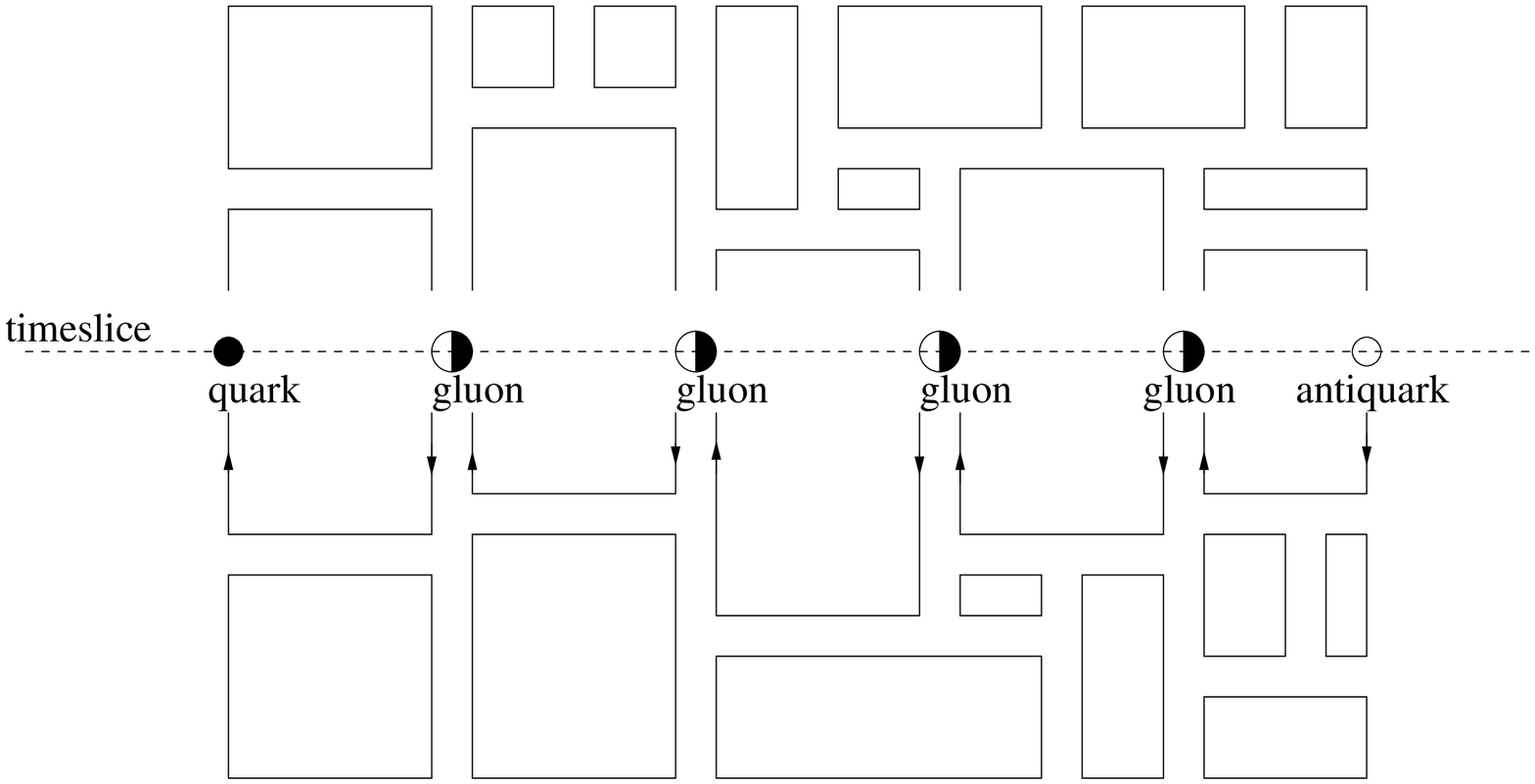}}}
\end{minipage}
\begin{minipage}[t]{16.5 cm}
\caption{The gluon chain as a time slice of a planar diagram
(shown here in double-line notation).  The solid (open) hemisphere
denotes a quark (antiquark) color index.}
\label{planar}
\end{minipage}
\end{center}
\end{figure}

   One of the reasons for being interested in the large N limit is that
this limit suggests a string description of the QCD flux tube.  The
possible interpretation of high-order planar diagrams as string worldsheets,
as indicated in Fig.\ \ref{planar},
was pointed out by 't Hooft in his seminal article on large N gauge theory
\cite{tH2}.  In more recent years, the AdS/CFT conjecture \cite{AdS}
has actually provided a precise string representation of Wilson loop
expectation values in the $N=\infty$ limit, at least
for large 't Hooft couplings $g^2 N \gg 1$.
On the other hand, the string of the
AdS/CFT approach is not quite the same thing as the QCD flux tube.  For
one thing, the AdS/CFT string lives in 10 dimensions, and for another,
there is also an AdS/CFT string representation for Wilson loops in the
non-confining case of unbroken ${\cal N}=4$ supersymmetric gauge theory, where
there is no flux tube at all.

   Returning to the planar diagram-as-worldsheet idea, note that a
time-slice of a high-order planar diagram for a  Wilson loop reveals a
sequence of gluons, with each interacting only with its nearest neighbors
in the diagram (Fig.\ \ref{planar}).  This picture suggests that the
QCD string might be regarded, in some gauge, as a ``chain'' of gluons,
with each gluon held in place by its attraction to its nearest neighbors
in the chain.  The linear potential in this {\bf gluon chain model}
comes about in the following way:  As heavy quarks separate, we expect that
at some point the interaction action energy increases rapidly due to the
running coupling.  Eventually, it becomes energetically favorable to insert
a gluon between the quarks, to reduce the effective color charge
separation. This fits in nicely with the result of ref.\ \cite{GO1},
where it is found that the Coulomb force of a quark-antiquark state, with
no constituent gluons, is much higher than the accepted asymptotic force.
As the quarks continue to separate, the process repeats,
and we end up with a chain of gluons.  The average
gluon separation $R$ along the axis joining the quarks is fixed, regardless of
the quark separation $L$, and the total energy of the chain is just the
energy per gluon times the number of gluons in the chain, i.e.\
\beq
       E_{chain} \approx n_{gluons} E_{gluon}
            = {E_{gluon} \over R} L = \s L
\eeq
In this picture, the linear growth in the number of gluons in the chain
is at the heart of the long-range static potential.

   The gluon-chain picture, originally proposed many years ago
by Thorn \cite{Thorn_chain} and the author \cite{Me_chain,GH}, and recently
revived in ref.\ \cite{gchain}, has quite a number of nice features.
For one thing, because of its large-N origin, the gluon chain model naturally
accounts for Casimir scaling in the large N limit.  For example, the string
tension in the adjoint representation is twice that of the fundamental
representation simply because there are two gluon chains stretching between
sources in the adjoint representation. But also the correct
N-ality dependence is obtained by ($1/N^2$ suppressed) string-breaking
interactions.   As discussed in detail
in ref.\ \cite{gchain}, the model also naturally accounts for the logarithmic
growth of the QCD flux tube with quark separation (roughening), and the
existence of a L\"{u}scher term.

   The gluon chain model is a representation of the QCD flux tube
in terms of particle (i.e.\ gluon) excitations, and is in some sense
``dual'' to the description of the QCD vacuum in terms of field
fluctuations (e.g.\ vortices and monopoles).  It is hoped that
this representation of the QCD string, in which gluon separations
(and therefore effective couplings) are bounded, may lend itself to
variational treatments of the QCD string and low-lying glueball states,
as outlined in ref.\ \cite{gchain}.  Another possible direction for
analytic work is the
approach initiated by Bardakci and Thorn \cite{BK}, in which planar diagrams
in light-cone gauge can be explicitly reformulated as string amplitudes.
In Coulomb gauge, the gluon-chain picture lends itself to tests via
Monte Carlo simulations, some of which were reported long
ago in the second article of ref.\ \cite{Me_chain}.

\section{Conclusions}

   I have emphasized center symmetry in this article
because there are good reasons to believe that center symmetry
is important.  To review briefly: The confinement and
deconfinement phases at finite
temperature are the symmetric and the spontaneously broken phases,
respectively, of a global $Z_N$ center symmetry.  The existence of a
finite string tension is related, by rigorous
arguments, to the behavior of the center vortex free energy.  Even in theories
with a trivial center subgroup, such as SO(3) lattice gauge theory, the vortex
free energy is a good order parameter.  When global $Z_N$ center
symmetry is explicitly broken by scalar fields, the phase transition from
a Higgs to a distinct confinement phase is lost.
Separate Higgs and confinement phases can exist only when the
scalar fields are in zero N-ality representations, preserving the
global center symmetry of the Lagrangian.  Finally, the asymptotic string
tension of static quarks depends on their color charge only through the
transformation properties of the quarks under the center subgroup.

   In view of these facts, a confinement mechanism based on center vortices
would seem to be the natural way of realizing global
center symmetry in the vacuum state.  In recent years it has become possible
to identify vortex locations in lattices generated by Monte
Carlo simulations, via center projection in an adjoint gauge.
Identification of objects in the vacuum by a gauge-fixing
procedure is always subject to ambiguities, but these worries can be at least
ameliorated by demonstrating the correlation of vortices with gauge-invariant
observables.  In particular: The number of vortices mod 2,
linked to a large Wilson loop in
SU(2) lattice gauge theory, is found to be strongly
correlated with the sign of the Wilson loop.
The vortex surface is associated with an average (gauge-invariant)
plaquette action which is well in excess of the vacuum average.
Removing vortices from unprojected lattice configurations
removes the string tension,
removes chiral symmetry breaking, and sends every configuration to the
trivial topological sector.  Apart from these correlations with gauge
invariant observables, it is found that vortices closed by lattice periodicity
in time explain the existence of a spatial string
tension across the deconfinement transition.  The density of center
vortices scales according to asymptotic freedom, and the string tension
of the projected lattice accounts, pretty nearly, for the accepted
asymptotic string tension.

    The abelian monopole theory, based on abelian projection, can also
claim some numerical success. Monte Carlo investigations of vortices
and monopoles should not be irreconcilable; I believe that the
relevant numerical fact is that abelian monopoles lie along vortex
lines in a monopole-antimonopole chain, as discussed in section 7.  In
general, abelian monopole mechanisms are most attractive
in models which contain an adjoint representation scalar field in the
Lagrangian, and therefore naturally single out a unique abelian
$U(1)^{N-1}$ subgroup.  QCD, on the other hand, does not single out a
unique abelian-projection subgroup, and the Casimir scaling of string
tensions found at intermediate distances, together with N-ality
dependence asymptotically, argue against the idea of an underlying
$U(1)^{N-1}$ mechanism.  It is possible, however, that monopole
worldlines on vortex sheets do play a role in the generation of
topological charge.

   We have some pieces of the puzzle, but they are not all in place.
It seems very likely that center symmetry and center vortices are an
important part of the picture.  But despite the numerical evidence
obtained in recent years, and despite the input from supersymmetry and
M-theory, quark confinement in QCD is imperfectly understood, and an
actual proof is elusive.  The confinement problem is still open, and
remains a major intellectual challenge in our field.

\vspace{33pt}

\ni {\Large \bf Acknowledgements}

\bigskip

   I would like to thank my co-workers, Manfried Faber and {\v S}tefan
Olejn\'{\i}k, for countless discussions about quark confinement, and
for a long and productive collaboration. This work is supported in
part by the U.S. Department of Energy under Grant No.\
DE-FG03-92ER40711.

\end{document}